\documentclass[11pt, a4paper]{article}
\usepackage{jheppub}

\usepackage{amsmath, amsfonts, amssymb}
\usepackage{slashed}
\usepackage{tensind}
\usepackage{mathrsfs}
\usepackage{enumerate}
\usepackage{booktabs}
\usepackage{multirow}
\usepackage{hyperref}

\newcommand{\otoprule}{\midrule[\heavyrulewidth]}

\newcommand{\cbottomrule}{\\[.5ex]\specialrule{\heavyrulewidth}{0ex}{0ex}}
\def\tablespacings{\renewcommand{\arraystretch}{1.2}\addtolength{\tabcolsep}{0pt}}

\newcommand{\ii}{i}

\newcommand{\bbR}{{\mathbb{R}}}
\newcommand{\bbC}{{\mathbb{C}}}
\newcommand{\bbZ}{{\mathbb{Z}}}

\newcommand{\loco}{\vert}

\newcommand{\trQ}{\text{\tiny Q}}
\newcommand{\trS}{\text{\tiny S}}

\newcommand {\cA}{{\cal A}}

\newcommand {\cC}{{\cal C}}
\newcommand {\cD}{{\cal D}}
\newcommand {\cE}{{\cal E}}
\newcommand {\cF}{{\cal F}}

\newcommand {\cJ}{{\cal J}}
\newcommand {\cK}{{\cal K}}
\newcommand {\cL}{{\cal L}}
\newcommand {\cM}{{\cal M}}
\newcommand {\cN}{{\cal N}}
\newcommand {\cO}{{\cal O}}

\newcommand {\cR}{{\cal R}}

\newcommand {\cV}{{\cal V}}

\newcommand {\cX}{{\cal X}}

\newcommand {\cZ}{{\cal Z}}

\def\a{\alpha}
\def\b{\beta}

\def\d{\delta}
\def\e{\epsilon}

\def\g{\gamma}

\def\l{\lambda}

\def\q{\theta}

\def\s{\sigma}

\def\u{\upsilon}

\def\z{\zeta}

\def\L{\Lambda}

\def\U{\Upsilon}

\newcommand{\ra}{{\mathrm a}}
\newcommand{\rb}{{\mathrm b}}
\newcommand{\rc}{{\mathrm c}}
\newcommand{\rd}{{\mathrm d}}

\newcommand{\ri}{{\mathrm i}}

\newcommand{\dalpha}{{\dot{\alpha}}}
\newcommand{\dbeta}{{\dot{\beta}}}
\newcommand{\dgamma}{{\dot{\gamma}}}

\newcommand{\dmu}{{\dot{\mu}}}

\newcommand{\1}{{\underline{1}}}
\newcommand{\2}{{\underline{2}}}

\newcommand{\veps}{\varepsilon}
\newcommand{\eps}{{\epsilon}}
\newcommand{\eol}{\notag \\}

\newcommand{\bsigma}{\bar{\sigma}}

\newcommand{\pa}{\partial}

\newcommand{\ccA}{\mathbb A}
\newcommand{\newG}{\cZ}

\def\kp{k_{\text{\tiny +}}}
\def\km{k_{\text{\tiny $-$}}}

\def\SL(#1){\ensuremath{\mathrm{SL}(#1)}}
\def\GL(#1){\ensuremath{\mathrm{GL}(#1)}}
\def\Sp(#1){\ensuremath{\mathrm{Sp}(#1)}}
\def\OSp(#1){\ensuremath{\mathrm{OSp}(#1)}}
\def\USp(#1){\ensuremath{\mathrm{USp}(#1)}}
\def\SU(#1){\ensuremath{\mathrm{SU}(#1)}}
\def\SUs(#1){\ensuremath{\mathrm{SU}^*(#1)}}
\def\U(#1){\ensuremath{\mathrm{U}(#1)}}
\def\SO(#1){\ensuremath{\mathrm{SO}(#1)}}
\def\ISO(#1){\ensuremath{\mathrm{ISO}(#1)}}

\newenvironment{formula}
{\begin{equation}\begin{aligned}}
{\end{aligned}\end{equation}\ignorespacesafterend}

\numberwithin{equation}{section}

\makeatletter
\g@addto@macro\bfseries{\boldmath}
\makeatother

\title{\boldmath{}Rigid 4D $\cN\!\!=\!2$ supersymmetric backgrounds\\and actions}
\author{Daniel Butter,}
\author{Gianluca Inverso,}
\author{and Ivano Lodato}
\affiliation{Nikhef Theory Group, \\
Science Park 105, 1098 XG Amsterdam, The Netherlands}
\emailAdd{dbutter@nikhef.nl}
\emailAdd{g.inverso@nikhef.nl}
\emailAdd{ilodato@nikhef.nl}

\preprint{Nikhef-2015-020}

\arxivnumber{1505.03500}

\abstract{We classify all $\cN=2$ rigid supersymmetric backgrounds
in four dimensions with both Lorentzian and Euclidean signature that preserve
eight real supercharges, up to discrete identifications.
Among the backgrounds we find specific warpings of $S^3\times\bbR$ and AdS$_3\times\bbR$, 
AdS$_2\times S^2$ and $H^2\times S^2$ with generic radii, and some more exotic geometries.
We provide the generic two-derivative rigid vector and hypermultiplet actions
and analyze the conditions imposed on the special K\"ahler and hyperk\"ahler target
spaces.
}

\begin{document}
\maketitle

\section{Introduction}
In the last few years, there has been a great deal of interest in supersymmetric field
theories on rigid curved backgrounds, beginning with the seminal work of
Pestun \cite{Pestun}. These efforts have exploited the principle of supersymmetric
localization to evaluate path integrals and compute certain supersymmetric observables
in various rigid backgrounds.

A systematic approach to addressing such curved spaces at the component level was initiated by
Festuccia and Seiberg \cite{FS}. Taking the point of view that a rigid supersymmetric
theory could be understood as a supergravity theory with the metric and other bosonic
components frozen to some background values, they
investigated the conditions required in both $4D$ $\cN=1$ old minimal and new minimal
supergravities so that four linear independent rigid supersymmetries existed.
Other aspects, such as the weaker requirements imposed by fewer supercharges in both signatures,
were addressed in later work \cite{Jia:2011hw, KTZ:SCSHolo, DFS:ECS, CKMTZ:SLCSHolo, Liu:2012bi, Samtleben:2012gy, DF:ECS2}.
In the case of extended $\cN=2$ supersymmetry in four dimensions,
conditions required for a single supercharge were analyzed
by Gupta and Murthy \cite{GM:AllSols} and by Klare and Zaffaroni \cite{KZ:ExtSUSY}.
The analysis of \cite{KZ:ExtSUSY} determined the main geometric criterion in either
Euclidean or Lorentzian signature: the spacetime must admit a conformal Killing
vector. In the presence of such a vector, one supercharge may be kept by turning on
background values for the $R$-symmetry gauge fields.

Our goal in this paper is to perform a complementary analysis to that of
\cite{GM:AllSols, KZ:ExtSUSY}. First, we derive the conditions imposed by requiring
\emph{full} $\cN=2$ supersymmetry --
eight linearly independent Killing spinors -- in both Lorentzian
and Euclidean signatures and classify the possible smooth backgrounds up to discrete
identifications.
Second, we construct the general vector and hypermultiplet actions on such
spaces and find the conditions on the allowed target spaces.

Our analysis, some of which appeared in a different context in \cite{BdWL},
leads to several interesting possibilities.
In addition to the geometries allowed in
the $\cN=1$ case \cite{FS} --
$\rm AdS_4$, $\mathbb R \times S^3$, ${\rm AdS}_3 \times \bbR$, and an Hpp-wave
arising as a Penrose limit of the last two cases --
we also find several backgrounds that support $\cN=2$ SUSY
with no admissible truncation to $\cN=1$.
The options in Lorentzian signature are summarized in Table \ref{tab:LBacks}
of Section~\ref{sec:LBacks}.
In brief, they involve the cases familiar from $\cN=1$ -- with the $S^3$, $\rm AdS_3$
and Hpp-wave admitting two alternative $\cN=2$ supersymmetry algebras with differing
$R$-symmetry groups -- and several cases requiring extended supersymmetry:
\begin{itemize}
\item a squashed $\mathbb R \times S^3$,
\item a timelike stretched, spacelike squashed, or null warped ${\rm AdS}_3 \times \bbR$,
\item a warped $S^3 \times \mathbb R$ where the $S^1$ fiber of $S^3$ is either timelike or null,
\item ${\rm AdS}_2 \times S^2$ with generic radii, with two different SUSY realizations
for each choice of radii,
\end{itemize}
as well as an Heis$_3\times\bbR$ space and Hpp-wave variants where the background fields become null.
(Some of these correspond to Penrose limits of other cases.)
Each of the resulting supersymmetry algebras can be identified as a massive
deformation of the super-Poincar\'e algebra, and indeed each
possesses a supercoset structure permitting the straightforward construction
of each of the Killing spinors, which we compute explicitly.\footnote{It is should be emphasized that
the relation between extended supersymmetry and at least some of these spaces is already
known. For example, the $\cN=2$ supersymmetry algebra on one of the unsquashed
$\bbR \times S^3$ cases was discussed in \cite{Sen:RS3}. More well-known is
the case of ${\rm AdS}_2 \times S^2$, related to the
one-parameter superalgebra $\rm D(2,1; \alpha)$ \cite{BILS:AdS};
this latter case includes for $\alpha=-1$
the Bertotti-Robinson spacetime relevant for the near
horizon geometry of BPS black holes.} 

The options in Euclidean signature are summarized in Table \ref{tab:EBacks}
of Section~\ref{sec:EBacks}. In addition to $S^4$, $H^4$, a two-sheeted $H^3\times\bbR$, and $S^3 \times \mathbb R$,
we find several geometries where extended supersymmetry plays a major role:
\begin{itemize}
\item a squashed or stretched $S^3 \times \mathbb R$,
\item a Heis$_3\times\bbR$ group manifold,
\item a warped $H^3 \times \mathbb R$, where the hyperboloid corresponds to $AdS_3$ spacetime with a Euclidean metric
\item $H^2 \times S^2$ with generic radii and two different SUSY realizations.
\end{itemize}
Aside from these, we find the possibility of flat Euclidean spaces where the
left-handed (or right-handed) supercharges are deformed. These include as particular
cases the full BPS limits of the Euclidean $\Omega$-background (corresponding to
$\eps_1 = \pm \eps_2$).

Because the spacetimes we discuss retain eight rigid supercharges, it is possible to
construct rigid $\cN=2$ superspaces for each. In fact, this will be the principle guiding
their classification. We follow the approach laid down by Kuzenko et al. in
\cite{Kuzenko:SCS, KNT-M:5DSCS, Kuzenko:SCSReview}, which applies the analysis of general (conformal)
isometries of curved superspaces \cite{BK} to geometries where full supersymmetry is
maintained. The presence of a rigid superspace enables the explicit construction of the component
Lagrangians for vector and hypermultiplets just as in a
Minkowski background. We present these
actions in their general form, applicable to any of the rigid $\cN=2$ backgrounds,
and find the constraints on the special K\"ahler and hyperk\"ahler target spaces
imposed by rigid supersymmetry. We also give the constraints on the
supersymmetric moduli spaces in these backgrounds and comment on how they differ
from the flat case.

The paper is laid out as follows. In Section~\ref{sec:L_SUSY}, we motivate and discuss
the general rigid Lorentzian $\cN=2$ supersymmetry algebra. The rigid backgrounds
allowed by this algebra are analyzed in Section~\ref{sec:LBacks}, while the corresponding
vector multiplet and hypermultiplet actions are given in Section~\ref{sec:L_Actions}.
In particular, we give the $\cN=2^*$ action in a general rigid Lorentzian background.
In Sections \ref{sec:E_SUSY} and \ref{sec:EBacks} we give the Euclidean supersymmetry
algebra and the possible rigid backgrounds.
The corresponding Euclidean actions are discussed in Section~\ref{sec:E_Actions}.

There are four technical appendices. Our conventions are discussed in Appendix \ref{App:Conv}.
The general action principles in rigid superspace are summarized in Appendix \ref{App:Actions}.
Explicit expressions for the geometric data of the rigid backgrounds including Killing spinors,
vielbeins, and background fields are provided in Appendices \ref{app:LBack} and \ref{app:EBack}.

\section{The general rigid Lorentzian supersymmetry algebra}\label{sec:L_SUSY}
We begin this section by describing the construction of the general rigid Lorentzian
supersymmetry algebra, which arises by freezing one of the $\cN=2$ supergravities.
As discussed in the introduction,
any off-shell supergravity corresponds to conformal supergravity coupled to
some compensating multiplet whose lowest component plays the role of the Planck
mass.

It helps to review the $\cN=1$ case.
As is well-known, in $\cN=1$ supergravity the simplest compensators
are a chiral multiplet or a linear multiplet, leading respectively
to old and new minimal Poincar\'e supergravity both with $12+12$ degrees
of freedom. Other options include a complex linear multiplet
(giving $20+20$), an unconstrained real superfield (giving $16+16$),
or an unconstrained complex superfield (giving $24+24$).
Each option eliminates the dilatation and $S$-supersymmetry and
several also eliminate the $R$-symmetry.\footnote{In
matter-coupled $\cN=1$ supergravities, the most natural description involves a
composite compensating multiplet corresponding to the K\"ahler cone from which the
Hodge-K\"ahler manifold descends.}

The $\cN=1$ conformal Killing spinor equation is given as
\begin{align}
(\delta_{\trQ} + \delta_{\trS}) \psi_{m \alpha} = 2 \cD_m \xi_\alpha + 2 i (\sigma_m \bar \eta)_\alpha = 0~.
\end{align}
$\xi_\alpha$ and $\eta_\alpha$ are respectively the local $Q$ and $S$-supersymmetry parameters,
$\cD_m$ carries the $R$-symmetry, dilatation, and spin connections,
and any solution $\xi$ of this equation is called a conformal
Killing spinor. Let $\Omega$ be a nowhere vanishing conformal compensator
of Weyl weight two, so that the physical Weyl-invariant metric is $\Omega \,g_{mn}$.
The lowest fermion $\psi_\alpha$ of $\Omega$
plays the role of an $S$-supersymmetry compensator.
The $S$-invariant gravitino is
$\psi_{m\alpha} + \frac{i}{2} (\sigma_m \bar \psi)_\alpha$.
Taking the $S$-supersymmetry gauge where $\psi_\alpha = 0$,
the Weyl gauge $\Omega = 1$, and the $K$-gauge where the dilatation connection vanishes,
then one finds
\begin{align}
(\delta_{\trQ} + \delta_{\trS}) \psi_\alpha = 4 \xi_\alpha \bar R - 2 G^b (\sigma_b \bar \xi)_\alpha 
	- 4 \eta_\alpha
\end{align}
where $\bar R$ and $G_a$ correspond to auxiliary components of $\Omega$ at the $\q^2$ level,
normalized to match the conventions of \cite{WB}.
By solving for $\eta$, one finds the Killing spinor equation of
$\U(1)$ supergravity,
\begin{align}
\cD_m \xi_\alpha = - i (\sigma_m \bar \xi)_\alpha R - \frac{i}{2} G^b (\sigma_m \bsigma_b \xi)_\alpha~.
\end{align}
Old minimal supergravity arises from
choosing $\Omega = \Phi \bar \Phi$ for a chiral compensator $\Phi$: then
the $\U(1)_R$ connection is fixed to $G_a$ after imposing the Weyl-U(1) gauge
$\Phi = 1$. Conversely, if we choose $\Omega = L$ for a
linear multiplet compensator, $R$ vanishes and $G_a$ is related to the dual
field strength of the two-form potential within $L$.
This is new minimal supergravity. However, it is possible to work purely
with $\rm U(1)$ supergravity, which simultaneously encompasses both minimal possibilities
while allowing more general supergeometries.

A corresponding story holds for $\cN=2$ supergravity.
The conformal Killing spinor equation is\footnote{We have relabelled
the auxiliary field $T_{ab}^\pm$ as $4\,W_{ab}^\pm$.}
\begin{align}
(\delta_{\trQ} + \delta_{\trS}) \psi_{m \alpha i}
	= 2 \cD_m \xi_{\alpha i} - i  W_{mn}^- (\sigma^n \bar\xi_i)_\alpha
	+ 2 i (\sigma_m \bar \eta_i)_\alpha = 0~.
\end{align}
Introducing a generic real compensator $\Omega$ of dimension two and
performing the analogous conformal gauge-fixings results in\footnote{To simplify the resulting
supergravity algebra, \label{foot:RConns}
we have redefined the $\U(1)_R$ and $\SU(2)_R$ connections as
$A_m \rightarrow A_m +  G_m$ and
$\cV_m{}^i{}_j \rightarrow \cV_m{}^i{}_j + 2 G_m{}^i{}_j$.}
\begin{align}\label{eq:KillingSpinor}
\cD_m \xi_{\alpha i} =
	\frac{i}{2} \bar S_{ij} (\sigma_m \bar \xi^j)_\alpha
	- \frac{i}{2} (Y_{mn}^+ - W_{mn}^-) (\sigma^n \bar\xi_i)_\alpha
	- 2i G^n (\sigma_{nm} \xi_i)_\alpha
	+ G^n{}^j{}_i (\sigma_n \bsigma_m \xi_j)_\alpha\,.
\end{align}
We have introduced $\bar S_{ij}$, $G_a$, $G_a{}^i{}_j$, and $Y_{ab}^+$
corresponding to $\q^2$ components of the real compensator $\Omega$.

The unconstrained superfield $\Omega$ corresponds to a ``maximal''
supergravity with $152+152$ degrees of freedom: we refer to this option
as $\U(2)$ supergravity because it retains the full $R$-symmetry structure group.
The obvious ``minimal'' choices are a vector multiplet or a
tensor multiplet, both leading to a $32+32$ supergravity
multiplet.
The lowest component $X$ of a vector multiplet carries $\U(1)_R$ charge,
breaking the $R$-symmetry $\U(2)_R$ to $\SU(2)_R$.
Its graviphoton field strength $F_{ab}$ and pseudoreal auxiliary field $Y^{ij}$
generate $Y_{ab}^\pm$ and $S^{ij} = \bar S^{ij}$, while the auxiliary
$G_a{}^i{}_j$ vanishes and $G_a$ is related (as with the $\cN=1$ chiral multiplet)
to the Higgsed $\U(1)_R$ connection.
The tensor multiplet is a bit more interesting. Its lowest component $L^{ij}$
fixes dilatations and breaks $\SU(2)_R$ to $\SO(2)_R$. 
Its three-form field strength $H_{abc}$ and complex auxiliary scalar $F$ contribute
respectively to $G_a{}^{ij}$ and $S^{ij}$ via $G_a{}^{ij} \sim \eps_{abcd} H^{bcd} \, L^{ij}$
and $S^{ij} \sim F \, L^{ij}$, while $Y_{ab}^\pm$ and $G_a$ vanish.
However, it will be more convenient for us to remain with the generic $\U(2)$ supergravity.

Now if we assume that the Killing spinors $\xi_{\alpha i}$ and $\bar \xi^{\dalpha i}$
are linearly independent at each point in spacetime,
we can in principle derive integrability conditions that impose
constraints on the background fields of the supergravity multiplet consistent with
the existence of eight supercharges.
In addition, there are also covariant fermions which must be invariant under
supersymmetry, leading to constraints on the other auxiliary fields.
Such a procedure was actually followed in \cite{BdWL}.

Following \cite{Kuzenko:SCS}, we will analyze the problem directly in curved superspace. The supergeometry corresponding to an unconstrained real compensator is the $\U(2)$ superspace of \cite{Howe} (we follow similar conventions as \cite{KLRT-M2} and \cite{Butter:CSG4d_2}). It involves a supermanifold $\cM^{4|8}$ with local coordinates $z^M = (x^m, \q^\mu{}_\imath, \bar \q_\dmu{}^\imath)$ consisting of four bosonic and eight Grassmann coordinates. It is equipped with a non-degenerate supervielbein $E_M{}^A$ and local Lorentz, $\SU(2)_R$ and $\U(1)_R$ connections $\Omega_M{}^{ab}$, $\cV_M{}^i{}_j$ and $\cA_M$, respectively. The covariant derivative $\cD_A = (\cD_\alpha{}^i, \bar \cD^\dalpha{}_i, \cD_a)$ is defined as
\begin{align}
\cD_A = E_A{}^M \Big(
	\pa_M - \frac{1}{2} \Omega_M{}^{ab} M_{ab}
	- \cV_M{}^i{}_j I^j{}_i
	- \cA_M \mathbb A
\Big)~.
\end{align}
The symbols $M_{ab}$, $I^i{}_j$ and $\mathbb A$ denote respectively the Lorentz,
$\SU(2)_R$, and $\U(1)_R$ generators, whose action is given by
\begin{align}\label{eq:MIAalg}
[M_{ab}, \cD_c] = 2 \eta_{c[a} \cD_{b]}~, \qquad
[M_{ab}, \cD_\alpha{}^i] = -(\sigma_{ab})_\alpha{}^\beta \cD_\beta{}^i~, \eol{}
[I^i{}_j, \cD_\alpha{}^k] = \delta^k_j \cD_\alpha{}^i - \frac{1}{2} \delta^i_j \cD_\alpha{}^k~, \qquad
[\mathbb A, \cD_\alpha{}^i] = -i \cD_\alpha{}^i~.
\end{align}
We denote by $w$ the $R$-charge of fields and operators, e.g. $w(\cD_\alpha{}^i) = -1$.
The graded algebra of covariant derivatives involves torsion and curvature tensors,
\begin{align}\label{eq:CovDAlg}
[\cD_A, \cD_B\} = -T_{AB}{}^C \cD_C
	- \frac{1}{2} R_{AB}{}^{cd} M_{cd}
	- R(\cV)_{AB}{}^i{}_j I^j{}_i
	- R(A)_{AB} \,\mathbb A~.
\end{align}
For the case of $\U(2)$ supergravity, these tensors are given by the
superfields $S^{ij}$, $G_a$, $G_a{}^i{}_j$, $Y_{ab}^-$, and $W_{ab}^-$ and
their covariant derivatives.

Recall that the algebra of spinor derivatives $\cD_\alpha{}^i$ and $\bar \cD^\dalpha{}_i$
encodes the structure of $\U(2)$ supergravity. Any
component bosonic or fermionic field $\phi$ is the lowest component of some
superfield $\Phi$, denoted $\phi = \Phi \loco$. The action of supersymmetry
on $\phi$ derives from a covariant Lie derivative of $\Phi$,
\begin{align}
\delta_{\trQ} \phi \equiv \delta \Phi \loco =  \xi^\alpha{}_i (\cD_\alpha{}^i \Phi)\loco
	+ \bar \xi_\dalpha{}^i (\bar\cD^\dalpha{}_i \Phi)\loco
\end{align}
with $\xi_{\alpha i}$ and $\bar \xi^\dalpha{}_i$ the local supergravity parameters.
The local supersymmetry algebra is equivalent to the spinor derivative algebra of \eqref{eq:CovDAlg},
which is explicitly given by
\begin{align} \label{eq:SUSYQQ}
\{\cD_\alpha{}^i, \cD_\beta{}^j\}
	&= 4 S^{ij} M_{\alpha\beta}
	+ \eps^{ij} \eps_{\alpha\beta} \,(Y^{cd-} - W^{cd+}) M_{cd}
	+ 2 \eps^{ij} \eps_{\alpha\beta} \,S^k{}_l I^l{}_k
	- 4 \,Y_{\alpha\beta} I^{ij}~, \eol
\{\bar \cD^\dalpha{}_i, \bar \cD^\dbeta{}_j\}
	&= 4 \bar S_{ij} \bar M^{\dalpha \dbeta}
	- \eps_{ij} \eps^{\dalpha \dbeta} \,(Y^{cd+} - W^{cd-}) M_{cd}
	- 2 \eps_{ij} \eps^{\dalpha\dbeta} \bar S^k{}_l I^l{}_k
	- 4 \bar Y^{\dalpha \dbeta} I_{ij}~,  \\[0.3ex]
\{\cD_\alpha{}^i, \bar \cD_{\dbeta j}\}
	&= -2i \,\delta^i_j \cD_{\alpha\dbeta}
	+ 2 (\sigma^{cd} \sigma^b + \sigma^b \bsigma^{cd})_{\alpha\dbeta}
		(\delta^i_j G_b + i G_b{}^i{}_j) M_{cd}
	- 8 G_{\alpha\dbeta} I^i{}_j
	+ 2 G_{\alpha \dbeta}{}^i{}_j \ccA~. \notag
\end{align}
We remind the reader that our spinor conventions are summarized in Appendix \ref{App:Conv}.
From the algebra, one may read off the $R$-charges
\begin{align}
w(S^{ij}) = w (Y_{\alpha\beta}) = w(\bar W_{\dalpha\dbeta}) = -2~, \qquad
w(G_a) = w (G_a{}^i{}_j) = 0~.
\end{align}
The reality properties of the superfields and spinor derivatives are
\begin{gather}
(W_{ab}^-)^* = W_{ab}^+~, \qquad
(Y_{ab}^-)^* = Y_{ab}^+~, \qquad (S^{ij})^* = \bar S_{ij}~, \qquad
(G_a)^* = G_a~, \eol
(G_a{}^i{}_j)^* = -G_a{}^j{}_i~, \qquad
(\cD_\alpha{}^i)^* = \bar \cD_{\dalpha i}~.
\end{gather}

We are interested in fermionic isometries of a fixed background manifold, that is, local covariant diffeomorphisms that leave its supervielbein, connections and the associated torsion and curvatures invariant.
If we restrict to a bosonic background (i.e. all background fermions
set to zero), to any such fermionic isometry is associated a Killing spinor satisfying \eqref{eq:KillingSpinor}.
In a fully supersymmetric background, however, we do not need to solve this equation explicitly.
Following \cite{KNT-M:5DSCS}, we observe that
if eight linearly independent Killing spinors $\xi$ exist, then requiring $\delta_{\trQ} \phi$ to vanish for any background field $\phi$ implies the background superfield $\Phi$ must be annihilated by the spinor derivatives.
In particular, we have
\begin{equation}\label{eq:SuperC}
\cD_\alpha{}^i W_{ab}^- = \cD_\alpha{}^i Y_{ab}^- = \cdots = 0~.
\end{equation}
The integrability conditions
\begin{align}\label{eq:SuperIC}
\{\cD_\alpha{}^i, \cD_\beta{}^j\} W_{ab}^- = \{\cD_\alpha{}^i, \cD_\beta{}^j\} Y_{ab}^- = \cdots = 0~,
\end{align}
together with closure of the algebra of covariant derivatives imply integrability of the Killing spinor equation and invariance under $\delta_{\trQ}$ of the background fermionic fields.
It is thus sufficient to classify the solutions to these conditions.

Several of the equations \eqref{eq:SuperIC} imply that products of
various pairs of fields must vanish:
\begin{gather}
G_a{}^{ij} Y_{bc}^\pm = G_a{}^{ij} W_{bc}^\pm = G_a{}^{ij} S^{kl}
	= G_a{}^{ij} \bar S_{kl} = G_a{}^{ij} G_b = 0~, \eol
S^{ij} G_a = \bar S_{ij} G_a = 0~, \qquad
S^{ij} Y_{ab}^\pm = \bar S_{ij} Y_{ab}^\pm = 0~, \qquad
S^{ij} W_{ab}^- = \bar S_{ij} W_{ab}^+ = 0~.
\label{eq:LProductConditions}
\end{gather}
In Lorentzian signature, the last condition is strengthened to
$S^{ij} W_{ab}^\pm = \bar S_{ij} W_{ab}^\pm = 0$.
From these relations we may identify four broad cases:
\begin{enumerate}[(I)]
\item $S^{ij} \neq 0$, all other fields vanishing;
\item $G_a{}^i{}_j \neq 0$, all other fields vanishing;
\item $G_a \neq 0$, perhaps with $Y_{ab}^\pm$ and/or $W_{ab}^\pm$ nonzero;
\item $Y_{ab}^\pm$ and/or $W_{ab}^\pm$ nonzero, but all other fields vanishing.
\end{enumerate}
Additional integrability conditions depend on which case we are in.
For (I), these amount to
$S^{ij} \propto \bar S^{ij}$ and $\cD_a S^{ij} = 0$.
For (II), we find
$G_{[a}{}^{ij} G_{b]}{}^{kl} = 0$ and $\cD_a G_b{}^{ij} = 0$,
implying that $G_a{}^{ij}$ can be decomposed as a product of
a covariantly constant vector and an $\SU(2)$ tensor.
For (III) and (IV), the additional conditions are
\begin{gather}
G^b W_{ba}^\pm = G^b Y_{ba}^\mp~, \qquad
Y_{ab}^\pm \propto W_{ab}^\pm~, \qquad Y^{ab-} W_{ab}^- = Y^{ab+} W_{ab}^+~, \eol
\cD_a G_b = 0~, \qquad
(\cD_a + \eps_a{}^{bcd} G_b M_{cd}) W_{ef}^\pm = 0~, \qquad
(\cD_a + \eps_a{}^{bcd} G_b M_{cd}) Y_{ef}^\pm = 0~.
\end{gather}
The last equations above imply that $W_{ab}^\pm$ and $Y_{ab}^\pm$
are covariantly constant with respect to a torsionful spin connection,
$\widetilde \omega_{m ab} = \omega_{m ab} - \eps_{mabc} G^c$.

We can now construct the entire algebra of covariant derivatives
for the general $\cN=2$ rigid superspace, treating all cases simultaneously.
It is convenient to combine $Y_{ab}^-$ and $W_{ab}^+$ into a single complex two-form
$\cZ_{ab}$ with $R$-charge $w=-2$,
\begin{align}\label{eq:DefZ}
\newG_{ab} := Y_{ab}^- - W_{ab}^+~, \qquad
\bar \newG_{ab} := Y_{ab}^+ - W_{ab}^-~.
\end{align}
The full superspace algebra can then be compactly written as
\begin{align}
\{\cD_\alpha{}^i, \cD_\beta{}^j\}
	&= 4 S^{ij} M_{\alpha\beta}
	+ \eps^{ij} \eps_{\alpha\beta} \,\newG^{cd} M_{cd}
	+ 2 \eps^{ij} \eps_{\alpha\beta} \,S^k{}_l I^l{}_k
	- 4 \,\newG_{\alpha\beta} I^{ij}~, \eol{}
\{\bar \cD^\dalpha{}_i, \bar \cD^\dbeta{}_j\}
	&= 4 \bar S_{ij} \bar M^{\dalpha \dbeta}
	- \eps_{ij} \eps^{\dalpha \dbeta} \,\bar \newG^{cd} M_{cd}
	- 2 \eps_{ij} \eps^{\dalpha\dbeta} \bar S^k{}_l I^l{}_k
	- 4 \bar \cZ^{\dalpha \dbeta} I_{ij}~,\eol{}
\{\cD_\alpha{}^i, \bar \cD_{\dbeta j}\}
	&= -2i \,\delta^i_j (\sigma^a)_{\alpha\dbeta} \cD_a
	- 2 i (\sigma_a)_{\alpha\dbeta} \eps^{abcd} (\delta^i_j G_b + i G_b{}^i{}_j) M_{cd}
	- 8 G_{\alpha\dbeta} I^i{}_j
	+ 2 G_{\alpha \dbeta}{}^i{}_j \ccA~, \eol{}
[\cD_a, \cD_\beta{}^j]
	&=
	\frac{i}{2} (\sigma_a)_{\beta \dgamma} S^{jk} \bar \cD^\dgamma{}_k
	- \frac{i}{2} \newG_{ab} (\sigma^b)_{\beta \dgamma} \bar \cD^{\dgamma j}
	-2 i G^b  (\sigma_{ba})_\beta{}^{\gamma} \cD_\gamma{}^j
	- G_b{}^j{}_k (\sigma_a \bsigma^b)_\beta{}^\gamma \cD_\gamma{}^k~, \eol {}
[\cD_a, \bar\cD^\dbeta{}_j]
	&=
	\frac{i}{2} (\bsigma_a)^{\dbeta \gamma} \bar S_{jk} \cD_\gamma{}^k
	+ \frac{i}{2} \bar\newG_{ab} (\bsigma^b)^{\dbeta \gamma} \cD_{\gamma j}
	+ 2 i G^b (\bsigma_{ba})^\dbeta{}_{\dgamma} \bar\cD^\dgamma{}_j
	+G_b{}^k{}_j (\bsigma_a \sigma^b)^\dbeta{}_\dgamma \bar\cD^\dgamma{}_k~,\eol {}
[\cD_a, \cD_b] &= -\frac{1}{2} R_{ab}{}^{cd} M_{cd}~.
\label{eq:LSusyAlg}
\end{align}
The Riemann tensor is explicitly determined to be
\begin{align}
R_{ab}{}^{cd} &=
	- \frac{1}{2} (\newG_{ab} \bar \newG^{cd} + \bar \newG_{ab} \newG^{cd})
	+ 8 \,G^2 \delta_a{}^{[c} \delta_b{}^{d]}
	- 16 \,G_{[a} G^{[c} \delta_{b]}{}^{d]}
	\eol & \quad
	+ 4  G^f_{ij} G_{f}^{ij} \delta_a{}^{[c} \delta_b{}^{d]}
	- 8 G_{[a}^{ij} G^{[c}_{ij} \delta_{b]}{}^{d]}
	+ S^{ij} \bar S_{ij} \delta_a{}^{[c} \delta_b{}^{d]}~.
\end{align}
One can show that the spacetime is conformally flat when either
$Y_{ab}^- \equiv \cZ_{ab}^-$ or $W_{ab}^- \equiv -\bar \cZ_{ab}^-$ vanishes,
and superconformally flat when $W_{ab}^- = 0$.

In terms of $\cZ_{ab}$, the integrability conditions read
\begin{align}
G^a \cZ_{ab} = 0~, \quad \eps^{abcd} \cZ_{ab} \bar \cZ_{cd} = 0~, \quad
\cZ_{ab}^\pm \propto \bar \cZ_{ab}^\pm~, \quad
(\cD_a + \eps_a{}^{bcd} G_b M_{cd}) \cZ_{ef}= 0~.
\end{align}
It follows that $\cZ_{ab}$ is a closed complex two-form, so it possesses a complex
locally-defined one-form potential
\begin{align}
\cZ_{(2)} = \rd\, C_{(1)}~.
\end{align}
However, the dual of $\cZ_{ab}$ is not closed unless $G_a = 0$.
In contrast, the dual three-forms of $G_a$ and $G_a{}^i{}_j$ are always closed.
We denote these by $H_{abc} = \eps_{abcd}\, G^d$ and
$H_{abc}{}^i{}_j = \eps_{abcd} \,G^d{}^i{}_j$ and introduce their two-form potentials
\begin{align}
H_{(3)} = \rd B_{(2)}~, \qquad
H_{(3)}{}^i{}_j = \rd B_{(2)}{}^i{}_j~.
\end{align}
These potentials will play a role in the vector and hypermultiplet actions.

\section{Lorentzian backgrounds}\label{sec:LBacks}

\subsection{General comments}
Let us make a few modifications to the supersymmetry algebra \eqref{eq:LSusyAlg}.
If one introduces a redefined vector derivative
\begin{align}
\label{torsionful vector derivative}
\widetilde \cD_a := \cD_a + \eps_a{}^{bcd} G_b M_{cd}~,
\end{align}
corresponding to a spin connection with $G$-dependent torsion,
then the algebra of covariant derivatives becomes
\begin{align}
\label{eq:general algebra}
\{\cD_\alpha{}^i, \cD_\beta{}^j\}
	&= 4 S^{ij} M_{\alpha\beta}
	+ \eps^{ij} \eps_{\alpha\beta} \,\newG^{cd} M_{cd}
	+ 2 \eps^{ij} \eps_{\alpha\beta} \,S^k{}_l I^l{}_k
	- 4 \,\newG_{\alpha\beta} I^{ij}~, \eol{}
\{\bar \cD^\dalpha{}_i, \bar \cD^\dbeta{}_j\}
	&= 4 \bar S_{ij} \bar M^{\dalpha \dbeta}
	- \eps_{ij} \eps^{\dalpha \dbeta} \,\bar \newG^{cd} M_{cd}
	- 2 \eps_{ij} \eps^{\dalpha\dbeta} \bar S^k{}_l I^l{}_k
	- 4 \bar \newG^{\dalpha \dbeta} I_{ij}~,\eol{}
\{\cD_\alpha{}^i, \bar \cD_{\dbeta j}\}
	&= -2i \,\delta^i_j (\sigma^a)_{\alpha\dbeta} \widetilde\cD_a
	+ 2 (\sigma_a)_{\alpha\dbeta} \eps^{abcd} G_b{}^i{}_j M_{cd}
	- 8 G_{\alpha\dbeta} I^i{}_j
	+ 2 G_{\alpha \dbeta}{}^i{}_j \ccA~, \eol{}
[\widetilde \cD_a, \cD_\beta{}^j]
	&=
	\frac{i}{2} (\sigma_a)_{\beta \dgamma} S^{jk} \bar \cD^\dgamma{}_k
	- \frac{i}{2} \newG_{ab} (\sigma^b)_{\beta \dgamma} \bar \cD^{\dgamma j}
	- 4 i G^b  (\sigma_{ba})_\beta{}^{\gamma} \cD_\gamma{}^j
	- G_b{}^j{}_k (\sigma_a \bsigma^b)_\beta{}^\gamma \cD_\gamma{}^k~, \eol {}
[\widetilde \cD_a, \bar\cD^\dbeta{}_j]
	&=
	\frac{i}{2} (\bsigma_a)^{\dbeta \gamma} \bar S_{jk} \cD_\gamma{}^k
	+ \frac{i}{2} \bar\newG_{ab} (\bsigma^b)^{\dbeta \gamma} \cD_{\gamma j}
	+ 4 i G^b (\bsigma_{ba})^\dbeta{}_{\dgamma} \bar\cD^\dgamma{}_j
	+G_b{}^k{}_j (\bsigma_a \sigma^b)^\dbeta{}_\dgamma \bar\cD^\dgamma{}_k~,\eol {}
[\widetilde \cD_a, \widetilde \cD_b] &= - \widetilde T_{ab}{}^c \widetilde\cD_c
	-\frac{1}{2} \widetilde R_{ab}{}^{cd} M_{cd}~,
\end{align}
where the torsion and Lorentz curvature tensors are given by
\begin{align}
\label{torsion and torsionful curvature}
\widetilde T_{ab}{}^c &= -4 \,\eps_{ab}{}^{cd} G_d~, \eol
\widetilde R_{ab}{}^{cd} &=
	- \frac{1}{2} (\newG_{ab} \bar \newG^{cd} + \bar \newG_{ab} \newG^{cd})
	+ 4 \,G^f_{ij} G_{f}^{ij} \delta_a{}^{[c} \delta_b{}^{d]}
	- 8 \,G_{[a}^{ij} G^{[c}_{ij} \delta_{b]}{}^{d]}
	+ S^{ij} \bar S_{ij} \delta_a{}^{[c} \delta_b{}^{d]}~.
\end{align}
The utility of this redefinition lies in the fact that the background superfields
are now each covariantly constant with respect to
$\widetilde \cD_a$, $\cD_\alpha{}^i$ and $\bar \cD^\dalpha{}_i$.
This eliminates any algebraic obstruction to taking the background superfields
(with tangent space indices) to be
actually constant.\footnote{For example, $G^2$ is constant so
by a suitable Lorentz transformation, $G^a$ can be taken
constant when it is timelike, spacelike, or null but never vanishing.
The case where $G^a$ is null and vanishes somewhere is precluded
because $\cD_b G_a=0$ implies the spacetime is not smooth.}
Let us summarize the gauge choices for each case.

\begin{itemize}
\item We choose $S^{ij} = \bar S^{ij} = \mu \,v^{ij}$ with $\mu$ real and
$v^{ij}$ pseudoreal, $(v^{ij})^* = v_{ij}$, and normalized,
$v^{ij} v_{jk} = \delta^i_k$. Because the superalgebra possesses only Lorentz and
$\rm SO(2) \subset \rm SU(2)$ $R$-symmetry, there is no local obstruction to eliminating
the $\U(1)_R$ connection and aligning the $\SU(2)_R$ connection along $v^{ij}$,
consistent with $\cD_A S^{ij} = 0$.

\item We take $G_a{}^i{}_j = g_a\, v^i{}_j$, with $v^{ij}$ pseudoreal.
The real constant $g_a$ may be timelike, spacelike, or null and partially breaks Lorentz symmetry.
The $\SU(2)_R$ connection is again aligned along $v^i{}_j$.

\item We choose $G_a$ constant and timelike, spacelike, or null.
The $\U(1)_R$ connection is pure gauge and can be taken to vanish.
A non-zero $\newG$ can be turned on in several configurations and thanks to the constraint $G^a\cZ_{ab}=0$ we can adopt a gauge where it is purely constant.
The residual global $\U(1)_R$ is broken by $\cZ$.

\item If only $\cZ$ is non-vanishing, the same discussion as above holds, but $\cZ$ is less constrained.
\end{itemize}
In each of these backgrounds, the algebra of covariant derivatives implies
that at least some of the $R$-symmetry curvature tensors $R(V)_{AB}{}^i{}_j$ and $R(A)_{AB}$
are non-vanishing, and so the superspace connections $\cV_M{}^i{}_j$ and $\cA_M$
cannot all vanish. However, the vanishing of the component curvatures $R(V)_{ab}{}^i{}_j$
and $R(A)_{ab}$ implies that the component connections $\cV_m{}^i{}_j$ and $A_m$ are
always (at least locally) pure gauge. We will make use of this property in our constructions
below by always choosing a gauge where the component $\SU(2)_R$ connection vanishes.
However, in a few of the cases, the $\U(1)_R$ connection will possess a non-trivial holonomy.

We summarize in Table \ref{tab:LBacks} the resulting consistent Lorentzian backgrounds.

\begin{table}[th!]
\def\twocol{\multicolumn{2}{l}}
\tablespacings
\centering
\begin{tabular}{ll@{\hskip 4em}l}
\toprule
\twocol{Background fields} & Geometry\\
\otoprule
\twocol{$S^{ij}\neq0$}                                         & AdS$_4$                                \\
\midrule
\twocol{$G_a{}^i{}_j$ timelike}                                & $\mathbb R\times S^3$                  \\
\twocol{$G_a{}^i{}_j$ null}                                    &  plane wave                               \\
\twocol{$G_a{}^i{}_j$ spacelike}                               & AdS$_3\times\bbR$                      \\
\midrule
\twocol{$G_a$ timelike}                                        & $\mathbb R\times S^3$                  \\
&$\newG_a{}^b$ elliptic                                        & $\mathbb R\times S^3$ squashed        \\
\midrule
\twocol{$G_a$ null}                                            & plane wave                                  \\
&$\newG_a{}^b$ elliptic                                        & `lightlike' $S^3\times \bbR$           \\
&$\newG_a{}^b$ parabolic                                       & plane wave                                \\
\midrule
\twocol{$G_a$ spacelike}                                       & AdS$_3\times\bbR$                      \\
&$\newG_a{}^b$ elliptic,                                       &                                        \\
&$\quad0<|\newG|^2 < 32 \,G^2$                                   & timelike stretched AdS$_3\times\bbR$   \\
&$\quad|\newG|^2=32 \,G^2$                                     & Heis$_3\times\bbR$  \\
&$\quad|\newG|^2 > 32 \,G^2$                                   & warped `Lorentzian' $S^3\times\bbR$           \\[.5ex]
&$\newG_a{}^b$ parabolic                                       & null warped AdS$_3\times\bbR$          \\
&$\newG_a{}^b$ hyperbolic                                      & spacelike squashed  AdS$_3\times\bbR$  \\
\midrule
\twocol{$\newG_{a}{}^{b}\neq 0$  but $G_a = 0$}                &                                        \\
&$\newG_a{}^b$ elliptic                                        & $\mathbb R^{1,1}\times S^2$            \\
&$\newG_a{}^b$ hyperbolic                                      & AdS$_2\times\mathbb R^2$               \\
&$\newG_a{}^b\sim \text{elliptic} + \text{hyperbolic}$         & AdS$_2\times S^2$                      \\
&$\newG_a{}^b$ parabolic                                       & plane wave
\cbottomrule
\end{tabular}
\caption{Consistent Lorentzian backgrounds. As $\cZ_a{}^b$ can be decomposed as a complex
linear combination of Lorentz generators, we distinguish its values in terms of orbits of
the latter.}
\label{tab:LBacks}
\end{table}

\subsection{The supercoset construction}

The full BPS geometries are at least locally isomorphic to supercoset spaces, generated by an algebra of bosonic and fermionic isometries that is isomorphic to the algebra of covariant derivatives. The cosets are obtained by quotienting out the Lorentz and $R$-symmetries remaining after gauge fixing the background fields to constant values.
If we take these supercoset spaces as global solutions, existence of eight Killing spinors is automatically guaranteed by transitivity of the supergroup action on the background, and explicit expressions are straightforward to derive.
Of course, solutions that are only locally isomorphic to supercoset spaces may be available, too, for instance in terms of orbifolds of homogeneous spaces.

Let us briefly review how the supercoset space construction works in practice and how it can be used to obtain explicit expressions for the metric and Killing spinors in particular (See for instance \cite{A-AL-TO:KillingSpinors,Kleppe:2006ys}).
We can consider the reciprocal superalgebra of formal generators $P_a$, $Q_\alpha{}^i$, $\bar Q^{\dalpha}{}_i$ satisfying the same commutation relations as $\widetilde \cD_a$, $\cD_\alpha{}^i$, $\bar \cD^{\dalpha}{}_i$.
A full supergroup is generated by $P_a$, $Q_\alpha{}^i$, and $\bar Q^{\dalpha}{}_i$ together with any residual Lorentz and $R$-symmetries. Quotienting by these, we obtain a supercoset space.
Representatives $\mathbb L$ can be defined by translation and supersymmetry generators:
\begin{equation}
\mathbb L = \exp\left(
x^a P_a + \theta^\alpha{}_i Q_\alpha{}^i + \bar\theta_\dalpha{}^i \bar Q^\dalpha{}_i
\right),
\end{equation}
although in practice we will make different coordinate and local gauge choices on a case by case basis.
Under a supergroup transformation $g$, the coset representatives transform as
\begin{equation}
g \mathbb L(x,\theta,\bar\theta) = 
\mathbb L(x',\theta',\bar\theta') h(x,\theta,\bar\theta,g),
\end{equation}
with $h(x,\theta,\bar\theta,g)$ a local Lorentz and $R$-symmetry transformation.
In particular, $P_a$, $Q_{\alpha}{}^i$, and $\bar Q^\dalpha{}_i$
generate part of the (left) isometry group.
Since we have assumed absence of fermionic background fields, we can always restrict to the bosonic submanifold $\theta=\bar\theta=0$, which is itself a coset space with a natural choice of representatives $L(x) \equiv \mathbb L\loco$.

The Cartan--Maurer form $\mathbb L^{-1} \rd\mathbb L$ can then be expanded in the superalgebra generators.
The coefficients of the translation and supersymmetry generators will be chosen as the (super)vielbein, and the remaining terms give rise to spin and $R$-symmetry connections.
Constructing covariant derivatives from these objects then reproduces \eqref{eq:general algebra}.

The advantage of this procedure is that we are allowed to compute the Cartan--Maurer form in any convenient (faithful) representation of the superalgebra, which means we can calculate explicit expressions for any relevant object with ease.
In particular, Killing spinors for a supercoset space are computed similarly to Killing vectors.
One constructs a supersymmetric isometry $\delta_{\trQ}$ in terms of eight constant spinors $\eps_{\alpha i}$ and $\bar\eps^{\dalpha i}$ via
\begin{align}
\delta_{\trQ} &= \mathbb L_{}^{-1} (\eps_i Q^i + \bar \eps^i \bar Q_i) \mathbb L
	= \xi_i Q^i + \bar \xi^i \bar Q_i + \xi^a P_a + \frac{1}{2} \l^{ab} M_{ab}
	+ \l^i{}_j I^j{}_i + \l \mathbb A~.
\end{align}
The local parameters $\xi^A =(\xi^{\alpha}{}_i, \bar \xi_{\dalpha}{}^i, \xi^a)$, $\l^{ab}$,
$\l^i{}_j$, $\l$ depend on $x$, $\q$, and any background fields and are parametrized
linearly in terms of $\eps_{\alpha i}$ and $\bar \eps^{\dalpha i}$.
The operation $\delta_{\trQ}$ acts on superfields as a combination of covariant superdiffeomorphisms and local gauge transformations,
\begin{align}
\delta_{\trQ} = \xi^A \cD_A + \frac{1}{2} \l^{ab} M_{ab}
	+ \l^i{}_j I^j{}_i + \l \mathbb A~.
\end{align}
The condition that $\delta_\trQ$ generates a superisometry is equivalent to
$[\delta_\trQ, \cD_A] = 0$; this leads to the Killing supervector condition
(see \cite{BK} for a general discussion). As a consequence of the supercoset structure,
this is satisfied automatically.

In practice, however, one only cares about how supersymmetry manifests on the
bosonic manifold. Because of the absence of background fermions in the algebra,
at $\theta=0$ we have
$\xi^a= \l^i{}_j = \l^{ab}= \l = 0$ and
\begin{align}
\delta_{\trQ} &=  L^{-1}(\eps_i Q^i + \bar \eps^i \bar Q_i)  L
	= \xi_i Q^i + \bar \xi^i\bar Q_i~.
\end{align}
The functions $\xi_{\alpha i}(x)$ and $\bar\xi^{\dalpha i}(x)$
are the Killing spinors.
Because (four-component) supercharges transform in some representation $R[\,\cdot\,]$ of the bosonic isometries,
\begin{equation}
\xi(x) = \epsilon \,R[L(x)],
\end{equation}
where $\xi$ and $\eps$ are given by their appropriate (four-component) Majorana conjugates.
The representation $R[L]$ can be deduced from the vector-spinor commutators of \eqref{eq:general algebra}.
Notice that in general it also includes a non-trivial action on $R$-symmetry indices.

In the next sections we provide an exhaustive classification of the supercoset spaces that give consistent global $\cN=2$ supersymmetric rigid backgrounds in Lorentzian signature.

\subsection{$\rm AdS_4$ spacetime}
When $S^{ij} \equiv \mu\, v^{ij}$ is the only non-vanishing background field, we recover the $\cN=2$ $\rm AdS_4$ superspace
$\rm AdS^{4|8} = \rm OSp(2|4) / SO(3,1) \times SO(2)$.
The bosonic body of the superalgebra is $\rm SO(3,2) \times \rm SO(2)$, corresponding to the product of the $\rm AdS_4$ algebra and the residual $R$-symmetry $\rm SO(2) \subset SU(2)$ generated by $S^{ij}$.
The properties of this superalgebra and its Killing spinors were discussed long ago in \cite{Breitenlohner:1982jf} (see \cite{KT-M:CFlat} and references therein for a superspace discussion), so we will not dwell on it here, except to remind that it possesses an $\cN=1$ subalgebra, which is evident upon performing a global $\SU(2)_R$ transformation to adopt the gauge
$v^{ij} = - \delta^{ij}$.

\subsection{Round and squashed spatial three-spheres ($\mathbb R \times S^3$)}

As a first non-trivial case we consider backgrounds locally isomorphic to $\mathbb R\times S^3$, where $\mathbb R$ corresponds to the direction of a timelike Killing vector and the metric on the $S^3$ is static.
We will show that full $\cN=2$ supersymmetry is compatible with a specific squashing of the $S^3$.
This case shall serve as a pedagogical example.

\subsubsection{The round $\bbR\times S^3$ and $\rm SU(2|2)$ supersymmetry}
\label{sec:round S3 lorentzian signature}

For the moment we assume that the $S^3$ possesses its round metric so that
its full $\rm SU(2) \times \rm SU(2)$ isometry group is intact.
This space admits two possible realizations of $\cN=2$ supersymmetry, corresponding to
extending the isometry group $\rm SU(2)\times SU(2)$ either to
$\rm SU(2|2)\times SU(2)$ or to $\rm SU(2|1)\times SU(2|1)$.

The first realization of $\cN=2$ supersymmetry on the round $S^3$ involves
the superalgebra $\rm SU(2|2)$. This corresponds to turning on a timelike $G^a$ only.
We can adopt the Lorentz gauge $G^a =(-g,0,0,0)$ for constant $g$,
then the bosonic part of the superalgebra
contains the generators $P_a = (P_0, P_I)$ with $I=1,2,3$ and commutation relations
\begin{align}
[ P_0, P_I ] = 0~, \qquad 
[P_I, P_J] = 4 g\, \veps_{IJK} P_K~.
\end{align} 
In addition, there are residual Lorentz generators $M_{IJ}$ acting as
\begin{align}
[M_{IJ}, P_K] = \delta_{IK} P_J - \delta_{JK} P_I~.
\end{align}
Finally, there are the SU(2)$_R$ generators $I_{ij}$.
It will be convenient to introduce the (dimensionless) $\rm SU(2)$ generators 
\begin{align}
\label{eq:SU(2) T generators}
T_I &\equiv \frac{1}{4g} P_I~, \qquad
[T_I, T_J] = \veps_{IJK} T_K~.
\end{align}
It is apparent that the spacetime is diffeomorphic to
$\bbR \times {\rm SU}(2) $, the $\bbR$ direction being generated by $P_0$
and the $S^3$ by $T_I$.

We can construct all relevant objects on this background exploiting its group manifold structure.
Introducing a generic element $L$ of $\mathbb R \times \rm SU(2)$,
the Maurer--Cartan form reads
\begin{equation}
\Omega = L^{-1} \rd L = \rd t\, P_0 + e^I P_I = \rd t\, P_0 + E^I T_I~.
\end{equation}
The $E^I$ are a dimensionless left-invariant vielbein related to the physical vielbein $e^I = \frac{1}{4g}\, E^I$.
They obey the structure equations
$\rd E^I = -\frac{1}{2} \veps_{IJK} E^J \wedge E^K$.
Choosing the explicit parametrization
$L=e^{t P_0} e^{\phi T_3} e^{\theta T_2} e^{\omega T_3}$
leads to the canonical choice of Euler angle coordinates on the $S^3$,
\begin{equation}
\label{eq:S3RoundV}
E^1=-\sin\theta\cos\omega\,\rd\phi+\sin\omega\,\rd\theta,\quad
E^2=\sin\theta\sin\omega\,\rd\phi+\cos\omega\,\rd\theta~,\quad
E^3=\rd\omega+\cos\theta\,\rd\phi,
\end{equation}
with a round metric
\begin{equation}
\label{S3 round metric}
\rd s^2 = -\rd t^2+\frac1{16g^2}
\left[    
(\rd\omega+\cos\theta\,\rd\phi)^2+
\rd\theta^2+\sin^2\theta\,\rd\phi^2
\right]~.
\end{equation}
We choose standard ranges so that $\theta \in [0, \pi]$, $\phi \in [0, 2\pi)$ and $\omega \in [0, 4\pi)$ with $\omega$ periodic with period $4\pi$ so that the element $L$ covers all of $\mathbb R \times \rm SU(2)$ exactly once.
It will be useful to decompose $L$ in terms of embedding coordinates for $S^3$ of unit radius.
Defining
\begin{align}
\begin{split}
\label{S3 embedding coordinates}
X = \cos\tfrac\theta2 \cos\tfrac{\omega+\phi}2,  \quad 
Y = \sin\tfrac\theta2 \cos\tfrac{\omega-\phi}2,  \quad 
V = \sin\tfrac\theta2 \sin\tfrac{\omega-\phi}2,  \quad 
W = \cos\tfrac\theta2 \sin\tfrac{\omega+\phi}2,   
\end{split}
\end{align}
we can write $L = e^{t P_0} [X +2(V T_1 + YT_2 +W T_3)]$ in any spinorial representation of SU(2), where the generators are constructed from a Clifford algebra and obey $T_I T_J =-\frac14\delta_{IJ}+\frac12 \epsilon_{IJK} T_K$.

The group manifold construction extends to a full supercoset construction.
From \eqref{eq:general algebra} and \eqref{eq:SU(2) T generators} we identify the superalgebra to be a central extension of $\rm SU(2|2)$, on which the residual Lorentz group acts as external automorphisms and the central charge is $P_0$.
We stress that this central charge is not an internal symmetry, but rather the timelike isometry of this spacetime.
Quotienting out the $R$-symmetry $\SU(2)$ we arrive at the supermanifold%
\footnote{In principle, we could also include the \SU(2) Lorentz symmetries in
the numerator, but they would be factored out in the denominator.}
\begin{equation}
\label{eq:SU(2|2) round supercoset}
\frac{\SU(2|2)_{(P_0)}}
{ \SU(2)_R }~.
\end{equation}

We can now construct the Killing spinors. The bosonic part of the
supercoset \eqref{eq:SU(2|2) round supercoset} has numerator $\mathbb R \times \SU(2) \times \SU(2)_R$.
The generators $Q_\alpha{}^i$ transform in the $(\mathbf 2, \bar{\mathbf 2})_0$
representation; this is evident by noting
$
[T_I, Q_\alpha{}^i]
	= -\frac12\epsilon_{IJK} (\sigma_{JK})_\alpha{}^\beta Q_\beta{}^i
$
and similarly for $\bar Q^\dbeta{}_j$.
Hence we find the appropriate representation to construct the Killing spinors:
\begin{align}
\label{eq:round S3 spinorial repr}
R[T_I] = \frac12\epsilon_{IJK} (\sigma_{JK})_\alpha{}^\beta \delta^i{}_j \qquad
R[P_0] = 0~.
\end{align}
They can be written directly in embedding coordinates \eqref{S3 embedding coordinates}, and they are given by
\begin{align}
\label{round S3 killing spinors}
\xi_{\alpha\,i} &= [X -2(W \sigma^{12} + V \sigma^{23} + Y\sigma^{31})]_\alpha{}^\beta \epsilon_{\beta\,i}  \eol
\bar\xi^{\dalpha}{}_i &= 
[X -2(W\bar\sigma^{12}+V\bar\sigma^{23}+Y\bar\sigma^{31})]^\dalpha{}_\dbeta\bar\epsilon^{\dbeta}{}_i \ .
\end{align}
These are just the well-known Killing spinors on $S^3$ (see for instance \cite{Lu:1998nu} for a general construction of Killing spinors on spheres).
The $R$-symmetry index is untouched by the Killing spinor equation in this case.
Equivalently, the superalgebra possesses an obvious $\cN=1$ subalgebra.
Notice that the $R$-symmetry connections are vanishing, and all Killing spinors are time-independent.%
\footnote{We have redefined the $R$-symmetry connections with respect to conformal supergravity, as described in footnote~\ref{foot:RConns}.
Our statement is unambiguous in the sense that a theory coupled to this background need not have (full) $\U(2)_R$ symmetry.}
By a local Lorentz transformation we can actually take the Killing spinors to be entirely constant.
This corresponds to generating the supercoset using the right isometries of $S^3$.

\subsubsection{The squashed $S^3$}

The background we have just discussed has full superisometry group $\SU(2|2)_{(P_0)}\times \SU(2)$.
The latter factor is generated by linear combinations of $P_I$ and $M_{IJ}$ and corresponds to the right isometries of $S^3$.
Since they do not affect the supercharges, we may ask if it is possible  to deform the $S^3$ geometry in a way that breaks only its right isometries, obtaining a squashed $\bbR\times S^3$ with full $\cN=2$ supersymmetry.

If we try to break all the right isometries, the requirement of a smooth limit to a flat space supersymmetry algebra uniquely fixes the normalizations of the $P_I$ generators to match those of the previous section. 
We hence end up back to the round $S^3$ geometry.

The only possibility is therefore to break the right \SU(2) to \U(1).
In fact, such a background is realized by turning on $\cZ$ along the $S^3$.
The resulting supercoset space is generated by  \SU(2|2) with the addition of a second central charge.
The $P_I$ generators turn out to be a linear combination of the left $S^3$ isometries and this central charge, which results in a change in the normalizations of the physical vielbein that gives rise to a squashed $S^3$.
Explicitly, we have $G^a=(-g,0,0,0)$ and gauge fix $\cZ_{ab} = 4 \lambda\delta_{ab}^{12}$ so that $|\cZ|^2\equiv\cZ_{ab}\bar\cZ^{ab}=8\l^2$. 
The bosonic part of the supercoset space is now generated by $P_a$ and $M_{12}$, in particular
\begin{align}
\big[ P_1,\, P_2 \big] = 4g P_3 + 4 \lambda^2 M_{12}\ ,\quad
\big[ P_3, P_1 \big] = 4 g P_2~, \quad
\big[ P_2, P_3 \big] = 4 g P_1~. 
\label{MD commutator}
\end{align}
The time translation $P_0$ commutes with all generators.
We define $\upsilon = \left(1+\frac{\lambda^2}{4g^2}\right)^{-1}$ and introduce dimensionless generators $T_I$ satisfying the $\rm SU(2)$ algebra as before, $[T_I,T_J]=\veps_{IJK} T_K$, as well as a second central charge $U$:
\begin{align}
\label{S3 T generators}
T_{1,2} \equiv \frac{\sqrt\u}{4g}P_{1,2},\quad
T_3&\equiv \upsilon\left( \frac1{4g}P_3 + \frac{\lambda^2}{4g^2} M_{12} \right),\quad 
U \equiv \frac1{4g} P_3 - M_{12} ,
\end{align}
The bosonic background is still topologically $\bbR\times S^3$, as we can use $P_0$ and $T_I$ to generate it.
However, since $P_I$ do not close onto themselves it is more convenient to consider the whole isometry group and quotient out the only residual Lorentz generator $M_{12}$, leading to the coset space
\begin{equation}
\bbR\times \frac{{\rm SU}(2)\times \mathrm U(1)_U}{\mathrm U(1)_M}.
\end{equation}
The form of the quotient already suggests the nature of the squashing: regarding $S^3$ as the Hopf fibration of an $S^1$ over $S^2\simeq\SU(2)/\U(1)_M$, we rescale the metric along the fiber as determined by the ratio of $T_3$ and $U$ in $\U(1)_M$.
The Cartan--Maurer form can be expanded as earlier, recalling that the physical vielbein is given by the coefficients of $P_a$ so that
\begin{align}
\label{squashed S3 vielbein}
e^0     = dt,\qquad
e^3     = \upsilon\frac1{4g}E^3,\qquad
e^{1,2} = \sqrt\upsilon\frac1{4g}E^{1,2},
\end{align}
in terms of the `round' vielbein $E^I$ defined by $\Omega = \rd t\, P_0 + E^I T_I$.
We can use the same coset representative as in the unsquashed case.
As anticipated, the metric on the $S^3$ is squashed due to the different normalization of the physical vielbein along the fiber $S^1$.
For our choice of coordinates we obtain
\begin{align}
\label{metric squashed S3 lorentzian signature}
\rd s^2
	&= -\rd t^2+\frac{\upsilon}{16g^2}
	\big[ \rd\theta^2+\sin^2\theta\,\rd\omega^2 + \upsilon(\rd\phi+\cos\theta\,\rd\omega)^2\big]~.
\end{align}
Notice that $0<\upsilon\le1$, so that we are only allowed to squash, but not stretch, along the fiber.
This constraint is a consequence of $\cN=2$ supersymmetry in Lorentzian signature and will be relaxed in the Euclidean case.

Computing Killing spinors using the supercoset construction proceeds as before, except now the supergroup possesses two central charges, $P_0$ and $U$; we denote it $\SU(2|2)_{(U,P_0)}$.
The supercoset space is now
\begin{equation}
\label{squashed S3 supercoset}
\frac{{\mathrm{SU}}(2|2)_{(U,\,P_0)}}
{ \mathrm{SU}(2)_{\rm R} \times \mathrm{U}(1)_{M} }.
\end{equation}
The supersymmetry generators mix chirality under the action of $P_a$,
so it will be convenient to introduce a four-component spinor and associated Killing spinors:
\begin{equation}
\label{four component spinors}
Q_{\hat\alpha}{}^i =
\begin{pmatrix}
	Q_\alpha{}^i \\
	\epsilon^{ij} \bar Q^{\dalpha}{}_j
\end{pmatrix}\ ,\qquad
\xi^{\hat\alpha}{}_i =
\begin{pmatrix}
	\xi^\alpha{}_i\,, & 
	-\epsilon_{ij}\bar\xi_{\dalpha}{}^j
\end{pmatrix}\ .
\end{equation}
Then the bosonic generators are represented as
\begin{align}
\label{eq:spinorial representation for G and Z case}
R[P_a] = -2\ii \,G^b \g_5\g_{ba}
	+ \frac{1}{4} \left(\cZ_{ab} (1+\g_5) - \bar\cZ_{ab} (1-\g_5) \right)\g^b,\quad
R[M_{12}] = -\frac{1}{2}\gamma_{12}
\end{align}
and the Killing spinors are 
$
\xi^{\hat\alpha}{}_i =
\epsilon^{\hat\beta}{}_i~R[L]_{\hat\beta}{}^{\hat\alpha}
$.
Notice that we can still write $R[L] = R[X +2(V T_1 + YT_2 +W T_3)]$ with embedding coordinates defined in \eqref{S3 embedding coordinates} for the round $S^3$.
Using standard Weyl notation the Killing spinors can then be expressed as
\begin{align}
\xi_{\alpha\,i} &= 
[X\mathbb I_2 -2(W\sigma^{12}+\sqrt\upsilon V\sigma^{23}+\sqrt\upsilon Y\sigma^{31})]_\alpha{}^\beta \epsilon_{\beta\,i}
+\ii\sqrt\upsilon\frac{\lambda}{2g}(Y\sigma^1-V\sigma^2)_{\alpha\dbeta}\bar\epsilon^{\dbeta}{}_i\ , 
\end{align}
with $\bar\xi^\dalpha{}_i$ given by complex conjugation.
When $\l$ is non-vanishing, there is non-trivial mixing between the
different $R$-symmetry components of the Killing spinors. This implies
that there is no truncation to an $\cN=1$ subalgebra for the case
of a squashed $S^3$.
The right isometries of $S^3$ being partially broken, we cannot make the Killing spinors constant by a local gauge transformation.
However, notice that they are still time-independent with vanishing $R$-symmetry connections.

For later use, we will need explicit expressions for the potentials of the background fields.
The $\cZ$ two-form admits a globally defined potential $C_{(1)}$, given by
\begin{align}
C_{(1)} = -\frac{1}{2} \frac{\l}{4g^2 + \l^2} \Big(\cos\theta\, \rd\phi + \rd \omega\Big)
	= -\frac{\l}{2g} e^3~.
\end{align}
Associated with $G^a$ we have its dual two-form potential
\begin{align}
\label{two form potential S3}
B_{(2)} &= \frac{\u^2}{64g^2} \times
\begin{cases}
(\cos\theta - 1) \,\rd\omega\,\rd\phi & \theta \in [0, \pi/2]~, \\
(\cos\theta + 1) \,\rd\omega\,\rd\phi & \theta \in [\pi/2, \pi]~.
\end{cases}
\end{align}

\subsubsection{$\rm SU(2|1)\times SU(2|1)$ supersymmetry}

As anticipated, there is a second way to realize $\cN=2$ supersymmetry on $\bbR\times S^3$.
This is associated with a nonvanishing $G_a{}^i{}_j$ field, breaking the $\SU(2)_R$ symmetry and possibly turning on a non-trivial axial $\U(1)_R$ connection.
Let us choose $G_a{}^i{}_j \equiv \ii g_a(\sigma_3)^i{}_j$ with $g^a=(-g,0,0,0)$.
Noting that $(\sigma_3)^i{}_j = (-)^{i+1} \delta^i{}_j$,
the fermionic part of the superalgebra is
\begin{align}
\{Q_\alpha{}^i,Q_\beta{}^j\} = \{\bar Q^\dalpha{}_i,\bar Q^\dbeta{}_j\}=0, \qquad
\{Q_\alpha{}^i,\bar Q_\dbeta{}_j\}  = - 2i \,\delta^i{}_j \Delta_{\alpha \dbeta}^{(i)}
\end{align}
where the eight generators $\Delta_a^{(i)}$ with $a=0,\cdots,3$ and $i=\1,\2$
can be written
\begin{equation}
\Delta_a^{(i)} \equiv P_a +(-)^i g \, \big(\epsilon_{0abc} M^{bc} + \delta^0_a\mathbb A\big)~.
\end{equation}
We see that the rest of the superalgebra similarly decomposes,
\begin{align}
[\Delta^{(i)}_a,Q_\alpha{}^j] &= (-)^{i+1} \,\delta^{ij}\,2\ri \,g\left(
	\delta^0_a\delta_\alpha^\beta
	- 2(\sigma_{0a})_\alpha{}^\beta\right) Q_\beta{}^i,\eol{}
[\Delta_a^{(i)},\Delta_b^{(j)}] &= (-)^i\delta^{ij} 4g \,\epsilon_{0ab}{}^c\Delta_c^{(i)}.
\end{align}
This superalgebra is just $\rm SU(2|1) \times \rm SU(2|1)$ with each copy labeled by $i$.
Up to normalizations, each supergroup is generated by bosonic elements
$\Delta_a^{(i)}$ corresponding to $\rm SU(2) \times U(1)$
and odd elements $Q_\alpha{}^i$ and $\bar Q^\dalpha{}_i$.
The temporal
generator $P_0$ and the $\U(1)_R$ generator $\mathbb A$ correspond respectively to the
antidiagonal and diagonal combinations of the two $\rm U(1)$ factors $\Delta_0^{(i)}$.
The surviving $\rm SO(3)$ generators of the Lorentz group correspond to
the diagonal $\rm SU(2)$ generated by $-\Delta_I^{(1)} + \Delta_I^{(2)}$.

In order to construct the spatial part of the coset space, it is sufficient to take
group elements of either $\rm SU(2)$ factor as coset representatives. We will choose the
second $\rm SU(2)$, so that
\begin{align}
T_I &\equiv \frac{1}{4g} \Delta_I^{(2)}~, \qquad
[T_I, T_J] = \veps_{IJK} T_K~, \qquad [\Delta_0^{(2)}, T_I] = 0~,
\end{align}
and construct the coset representatives as in the previous sections, but using $\Delta^{(2)}_0$ rather than $P_0$ for the time direction for reasons that will be apparent soon.
This introduces a U$(1)_R$ transformation that can be trivially quotiented away at the bosonic level, but will become relevant in the supercoset construction.
We obtain the same round vielbein and metric as in the previous sections.

The situation with the Killing spinors is quite interesting.
Let us denote them $\hat\xi$ in this section.
The odd elements of the supergroup transform in the representation
\begin{equation}
(\mathbf2,\mathbf1)_{+1/2,0} + (\bar{\mathbf{2}},\mathbf1)_{-1/2,0}
+(\mathbf1,\mathbf2)_{0,+1/2} + (\mathbf1,\bar{\mathbf{2}})_{0,-1/2}
\end{equation}
of $\rm U(2)_{(1)}\times \rm U(2)_{(2)}$, with the natural Majorana condition giving eight supercharges.
The same representations hold for the Killing spinors, so if we generate the coset representatives using only $\rm U(2)_{(2)}$ generators, the Weyl spinors $\hat\xi_\1$ and their complex conjugates will be entirely constant.
We will shortly prove this explicitly by construction, but it helps to
motivate this first by taking a look at the Killing spinor equations,
\begin{equation}
(\cD_a + g\epsilon_{0acd}\sigma^{cd}+\ii g\delta^0_a)\hat\xi_\1=0~,\qquad
(\cD_a - g\epsilon_{0acd}\sigma^{cd}-\ii g\delta^0_a)\hat\xi_\2=0~.
\end{equation}
The background field $G_a{}^i{}_j$ couples asymmetrically to the two spinors.
From the Maurer--Cartan form we find that the $\mathrm U(1)_R$ connection is given by
\begin{equation}
A = g \, e^0 = g\, \rd t~,
\end{equation}
therefore the $\rm U(1)$ connection within $\cD_a$ cancels the additional $i g$ factor in
the $\hat\xi_\1$ equation. Similarly, we find that our choice of vielbein has led to a
spin connection within $\cD_a$ that cancels the additional $g \eps_{0acd} \sigma^{cd}$
term.

We can now see this explicitly from the coset construction. The generators $T_I$ and $\Delta_0^{(2)}$
commute with $Q_\alpha{}^\1$, implying that the associated Killing
spinors $\hat\xi$ are constants,
\begin{align}
\hat\xi_{\alpha \1} = \eps_{\alpha \1}~, \qquad
\bar{\hat \xi}^{\dalpha \1} = \eps^{\dalpha \1}~.
\end{align}
However, the second set of Killing spinors are non-trivial. 
Noting that now the time generator $\Delta^{(2)}_0$ is non-trivially represented, we obtain Killing spinors
\begin{align}
\hat\xi_{\alpha\,\2} = e^{2 i g t} \xi_{\alpha\,\2},\qquad
\bar{\hat{\xi}}^{\dalpha\,\2} = e^{-2 i g t} \bar\xi^{\dalpha\,\2},
\end{align}
with $\xi_{\alpha\,\2}$, $\bar\xi^{\dalpha}{}_\2$ defined in \eqref{round S3 killing spinors}.
It is easy to see that the $t$-dependence in the second set of Killing spinors can be shuffled into the first set via a $\U(1)_R$ gauge transformation, and the same is true for the additional Lorentz factors.

Despite the fact that we are working in Lorentzian signature, it is instructive to see what happens if we compactify the time direction.
Periodic Killing spinors are allowed only for $t\simeq t+n \pi / g$ for integer $n$.
A non-trivial Wilson line for $A$ is generated for odd $n$.
For even $n$, the gauge field could be turned off by a gauge transformation, leaving $\hat\xi_\1$ and $\hat\xi_\2$ both with $t$-dependent factors $e^{-i g t}$ and $e^{+ i g t}$, respectively.
These conditions are similar to the findings of \cite{FS,DFS:ECS} in $\cN=1$.
Analogous conditions arise for the other backgrounds generated by $G^i{}_j$.

The dual potential associated with $G^i{}_j$ is simply $B_{(2)}{}^i{}_j = B_{(2)}\, (\ii\sigma_3)^i{}_j$, with $B_{(2)}$ given by \eqref{two form potential S3}.

\subsection{Warped AdS$_3$ spaces ($\mathrm{wAdS}_3\times\bbR$)}
\label{sec:AdS3}

Most of our discussion of the three-sphere can be repeated for backgrounds locally isomorphic to ${\rm AdS}_3\times \mathbb R$, so we defer most of the details to Appendix~\ref{app:LBack} and discuss mainly the differences.
These geometries are sourced by spacelike $G$ or $G^i{}_j$ and give rise to \SU(1,1|2) and $\SU(1,1|1)^2$ superalgebras respectively.
In the former case, the AdS$_3$ space can be warped by $\cZ$.
Without warping the two supercoset spaces are
\begin{equation}
\frac{\SU(1,1|2)}{\SU(2)_R}\qquad\text{and}\qquad
\frac{\SU(1,1|1)\times\SU(1,1|1)}{\U(1)_R}\,.
\end{equation}
The former case includes a central charge isometry generating the flat $\bbR$ direction.
In the latter case, as with the $\SU(2|1)^2$ sphere, the spacetime can be entirely generated by the even part of one \SU(1,1|1) factor, leaving half of the Killing spinors entirely constant.
Also in this case, if we compactify the flat direction to a circle we find similar restrictions on its radius, and a non-trivial connection for the axial $\U(1)_R$ might be required in order to have globally defined Killing spinors.
No such restriction is necessary for \SU(1,1|2) supersymmetry, where Killing spinors can be made entirely constant by a choice of local Lorentz gauge and for vanishing $R$-symmetry connections, analogously to the $\SU(2|2)$ sphere.

In each of these situations, \SU(1,1) generators $T_I$ with $I = 0,1,2$ can be defined and we can pick as group or coset representative
\begin{equation}
L = e^{\phi T_2} e^{\rho T_1} e^{\tau T_0}.
\end{equation}
We can regard AdS$_3\simeq\SU(1,1)$ either as a timelike $S^1$ fibered over an $H^2$ or as an $H^1$ fibered over AdS$_2$.
These two possible fibers correspond to the rightmost and leftmost factors of $L$ respectively.
Warped AdS$_3$ comes in different kinds depending on which fiber we deform, which is reflected in the orientation of $\cZ$ along the spacetime.
The unwarped metric reads
\begin{equation}
\label{metric AdS3}
\rd s^2 = \frac{1}{16g^2} ( -\rd\tau^2 +\rd\phi^2 +\rd\rho^2 -2\rd\tau\,\rd\phi\, \sinh\rho ) ~+\rd z^2,
\end{equation}
where $g^2$ is the norm of the background vector.
Two convenient orthonormal frames can be constructed from the left- and right-invariant Cartan--Maurer forms on AdS$_3$, defined as $\Omega = L^{-1}\rd L$ and $\Omega'=L\,\rd(L^{-1})$: they read
\begin{align}
\begin{aligned}
\label{SU(1,1) invariant vielbein}
E^0 &= \rd\tau+\sinh\rho\,\rd\phi,                     \\
E^1 &= \cos\tau\,\rd\rho+\sin\tau\cosh\rho\,\rd\phi,   \\
E^2 &= -\sin\tau\,\rd\rho+\cos\tau\cosh\rho\,\rd\phi, 
\end{aligned}
&\quad&
\begin{aligned}
E'^0 &= -\cosh\phi\cosh\rho\,\rd\tau - \sinh\phi\,\rd\rho,\\
E'^1 &= -\cosh\phi\,\rd\rho - \sinh\phi\cosh\rho\,\rd\tau,\\
E'^2 &= -\rd\phi + \sinh\rho\,\rd\tau.
\end{aligned}
\end{align}
Both sets satisfy the same structure equations 
$2\,\rd E^I = - \eta^{IJ} \veps_{JKL} E^K \wedge E^L$ with $\eta_{IJ}=\text{diag}(-1,\,1,\,1)$.
Each choice privileges a different Hopf fibration of AdS$_3$.
The coordinates we use can be related to embedding coordinates defining AdS$_3$ as the surface $X^2-Y^2-V^2-W^2=1$, parameterized as
\begin{align}
\begin{split}
\label{eq:AdS3 embedding coordinates}
X &= \cos\tfrac{\tau}{2} \cosh\tfrac{\rho}{2} \cosh\tfrac{\phi}{2} - \sin\tfrac{\tau}{2} \sinh\tfrac{\rho}{2} \sinh\tfrac{\phi}{2}  \,, \\
Y &= \cos\tfrac{\tau}{2} \sinh\tfrac{\rho}{2} \sinh\tfrac{\phi}{2} + \sin\tfrac{\tau}{2} \cosh\tfrac{\rho}{2} \cosh\tfrac{\phi}{2}  \,, \\
V &= \cos\tfrac{\tau}{2} \sinh\tfrac{\rho}{2} \cosh\tfrac{\phi}{2} + \sin\tfrac{\tau}{2} \cosh\tfrac{\rho}{2} \sinh\tfrac{\phi}{2}  \,, \\
W &= \cos\tfrac{\tau}{2} \cosh\tfrac{\rho}{2} \sinh\tfrac{\phi}{2} - \sin\tfrac{\tau}{2} \sinh\tfrac{\rho}{2} \cosh\tfrac{\phi}{2}   \,.
\end{split}
\end{align}
The $\tau$ coordinate has periodicity $4\pi$ for global AdS$_3$.

\subsubsection{Timelike stretched AdS$_3\times \bbR$}

Let us now focus on \SU(1,1|2) supersymmetry and introduce warping.
This is allowed by breaking the right \SU(1,1) isometries of AdS$_3$ to a one-dimensional subgroup determined by $\cZ$.
The first case we consider corresponds to a timelike stretching of AdS$_3$, obtained by turning on $\cZ$ along the $H^2$ base of the fibration $S^1\hookrightarrow{\rm AdS}_3\to H^2$.
Defining $G^2=g^2$, $|\cZ|^2=8\l^2$, we impose $\l^2<4g^2$.
We will later discuss the geometries obtained for larger $\l$.
The supercoset space is
\begin{equation}
\label{supercoset timelike tretched AdS3}
{\frac{\SU(1,1|2)_{(U,\, G\cdot P)}}{\mathrm U(1)_M\times \SU(2)_R}}
\end{equation}
analogously to the squashed $\bbR\times S^3$ case, the denominator $\U(1)_M$ mixes the compact isometry of \SU(1,1) with a central charge $U$, generating the warping.
The other central charge $G\cdot P= G^aP_a$ corresponds to the flat $\bbR$ direction (we can gauge fix it to $P_3$ for definiteness).
The resulting metric reads
\begin{align}
\label{metric timelike stretched ads3}
\rd s^2 &= 
\frac{\upsilon}{16g^2}[ -\upsilon(\rd\tau + \sinh\rho\,\rd\phi)^2
+ \rd\rho^2 + \cosh^2\rho\,\rd\phi^2 ]\, + \rd z^2,
\end{align}
with warping parameter $\u\equiv1/\big(1-\frac{\l^2}{4g^2}\big)\ge1$.
As a bosonic background, timelike `squashed' AdS$_3$ with $0<\upsilon<1$ would be also possible.
Supersymmetry restricts us to timelike `stretching' only.
It is known that for any value of the stretching this space contains closed timelike curves.
In particular, $\upsilon=2$ corresponds to G\"odel spacetime \cite{Rooman:1998xf,Bengtsson:2005zj}.

\subsubsection{Spacelike squashed AdS$_3\times \bbR$}

Now let us take $\cZ$ along an AdS$_2$ subspace of AdS$_3$ and define $|\cZ|^2=-8\l^2$.
The right isometries are broken to the $\SO(1,1)$ that preserves $\cZ$ and the supercoset reads
\begin{equation}
{\frac{\SU(1,1|2)_{(U,\, G\cdot P)}}{\SO(1,1)_M\times \SU(2)_R}}\,.
\end{equation}
This time the spacetime is squashed along the non-compact fiber over AdS$_2$, with metric
\begin{align}
\label{metric spacelike squashed ads3}
\rd s^2
&=
\frac{\upsilon}{16g^2}
\big[ 
-\cosh^2\rho\,\rd\tau^2+\rd\rho^2+\upsilon(\rd\phi+\sinh\rho\,\rd\tau)^2
\big]\,+\rd z^2\,,
\end{align}
with $0<\u\equiv1/\big( 1+\frac{\l^2}{4g^2} \big)\le1$.
In this case supersymmetry only allows for squashing rather than stretching.

\subsubsection{Lightlike warped AdS$_3\times\bbR$}

In this final case, we take $\cZ$ along a null surface determined for instance as the coset space 
$ \mathrm{AdS}_3 / N_- $,
the denominator being an (everywhere) null isometry.
We use the notation $N_{T}$ for the monoparametric subgroup of a parabolic generator $T$.
Unsurprisingly, the supercoset space becomes
\begin{equation}
{\frac{\SU(1,1|2)_{(U,\, G\cdot P)}}{N_M \times \SU(2)_R}}\,,
\end{equation}
where both $U$ and the residual Lorentz generator are null.
In our usual set of global coordinates the metric reads
\begin{align}
\label{metric null warped ads3}
\rd s^2  &= \frac{1}{16g^2}[-\rd\tau^2+\rd\phi^2+\rd\rho^2-2\sinh\rho\,\rd\tau\rd\phi
-\frac{\l^2}{4g^2} e^{2\phi}(\rd\rho+\cosh\rho\,\rd\tau)^2
]+\rd z^2.
\end{align}
A shift in $\phi$, which corresponds to one of the isometries of AdS$_3$ broken by $\cZ$, can absorb the absolute value of the squashing parameter.
The sign of the warping, however, is fixed by supersymmetry.
It can be convenient to rewrite this metric in Poincar\'e coordinates:
\begin{equation}
\label{metric null warped ads3 Poincarè}
\rd s^2 = \frac1{4g^2}\left( \frac{\rd r^2}{r^2} +\frac{\rd x_+\,\rd x_-}{r^2} -\frac{\l^2}{4g^2}\frac{\rd x_+^2}{r^4} \right) +\rd z^2\,.
\end{equation}

\subsection{AdS$_2\times S^2$ spacetimes and D$(2,1;\alpha)$}
\label{sec:AdS2xS2}

Another rich and interesting spacetime geometry is AdS$_2\times S^2$.
These geometries are sourced by a complex $\cZ$ flux alone, excluding the case in which it is entirely supported on a null hypersurface.
Then $\cZ$ is generally a complex linear combination of two real forms wrapping a timelike and a spacelike hypersurface respectively.
These forms source the AdS$_2\times S^2$ background, and when either vanishes the associated factor in the geometry is flattened.
The $\cN=2$ supersymmetric extensions of AdS$_2\times S^2$ are real forms of the $\rm D(2,1;\alpha)$ Lie superalgebras, where $\alpha$ is basically the ratio of the radii of AdS$_2$ and $S^2$
\cite{BILS:AdS}.

It can be useful to see how the D$(2,1;\alpha)$ superalgebra arises.
Let us gauge-fix the $\U(1)_R$ and local Lorentz gauge so that
\begin{equation}
\cZ_{ab}=2\ii\lambda_+\delta^{12}_{ab}-2\lambda_-\delta^{03}_{ab}, \qquad\lambda_\pm\in\bbR.
\end{equation}
The Riemann tensor then reads $R_{ab}{}^{cd} = 4(\lambda^2_-\delta^{03}_{ab}\delta_{03}^{cd}-\lambda_+^2\delta^{12}_{ab}\delta_{12}^{cd})$, with curvature radii $1/|\lambda_-|$, $1/|\lambda_+|$ for AdS$_2$ and $S^2$ respectively.
The isometry algebra is
\begin{equation}
[P_a,\,P_b] = -2\lambda^2_-\delta^{03}_{ab}M_{03}+2\lambda^2_+\delta^{12}_{ab}M_{12},
\end{equation}
corresponding to $\rm SU(1,1)\times SU(2)$.
For definiteness, an appropriate choice of coordinates gives us the explicit metric
\begin{equation}
\label{metric AdS2xS2}
\rd s^2  =  \frac1{\l_-^2} (-\rd\tau^2\,\cosh\rho + \rd \rho^2)
+ \frac1{\l_+^2} \left(\rd\theta^2+\sin^2\theta\,\rd\phi^2\right).
\end{equation}

The supercharges transform in the $(\mathbf2,\mathbf2,\mathbf2)$ of ${\rm SU(1,1)\times SU(2)\times SU(2)}_R$.
This can be made explicit by a similarity transformation on the four-component spinor $Q_{\hat\a}{}^i$ of \eqref{four component spinors}:
\begin{equation}
Q_{\tilde a\,\tilde\alpha}{}^i\equiv S_{\tilde a\,\tilde\alpha}{}^{\hat\a}\, Q_{\hat\a}{}^i\,,
\qquad
S\equiv
\begin{pmatrix}
1 & 0& 0& 0\\
0 &0 &0 & 1\\
0 &0 &-1 & 0\\
0 &1 &0 & 0
\end{pmatrix}
\end{equation}
where $\tilde a,\,\tilde\alpha$ are fundamental indices of SU(1,1) and SU(2) respectively.
The new set of gamma matrices reads
\begin{equation}
\tilde\gamma_0=\ii\sigma_1\otimes\sigma_3,\quad
\tilde\gamma_1=-\mathbb I_2\otimes\sigma_2,\quad
\tilde\gamma_2=\mathbb I_2\otimes\sigma_1,\quad
\tilde\gamma_3=\sigma_2\otimes\sigma_3,\quad
\tilde\gamma_5=\sigma_3\otimes\sigma_3.
\end{equation}
After some algebra, we obtain the commutation relations
\begin{align}
\{ Q_{\tilde a\,\tilde\alpha\,i},\ Q_{\tilde b\,\tilde\beta\,j} \} &=
2(\lambda_++\lambda_-)\epsilon_{\tilde a\tilde b}\epsilon_{\tilde\alpha\tilde\beta}I_{ij}
-2\lambda_+\epsilon_{\tilde a\tilde b}T_{\tilde\alpha\tilde\beta}\epsilon_{ij}
-2\lambda_-T_{\tilde a\tilde b}\epsilon_{\tilde\alpha\tilde\beta}\epsilon_{ij}\,, \eol{}
[T_{\tilde a\tilde b},Q_{\tilde c\,\tilde\alpha\,i}] &= 
\epsilon_{\tilde c (\tilde a\vphantom{\tilde b}} Q_{\tilde b)\tilde\alpha\,i}
,\qquad
[T_{\tilde \a\tilde \b},Q_{\tilde a\,\tilde\gamma\,i}] = 
\epsilon_{\tilde\gamma (\tilde \a\vphantom{\tilde \b}} Q_{\tilde a|\tilde\b)i}\,,
\end{align}
where the conventions for all $\epsilon$ symbols are the same as for standard spinor indices and we have defined
\begin{align}
T_{\tilde a\tilde b}&\equiv 
\begin{pmatrix}
\frac{\ii}{\l_-} (P_0+P_3)   & M_{03}                        \\
M_{03}                       & \frac{\ii}{\l_-}(-P_0+P_3) 
\end{pmatrix}\,,
\quad
T_{\tilde\alpha\tilde\beta}\equiv 
\begin{pmatrix}
\frac{1}{\l_+} (P_2+\ii P_1) & \ii  M_{12}                   \\
\ii M_{12}                   & \frac{1}{\l_+}(P_2 -\ii P_1) 
\end{pmatrix}\,,
\end{align}
so that they all have a commutator with the supercharges analogous to $I_{ij}$.
The reality conditions on the generators are
\begin{align}
(Q_{\tilde a\,\tilde\alpha\,i})^* &= (\sigma_3)_{\tilde a}{}^{\tilde b} \e^{\tilde\a\,\tilde\b} Q_{\tilde b\,\tilde\b}{}^i\,,\quad
(T_{\tilde a\tilde b})^* = -(\sigma_3)_{\tilde a}{}^{\tilde c} (\sigma_3)_{\tilde b}{}^{\tilde d} T_{\tilde c\tilde d}\,,\quad
(T_{\tilde\alpha\tilde\beta})^* = \e^{\tilde\a\,\tilde\g}\e^{\tilde\b\,\tilde\d}  T_{\tilde\g\tilde\d}\,.
\end{align}

The (anti)commutation relations we have just uncovered are the ones of the real form of D$(2,1;\alpha)$ with Lie subalgebra $\SU(1,1)\times \SU(2)\times \SU(2)_R$.
The parameter $\a$ can be defined as $\alpha\equiv\l_+/\l_-$ for $\l_-\neq0$.
We will slightly abuse the notation and include the case $\l_-=0,\l_+\neq0$ as $\alpha=\infty$.
There are four special subcases of the general algebra (see the discussion in \cite{BILS:AdS}):%
\footnote{We use the same symbol for the real and complex form of D$(2,1;\alpha)$.
Other real forms exist, but we will not encounter them.
Because the algebra is real, $\alpha$ must be real as well.} 
\begin{itemize}
\item D$(2,1;1)\simeq\rm OSp(4^*|2)$ corresponds to a superspace built on AdS$_2\times S^2$ with equal radii.
The space is superconformally flat, as reflected by the vanishing of the tensor $W^+_{ab} = - \cZ^+_{ab}$. 
It can be embedded in the superconformal group SU$(2,2|2)$.
\item D$(2,1;0)\simeq \rm SU(1,1|2)$, where the Lie subalgebra reduces to ${\rm SU(1,1)\times SU}(2)_R$ and the space is AdS$_2\times\mathbb R^2$.
The original SU(2) isometries of the sphere contract to ISO(2), the compact generator acting as an external automorphism.
\item D$(2,1;-1)\simeq \rm SU(1,1|2)$, where this time the SU(2) factor corresponds to the $S^2$ isometries and AdS$_2\times S^2$ have the same radii.
The SU$(2)_R$ group acts as external automorphisms.
This is the superalgebra obtained as the near horizon limit of supersymmetric black holes in $\cN=2,\ D=4$ supergravity.
\item D$(2,1;\infty)\simeq \rm SU(2|2)$, where AdS$_2$ is flattened to $\mathbb R^{1,1}$ and SU(1,1) is contracted to ISO(1,1), an SO(1,1) subgroup acting as an external automorphism.
\end{itemize}
All other values of $\alpha$ correspond to AdS$_2\times S^2$ with different radii.%
\footnote{There are two other isomorphisms.
First, $\alpha=-2$ gives rise to an OSp($4^*|2$) algebra, however it exchanges the roles of the spatial SU(2) with $\SU(2)_R$.
As a consequence, this case is not superconformally flat, as is clear from the fact that $T^+_{ab}$ is non-vanishing.
Second, $\alpha=-1/2$ gives rise to an OSp$(4|2)$ algebra.
}
Notice that for each choice of radii, there are two distinct supersymmetry algebras
depending on the sign of $\alpha$.
The supercosets built from these algebras can be encoded in the general expression
\begin{equation}
\frac{\rm D(2,1;\alpha)}{{\rm SO(1,1)\times U(1)\times SU(2)}_R},
\end{equation}
with the exception of the cases $\alpha=0,-1,\infty$, where the numerator includes respectively the ISO(2), SU$(2)_R$ or ISO(1,1) external automorphisms as a semidirect product.
The coset construction allows us to define Killing spinors with ease as usual, see Appendix~\ref{app:LBack}.

\subsection{Other geometries}

\subsubsection{Warped Lorentzian $S^3\times\bbR$}
\label{sec:Lorentzian S3}
Our analysis of rigid backgrounds also gives rise to more exotic solutions beyond the spaces discussed so far.
A first interesting class of less conventional geometries is given by SU(2) group manifolds with non-Euclidean metrics.
Regarding $S^3\sim\rm SU(2)$ as a Hopf fibration, these geometries correspond to taking the fiber circle to be either timelike or lightlike.

The first case we analyze is an $S^3$ with Lorentzian metric.
Such a manifold is easily defined using the standard left-invariant one--forms of SU(2) \eqref{eq:S3RoundV} and treating the one associated with the fiber as timelike.
This space corresponds to a radial section of Taub--NUT \cite{Misner:TaubNUT}.

The background is sourced by the same field configuration as timelike stretched AdS$_3\times \bbR$, i.e. a spacelike $G$ and a $\cZ$ two-form along a timelike hypersurface.
However, we now take $\l^2>4g^2$.
This induces a change of topology from timelike stretched AdS$_3$ to a compact space with $\SU(2)\times\U(1)$ isometries.
The supercoset is formally identical to the standard squashed $\bbR\times S^3$:
\begin{equation}
\label{lorentzian S3 supercoset}
\frac{ \SU(2|2)_{(U,\,G\cdot P)} } { \SU(2)_R \times \U(1)_M },
\end{equation}
though now the deformed $S^1$ fiber is regarded as the timelike direction.
The metric reads
\begin{align}
\label{metric Lorentzian S3}
\rd s^2 = \frac\upsilon{4g^2} [-\upsilon(\rd\omega+\cos\theta\,\rd\phi)^2+\rd\theta^2+\sin^2\theta\,\rd\phi^2] + \rd z^2,\quad 
\u\equiv\left(\frac{\l^2}{4g^2}-1\right)^{-1}>0.
\end{align}
Such a geometry admits only $\rm SU(2)\times U(1)$ as an isometry group, regardless of the value of $\upsilon$, because two of the (right) isometries are broken by the choice of Lorentzian signature.

\subsubsection{Lightlike $S^3\times \bbR$}

In 4D it is also possible to treat the Hopf fiber of $S^3$ as a lightlike direction, obtaining a metric of the form
\begin{align}
\label{metric lightlike S3}
\rd s^2 = \frac1{4\lambda^2}\left(
2\rd u(\rd\omega+\cos\theta\,\rd\phi)+\rd\theta^2+\sin^2\theta\,\rd\phi^2
\right).
\end{align}
The resulting space is (locally) isomorphic to a `lightlike' $S^3\times\bbR$.
Such a space arises from our analysis if we take $G$ null (but everywhere non-vanishing) and $\cZ$ spacelike with $|\cZ|^2\equiv8\lambda^2$.
Setting $\cZ=0$ yields a different geometry which is not topologically a sphere.
The supercoset space is again \eqref{lorentzian S3 supercoset}.
Killing spinors are constructed in Appendix~\ref{app:LBack}.
A full analysis of the global properties of this space goes beyond the scope of this paper, but we note that a global lightlike $S^3\times \bbR$ has closed null curves but no closed timelike curves.

\subsubsection{`Overstretched' AdS$_3$ (Heis$_3\times\bbR$)}
\label{ovestretched AdS3}

There is a threshold case between timelike stretched AdS$_3\times\bbR$ and the Lorentzian $S^3\times\bbR$, obtained for spacelike $G$ and spacelike $\cZ$ with $\l^2=4g^2$.
It corresponds to a non-semisimple contraction of the \SU(1,1|2) and \SU(2|2) superalgebra with ${\rm Heis}_3\rtimes\U(1)_M\times\bbR_{G\cdot P}$ bosonic isometries.
The geometry is in fact the group manifold $\mathrm{Heis}_3\times\bbR$ where the central charge corresponds to the timelike isometry.
The metric is
\begin{equation}
\label{metric relativistic fluid}
\rd s^2 = -(\rd t +2g\,x\rd y-2g\,y\rd x)^2 +\rd x^2+\rd y^2 +\rd z^2.
\end{equation}
This metric can also be interpreted as a rotating relativistic fluid.

\subsubsection{Plane waves}

Finally, there is a class of plane wave geometries admitting full $\cN=2$ supersymmetry, generalizing the supergravity solution of Kowalski-Glikman \cite{KowalskiGlikman:1985im}.%
\footnote{Also see \cite{KowalskiGlikman:1984wv,Blau:2001ne} for similar solutions of 11d and Type IIB supergravity.}
Some of these can be obtained as Penrose limits of the geometries encountered so far, following e.g. \cite{Blau:2002dy,Blau:2002mw}, but it is more convenient (and more general) to study these backgrounds independently.

Not surprisingly, plane waves are obtained for null $G$ or $G^i{}_j$, giving rise to the same spacetime with different realizations of $\cN=2$ supersymmetry.
Reflecting their origin as Penrose limits, these superalgebras are contractions of those encountered for $S^3\times\bbR$ and AdS$_3\times\bbR$.
At the bosonic level we can just restrict to a null, everywhere non-vanishing $G$. Then the most general plane wave geometry is obtained turning on also $\cZ$ along some null hypersurface parallel to $G$.

Let us show how the bosonic background arises.
We take $G^a=\frac{1}{\sqrt2}(g,0,0,g)=g\d^a_+$ and $\cZ_{ab}=2\sqrt2\lambda_+\delta_{ab}^{-1} -2\sqrt2\ii \lambda_- \delta_{ab}^{-2}$, where we can fix $\lambda_+\ge 0$, $\lambda_-\in\bbR$.
The cases $g=0$ and $\l_\pm=0$  are included in our analysis as all objects depend smoothly on these parameters.
When $\l_\pm=0$ the same bosonic background can be realized by $G^i{}_j$.
Defining $P_{\pm}\equiv\frac1{\sqrt2}(P_3\pm P_0)$, 
the non-vanishing commutators of the bosonic isometry algebra read
\begin{align}
&[P_a,\,P_b] = -4g\epsilon_{+ab}{}^cP_c+4\lambda_-^2\delta^{-2}_{ab}M_{+2}+4\lambda_+^2\delta^{-1}_{ab}M_{+1},\nonumber
&& (\epsilon_{+-12}=+1)
\\[.5ex]
&[M_{+d},P_-]=P_d,\qquad[M_{+d},P_d]=-P_+,&& (d=1,2).
\end{align}
This algebra is the semi-direct product of $P_-$ and a five-dimensional Heisenberg algebra (with central charge $P_+$).
The former generates the null direction complementary to the three-dimensional null space defined by $G$.
The residual Lorentz algebra contains the two null generators $M_{+1},\ M_{+2}$.
Then the general plane wave that we find is a coset space
\begin{equation}
\frac{\mathrm \bbR_{P_-}\ltimes\mathrm{Heis}_5}{N^2_{M_{+1},\, M_{+2}}}.
\end{equation}
When $\cZ=0$ this simplifies to the group manifold $\bbR_{P_-}\ltimes\mathrm{Heis}_3$.
We obtain the metric
\begin{align}
\label{metric plane wave}
\rd s^2 &= 2\rd u\,\rd v+4g(y\rd x-x\rd y)\rd u-2(\lambda_-^2x^2+\lambda_+^2y^2)\rd u^2+\rd x^2+\rd y^2.
\end{align}
Another possibility is to use Brinkmann coordinates, which we obtain via\begin{align}
x\to\hat x = x \cos 2gu + y \sin 2gu~, \qquad
y\to\hat y = -x \sin 2gu + y \cos 2gu~.
\end{align}
The metric then takes the standard plane wave form
\begin{align}
\label{metric plane wave brinkmann coords}
\rd s^2 &= 2\rd u\,\rd v+A_{\sf mn}(gu)\hat x^{\sf m} \hat x^{\sf n}\rd u^2+\delta_{\sf mn}\rd \hat x^{\sf m}\rd \hat x^{\sf n},\qquad {\sf m}=1,2\ ,
\\[1ex]
A(gu) &= -(4g^2+\l_+^2+\l_-^2)\mathbb I_2
+(\l^2_+-\l^2_-)
\begin{pmatrix}
\cos4gu & -\sin4gu \\
-\sin4gu & -\cos4gu
\end{pmatrix}\ .
\end{align}


\section{Rigid supersymmetric actions} \label{sec:L_Actions}
Each of the spaces we have discussed admits eight Killing spinors and is
described by the same rigid $\cN=2$ supersymmetry algebra \eqref{eq:LSusyAlg},
parametrized by the background fields $S^{ij}$, $\cZ_{ab}$, $G_a$, and $G_a{}^i{}_j$.
This suggests it should be possible to construct in a unified way
the supersymmetric vector multiplet and hypermultiplet actions for these
spaces simultaneously: one should simply insert the relevant values
for the metric and the other background fields.
That will be the goal of this section. We describe first
off-shell vector multiplets and on-shell hypermultiplets in these geometries.
After this, we discuss how some of these actions could be derived directly from
conformal supergravity by taking a certain rigid limit. Finally, as an explicit
example, we describe the $\cN=2^*$ action in a rigid background.

\subsection{Vector multiplets}
We will begin our discussion of vector multiplets with the abelian case, where
the situation is comparably simpler, before gauging
the isometries of the special K\"ahler manifold.

\subsubsection*{Abelian vector multiplets}
An abelian $\cN=2$ vector multiplet consists of a complex scalar $X^I$, a Weyl
fermion $\l_{\alpha i}{}^I$, a pseudoreal auxiliary triplet $Y^{ij I}$, and a real
connection $A_m{}^I$. In Lorentzian signature,
\begin{align}
(X^I)^*= \bar X^I~, \qquad (\l_{\alpha i}{}^I)^* = \bar \l_{\dalpha}{}^i{}^I~, \qquad
(Y^{ij I})^* = Y_{ij}{}^I~.
\end{align}
In rigid superspace, these components are contained within a complex superfield $\cX^I$
which is chiral, $\bar \cD_{\dalpha i} \cX^I = 0$, and obeys a superspace Bianchi
identity \cite{GrimmSohniusWess, KLRT-M1}
\begin{align}\label{eq:VM_BI}
(\cD^{ij} + 4 S^{ij}) \cX^I = (\bar \cD^{ij} + 4 \bar S^{ij}) \bar \cX^I~.
\end{align}
The component fields are defined using the superfield via
\begin{gather}
X^I := \cX^I \loco~, \qquad  \bar X^I := \bar \cX^I \loco~, \eol
\l_{\alpha i}{}^I := -\cD_{\alpha i} \cX^I \loco~, \qquad
\bar\l^{\dalpha i \, I} := \bar\cD^{\dalpha i} \bar \cX^I \loco~, \eol
Y^{ij I} := -\frac{1}{2} (\cD^{ij} + 4 S^{ij}) \cX^I \loco~, \eol
F_{ab}{}^I = \frac{1}{4} (\sigma_{ab})^{\beta \alpha} \cD_{\beta\alpha} \cX^I\loco
	+ \frac{1}{4} (\bsigma_{ab})^{\dbeta \dalpha} \bar\cD_{\dbeta \dalpha} \bar \cX^I \loco
	- \newG_{ab} X^I 
	- \bar \newG_{ab} \bar X^I~,
\label{eq:VM_Comps}
\end{gather}
where the vertical bar denotes taking the $\q=0$ projection.

As a consequence of the superspace Bianchi identity,
$F_{mn}{}^I$ is a field strength for a one-form $A_m{}^I$.
Both of these can be lifted to superspace forms 
$\cA^I$ and $\cF^I$.
One must constrain the tangent space components
$\cF_{AB}{}^I$
so that it is entirely determined in terms of the superfield
$\cX^I$:
\begin{gather}
\cF_{\alpha}^i{}_\beta^j{}^I = -4 \,\eps_{\alpha\beta} \eps^{ij} \bar \cX^I~, \qquad
\cF^{\dalpha}_i{}^\dbeta_j{}^I = 4 \,\eps^{\dalpha\dbeta} \,\eps_{ij} \cX^I~, \eol
\cF_\alpha^i{}_{\beta \dbeta}{}^I = 2i \,\eps_{\alpha\beta} \,\bar \cD_\dbeta^i \bar \cX^I~, \qquad
\cF_{\dalpha i}{}_{\,\beta \dbeta}{}^I = 2i \,\eps_{\dalpha\dbeta} \cD_{\beta i} \cX^I~, \eol
\cF_{ab}{}^I = \frac{1}{4} (\sigma_{ab})^{\beta \alpha} \cD_{\beta\alpha} \cX^I
	+ \frac{1}{4} (\bsigma_{ab})^{\dbeta \dalpha} \bar\cD_{\dbeta \dalpha} \bar \cX^I
	- \newG_{ab} \cX^I 
	- \bar \newG_{ab} \bar \cX^I~.
\label{eq:VM_SuperF}
\end{gather}

Because the vector multiplet includes its auxiliary field, the supersymmetry algebra
closes off-shell. The variations of the component fields are
\begin{subequations}
\begin{align}
\delta X^I &= -\xi^\alpha_i \l_\alpha^i{}^I~, \qquad \qquad
\delta \bar X^I = \bar \xi_\dalpha^i \bar \l^\dalpha_i{}^I~, \\[0.5ex]
\delta \l_{\alpha i}{}^I &=
	(F_{ab}{}^I + \cZ_{ab} X^I + \bar \cZ_{ab}\bar X^I) (\sigma^{ab} \xi_i)_\alpha
	+ (Y_{ij}{}^I + 2 S_{ij} X^I) \xi_\alpha{}^j
	\eol & \quad
	- 2 i \,\cD_{a} X^I \, (\sigma^a \bar \xi_i)_\alpha
	+ 4i \,G_{a\, ij} X^I \, (\sigma^a \bar\xi^{j})_\alpha~, \\[0.5ex]
\delta \bar \l^{\dalpha i}{}^I &=
	(F_{ab}{}^I + \cZ_{ab} X^I + \bar \cZ_{ab}\bar X^I) (\bsigma^{ab} \xi^i)^\dalpha
	- (Y^{ij}{}^I + 2 \bar S^{ij} \bar X^I) \bar \xi^\dalpha{}_j
	\eol & \quad
	+ 2 i \,\cD_a \bar X^I \, (\bsigma^a \xi^i)^\dalpha
	+ 4i \,G_{a}{}^{ij} \bar X^I \,(\bsigma^a \xi_{j})^\dalpha ~, \\[0.5ex]
\delta Y_{ij}{}^I &=
	2i \,\xi_{(i} {\slashed{\cD}} \bar \l_{j)}{}^I
	- 4i G_a{}_{k (i} \,\xi_{j)} \sigma^a \bar\l{}^{k I}
	- 2 G_a \, \xi_{(i} \sigma^a \bar \l_{j)}{}^I
	\eol & \quad
	- 2i \,\bar\xi_{(i} {\slashed{\cD}} \l_{j)}{}^I
	+ 4i G_a{}_{k (i} \,\bar \xi_{j)} \bsigma^a \l{}^{k I}
	- 2 G_a \, \bar \xi_{(i} \bsigma^a \l_{j)}{}^I
	~, \\[0.5ex]
\delta A_m{}^I &= i (\xi_j \sigma_m \bar \l^j{}^I) + i (\bar \xi^j \bsigma_m \l_j{}^I)~.
\end{align}
\label{eq:VM_SUSYAb}
\end{subequations}

\subsubsection*{Abelian vector multiplet action}
The vector multiplet actions are straightforward to construct in superspace.
They are given by a chiral superspace integral of a holomorphic
prepotential function $F(\cX^I)$,
\begin{align}\label{eq:VM_SuperAction}
-i \int \rd^4x\, \rd^4\q\, \cE\, F(\cX)
+ i \int \rd^4x\, \rd^4\bar\q\, \bar\cE\, \bar F(\bar\cX)~.
\end{align}
In a Minkowski background, or indeed in any rigid $\cN=2$ background with
$G_a{}^{ij} = 0$, the prepotential needs to satisfy no further
restrictions. However, in a rigid background with nonvanishing $G_a{}^{ij}$,
the $\U(1)_R$ symmetry forces the prepotential to be homogeneous of weight two
-- that is, it must be a superconformal model.

Deriving the component action of \eqref{eq:VM_SuperAction} is a straightforward
application of superspace techniques (see e.g. \cite{BN:CR}).
As in the rigid Minkowski case, the sigma model is an affine special K\"ahler manifold
(see e.g. \cite{CRTvP:WhatIsSK}) with K\"ahler potential $K$ and metric $g_{IJ}$:
\begin{align}
K = i X^I \bar F_I -i \bar X^I F_I~, \qquad g_{IJ} = -i F_{IJ} + i \bar F_{IJ}~.
\end{align}
The leading bosonic kinetic terms are
\begin{align}
\cL = - g_{IJ}\, \cD_a X^I \cD^a \bar X^J
	+ \frac{i}{4} F_{IJ} F_{ab}^{- I} F^{ab - J}
	- \frac{i}{4} \bar F_{IJ} F_{ab}^{+I} F^{ab + J}
	+ \cdots~.
\end{align}
We will give the full Lagrangian, including gauged isometries,
in due course, but for now there are some
important features to discuss in the ungauged case.
Foremost is that the action admits the same group of duality transformations as
in flat space. These can be seen easily in the superspace description,
where the superfield equations of motion imply that the chiral superfield
$F_I(\cX)$ also obeys the constraint \eqref{eq:VM_BI}.
The duality transformations belong to the inhomogeneous symplectic group
${\rm ISp}(2n, \mathbb R)$,
\begin{align}
\begin{pmatrix}
\cX^I \\
F_I
\end{pmatrix} \longrightarrow
\begin{pmatrix}
U^I{}_J & Z^{IJ} \\
W_{IJ} & V_I{}^J
\end{pmatrix}
\begin{pmatrix}
\cX^J \\
F_J
\end{pmatrix}
+ 
\begin{pmatrix}
C^I \\
C_I
\end{pmatrix}
\label{eq:VM_ISP}
\end{align}
where the matrix is an element of $\rm Sp(2n, \mathbb R)$ and $C^I$ and $C_I$ are constant complex
numbers.\footnote{One can also consider a global ${\rm U}(1)_R$ transformation
as in the Minkowski case, but it will not play a role in what follows.}
When $S^{ij}$ is nonzero, $C^I$ must obey the extra constraint
$S^{ij} C^I = \bar S^{ij} \bar C^I$, and similarly for $C_I$, as a consequence
of \eqref{eq:VM_BI}.
One may decompose the $C^I$ and $C_I$ in terms of real parameters $U^I{}_{(j)}$ and $W_{I (j)}$ with $j = 1, 2$, so that
\begin{align}
C^I = U^I{}_{(1)} + i U^I{}_{(2)}~, \qquad
C_I = W_{I (1)} + i W_{I (2)}~.
\end{align}
This suggestive decomposition reflects the presence of so-called background vector
multiplets in the rigid supersymmetric geometries we have been discussing.

\subsubsection*{Background vector multiplets and central charges}
There is a close relationship between background vector multiplets and the possibility
of extending the supersymmetry algebra by a complex internal central charge $Z$.
Provided that $G_a{}^{ij} = 0$, we can deform the algebra with the extra terms
\begin{align}
\{\cD_\alpha{}^i, \cD_\beta{}^j\} \sim 4 \eps_{\alpha\beta} \eps^{ij} \bar Z~, \quad
\{\bar\cD^\dalpha{}_i, \bar \cD^\dbeta{}_j\} \sim -4 \eps^{\dalpha\dbeta} \eps_{ij} Z~, \quad
[\cD_a, \cD_b] \sim \cZ_{ab} Z + \bar \cZ_{ab} \bar Z~.
\end{align}
The covariant derivatives $\cD_M$ now carry a complex central charge
connection, with the background field $\cZ_{ab}$ playing the role of the bosonic
field strength for $Z$.
These observations can be clarified if we write the complex
central charge in terms of two real central charges, $Z = Z_{(1)} + i Z_{(2)}$, and 
interpret the algebra of covariant derivatives above as
\begin{align}
[\cD_A, \cD_B] \sim -\cF_{AB}{}^{(i)} Z_{(i)}
\end{align}
where $\cF_{MN}{}^{(i)}$ is the field strength for two real abelian one-forms
$\cA_M{}^{(i)}$. Comparing to \eqref{eq:VM_SuperF}, one can see that the field strengths
$\cF_{AB}{}^{(i)}$ are associated with constant vector multiplet superfields
$\cX^{(1)} = 1$ and $\cX^{(2)} = i$.
In particular, the background vectors possess non-vanishing field strengths
with one-form potentials (see \eqref{eq:VM_SuperF}),
\begin{alignat}{2}
F_{ab}{}^{(1)} &= - 2 \,\textrm{Re}\, \cZ_{ab}~, &\qquad
F_{ab}{}^{(2)} &= 2 \,\textrm{Im}\, \cZ_{ab}~, \eol
A_m{}^{(1)} &= - 2 \,\textrm{Re}\, C_{m}~, &\qquad
A_m{}^{(2)} &= 2 \,\textrm{Im}\, C_{m}~.
\end{alignat}
More generally, we can choose $\cX^{(1)}$
to possess any phase and take $\cX^{(2)} = i \cX^{(1)}$.
A subtlety emerges when $S^{ij}$ is non-vanishing: then \eqref{eq:VM_BI} implies
$S^{jk} \cX^{(i)} = \bar S^{jk} \bar \cX^{(i)}$, and so only one independent choice
of $\cX^{(i)}$ is possible. For pseudoreal $S^{ij}$,
we take $\cX^{(1)} = 1$ and drop $\cX^{(2)}$.

These observations readily admit a simple explanation of the inhomogeneous symplectic
transformation \eqref{eq:VM_ISP}. Rewriting that transformation as
\begin{align}
\begin{pmatrix}
\cX^I \\
F_I
\end{pmatrix} \longrightarrow
\begin{pmatrix}
U^I{}_J & Z^{IJ} \\
W_{IJ} & V_I{}^J
\end{pmatrix}
\begin{pmatrix}
\cX^J \\
F_J
\end{pmatrix}
+ 
\begin{pmatrix}
U^I{}_{(j)} \cX^{(j)} \\
W_{I (j)} \cX^{(j)}
\end{pmatrix}
\label{eq:VM_ISPnew}
\end{align}
one finds that it can be embedded into $\mathrm{Sp}(2n+4, \mathbb R)$ acting on
the $2n+4$ vector
$(\cX^{\hat I}, F_{\hat I})$
where $\cX^{\hat I} = (\cX^I, \cX^{(i)})$. The prepotential $F(\cX^{\hat I})$ should be taken as
a homogeneous prepotential with two constant vector multiplets $\cX^{(i)}$.\footnote{When
both constant vector multiplets are present, the lift of the original inhomogeneous
$F(\cX^I)$ to the homogeneous $F(\cX^{\hat I})$ is not unique.}
Because the background multiplets are never placed on-shell,
$F_{(i)}$ does not obey \eqref{eq:VM_BI} and must not
mix into the other multiplets: this zeroes out the entries
$Z^{I (j)}$ and $V_I{}^{(j)}$ of the $\mathrm{Sp}(2n+4, \mathbb R)$ matrix,
recovering \eqref{eq:VM_ISPnew}. Imposing in addition the invariance of
the constant $\cX^{(i)}$ determines the other entries.

\subsubsection*{Non-abelian (gauged) vector multiplets}
Until now we have been dealing with abelian (ungauged) vector multiplets. Before
discussing the action in detail, we should allow for the possibility of gauging
isometries of the special K\"ahler manifold.
It is sufficient to discuss the group of isometries on the one-forms $\cA^I$.
Taking into account the additional background one-forms $\cA^{(i)}$,
we are led to consider
\begin{align}
\delta \cA^I &= \rd \L^I
	+ \cA^J (\L^K f_{KJ}{}^I + \L^{(k)} f_{(k)J}{}^I)
	+ \cA^{(j)} (\L^K f_{K (j)}{}^I+ \L^{(k)} f_{(k) (j)}{}^I)
	~,\eol
\delta \cA^{(i)} &= \rd \L^{(i)}~.
\end{align}
Collectively
these can be written
$\delta \cA^{\hat I} = \rd \L^{\hat I} + \cA^{\hat J} \L^{\hat K} f_{\hat K \hat J}{}^{\hat I}$.
The Lie algebra here is 
\begin{align}
[T_I, T_J] = f_{IJ}{}^K T_K~, \qquad
[Z_{(i)}, T_J] = f_{(i) J}{}^K T_K~, \qquad
[Z_{(i)}, Z_{(j)}] = f_{(i) (j)}{}^K T_K~.
\end{align}
In the gauged case, $Z_{(i)}$ no longer commute with the gauge generators
but we will still occasionally refer to them as central charges.
The superfields $\cX^I$ and $\cX^{(i)}$ transform as
\begin{align}
\delta \cX^I =
	\cX^J (\L^K f_{KJ}{}^I + \L^{(k)} f_{(k)j}{}^I)
	+ \cX^{(j)} (\L^K f_{K (j)}{}^I + \L^{(k)} f_{(k) (j)}{}^I)~, \qquad
\delta \cX^{(i)} = 0~.
\end{align}
Collectively these can be written as
$\delta \cX^{\hat I} = \cX^{\hat J} \L^{\hat K} f_{\hat K \hat J}{}^{\hat I}$.
Comparing with \eqref{eq:VM_ISPnew}, one can see that this gauges a subgroup
of ${\rm ISp}(2n, \mathbb R)$ with infinitesimal elements
\begin{align}
u^I{}_J = \L^K f_{K J}{}^I + \L^{(k)} f_{(k) J}{}^I~, \qquad
u^I{}_{(j)} = \L^K f_{K (j)}{}^I + \L^{(k)} f_{(k) (j)}{}^I~.
\end{align}
The full embedding of the gauge group may also involve $w_{I \hat J}$, as we
will discuss shortly.

Now defining the superspace covariant derivatives to carry the connections $\cA^I$
and $\cA^{(i)}$, the SUSY transformations of the vector multiplets $\cX^I$ can be
determined; they differ only slightly from
\eqref{eq:VM_SUSYAb}. It is convenient to introduce the Killing vector $J_{\hat J}{}^I$
defined by
\begin{align}
\delta X^I = \L^{\hat J} J_{\hat J}{}^I~, \qquad
J_{\hat J}{}^I = - X^{\hat K} f_{\hat K\hat J}{}^I~, \qquad
\end{align}
Including a uniform coupling constant $g$ to track the gauging
terms, one finds
\begin{subequations}
\begin{align}
\delta X^I &= -\xi^\alpha_i \l_\alpha^i{}^I~, \qquad \qquad
\delta \bar X^I = \bar \xi_\dalpha^i \bar \l^\dalpha_i{}^I~, \displaybreak[1] \\[0.5ex]
\delta \l_{\alpha i}{}^I &=
	(F_{ab}{}^I + \cZ_{ab} X^I + \bar \cZ_{ab}\bar X^I) (\sigma^{ab} \xi_i)_\alpha
	+ (Y_{ij}{}^I + 2 S_{ij} X^I) \xi_\alpha{}^j
	\eol & \quad
	- 2 i \,\cD_{a} X^I \, (\sigma^a \bar \xi_i)_\alpha
	+ 4i \,G_{a\, ij} X^I \, (\sigma^a \bar\xi^{j})_\alpha
	- 2 g \,\bar X^{\hat J} J_{\hat J}{}^I\,\xi_\alpha{}_i 
	~, \displaybreak[1] \\[0.5ex]
\delta \bar \l^{\dalpha i}{}^I &=
	(F_{ab}{}^I + \cZ_{ab} X^I + \bar \cZ_{ab}\bar X^I) (\bsigma^{ab} \bar \xi^i)^\dalpha
	- (Y^{ij}{}^I + 2 \bar S^{ij} \bar X^I) \bar \xi^\dalpha{}_j
	\eol & \quad
	+ 2 i \,\cD_a \bar X^I \, (\bsigma^a \xi^i)^\dalpha
	+ 4i \,G_{a}{}^{ij} \bar X^I \,(\bsigma^a \xi_{j})^\dalpha
	- 2 g \,X^{\hat J} \bar J_{\hat J}{}^I \, \bar \xi^\dalpha{}^i 
	~, \displaybreak[1] \\[0.5ex]
\delta Y_{ij}{}^I &=
	2i \,\xi_{(i} {\slashed{\cD}} \bar \l_{j)}{}^I
	- 4i G_a{}_{k (i} \,\xi_{j)} \sigma^a \bar\l{}^{k I}
	- 2 G_a \, \xi_{(i} \sigma^a \bar \l_{j)}{}^I
	\eol & \quad
	- 2i \,\bar\xi_{(i} {\slashed{\cD}} \l_{j)}{}^I
	+ 4i G_a{}_{k (i} \,\bar \xi_{j)} \bsigma^a \l{}^{k I}
	- 2 G_a \, \bar \xi_{(i} \bsigma^a \l_{j)}{}^I
	\eol & \quad
	- 4 g\, \xi_{(i} \l_{j)}{}^J\, \bar J_{J}{}^I
	- 4 g\, \bar\xi_{(i} \bar \l_{j)}{}^J\, J_J{}^I
	~, \displaybreak[1] \\[0.5ex]
\delta A_m{}^I &= i (\xi_j \sigma_m \bar \l^j{}^I) + i (\bar \xi^j \bsigma_m \l_j{}^I)~.
\end{align}
\label{eq:VM_SUSYNAb}
\end{subequations}
The vector derivative carries the full set of connections, i.e.
\begin{align}
\cD_a X^I = e_a{}^m \Big(\pa_m X^I
	- 2 i A_m X^I
	+ g\, A_m{}^{\hat J} X^{\hat K} f_{\hat K \hat J}{}^I
	\Big)~,
\label{eq:VM_DX}
\end{align}
with the contributions from the constants $X^{(k)}$ gauging inhomogeneous
transformations.
Similarly, the field strengths $F_{mn}{}^I$ are now given by
\begin{align}
F_{mn}{}^I = 2 \pa_{[m} A_{n]}{}^I + g\, A_m{}^{\hat J} A_n{}^{\hat K} f_{\hat K \hat J}{}^I
\end{align}

\subsubsection*{Vector multiplet action with gauged isometries}
Now we will give the final form of the action, including gauged isometries.
Assuming the prepotential $F$ is gauge-invariant, we calculate
the component reduction of \eqref{eq:VM_SuperAction}:
\begin{align}
\cL &= -g_{IJ} \cD_a X^I \cD^a \bar X^J
	- \frac{i}{4} g_{IJ} \l_j^I \hat {\slashed{\cD}} \bar\l^{j J}
	- \frac{i}{4} g_{IJ} \bar\l^{j J} \hat {\slashed{\cD}} \l_j^I 
	\eol & \quad
	+ \frac{i}{4} F_{IJ} F_{ab}^{- I} F^{ab - J}
	- \frac{i}{4} \bar F_{IJ} F_{ab}^{+I} F^{ab + J}
	+ \frac{1}{8} g_{I J} Y^{ij I} Y_{ij}{}^J
	\eol & \quad
	+ 4 \,G^a \Big(F_I \cD_a \bar X^I+ \bar F_I \cD_a X^I\Big)
	+ Y_{ij}{}^I \cO^{ij}_I + F_{ab}{}^I \cO^{ab}_I
	+ \cL_{\rm pot} + \cL_{\rm ferm}~.
\label{eq:VM_Action1}
\end{align}
We have collected the moment couplings into $\cO^{ab}_I$ and
the terms linear in the auxiliary field into $\cO^{ij}_I$. Additional
terms contributing to the scalar potential are given in $\cL_{\rm pot}$
and additional fermionic terms appear in $\cL_{\rm ferm}$.
Before giving these expressions, let us comment on the leading terms
of \eqref{eq:VM_Action1}.
The covariant derivative of the scalars is given in \eqref{eq:VM_DX},
whereas the covariant derivative of the gauginos includes the special
K\"ahler connection,
\begin{align}
\hat \cD_a \l_{\alpha j}{}^I = \cD_a \l_{\alpha j}{}^I
	+ i g^{IJ}\, F_{JKL} \cD_a X^K\, \l_{\alpha j}{}^L~.
\end{align}
The terms involving the field strengths can be rewritten as usual as
\begin{align}
-\frac{1}{8} g_{IJ} F_{ab}{}^I F^{ab J}
	- \frac{1}{8} (F_{IJ} + \bar F_{IJ}) F_{ab}{}^I \tilde F^{ab J}
\label{eq:VM_F2}
\end{align}
where $F_{IJ} + \bar F_{IJ}$ describes a generalized $\theta$-term.
The expression involving $G^a$ in \eqref{eq:VM_Action1}
can be rewritten up to a total derivative using its dual two-form potential as
\begin{align}
4 G^a (F_I \cD_a \bar X^I + \bar F_I \cD_a X^I) &=
	\eps^{mnpq}\,B_{mn}\, \Big(2i \,g_{IJ} \, \cD_p X^I\, \cD_q \bar X^J
		- g\, F_{pq}{}^{\hat I} D_{\hat I}\Big) \eol
	&= 2i \, \eps^{mnpq}\,B_{mn}\, \pa_p X^I\, \pa_q \bar X^J\, g_{IJ} 
	+ \frac{2}{3} g \,\eps^{mnpq} H_{mnp} A_q{}^{\hat I} D_{\hat I}~.
\label{eq:VM_BF}
\end{align}
This involves a generalized $BF$ coupling between the background two-form $B$ and
the special K\"ahler two-form $-i\, g_{IJ}\, \rd X^I \wedge \rd \bar X^J$. Note that
we have dropped the $\U(1)_R$ connection $A_m$ above because when $G^a$ is non-zero
this connection is always taken to vanish.

The remaining terms in the Lagrangian involve
\begin{align}
\cO^{ab}_I &=
	\frac{1}{4} \eps^{abcd}\cZ_{cd} (F_I - \frac{1}{2} (F_{IJ} + \bar F_{IJ}) X^J)
	+ \frac{1}{4} \eps^{abcd} \bar\cZ_{cd} (\bar F_I - \frac{1}{2} (F_{IJ} + \bar F_{IJ}) \bar X^J)
	\eol & \quad
	- \frac{1}{4} g_{IJ} (X^J \cZ^{ab} + \bar X^J \bar \cZ^{ab})
	- \frac{i}{8} F_{IJK} \l^{k J} \sigma^{ab} \l_k^K
	+ \frac{i}{8} \bar F_{IJK} \bar \l^{k J} \bsigma^{ab} \bar\l_k^K~,
	\label{eq:VM_Oab} \displaybreak[1]\\[0.5ex]
\cO^{ij}_I &=
	\frac{i}{2} S^{ij} (F_I - F_{IJ} X^J)
	- \frac{i}{2} \bar S^{ij} (\bar F_I - \bar F_{IJ} \bar X^J)
	+ \frac{i}{8} F_{IJK} (\l^{i J} \l^{j K})
	- \frac{i}{8} \bar F_{IJK} (\bar \l^{i J} \bar \l^{j K})~, \displaybreak[1]\\[-2.0ex]
\cL_{\rm pot}
	&=
	- \frac{1}{8} g_{IJ} (X^I\cZ_{ab} + \bar X^I \bar\cZ_{ab})(X^J \cZ^{ab} + \bar X^J \bar \cZ^{ab})
	\eol & \quad
	- \frac{1}{8} (F_{IJ} + \bar F_{IJ}) \Big(
	X^I\cZ_{ab} + \bar X^I \bar \cZ^{ab}\Big)
	\Big(X^J \tilde \cZ^{ab} + \bar X^J \tilde{\bar \cZ}^{ab}\Big)
	\eol & \quad
	+ \frac{1}{2}\cZ_{ab} \tilde \cZ^{ab} (F_I X^I - F)
	+ \frac{1}{2} \bar\cZ_{ab} \tilde {\bar \cZ}^{ab} (\bar F_I \bar X^I - \bar F)
	\eol & \quad
	+ i S^{ij} S_{ij} (2 F_I X^I - \frac{1}{2} X^I X^J F_{IJ} - 3 F)
	- i \bar S^{ij} \bar S_{ij} (2 \bar F_I \bar X^I - \frac{1}{2} \bar F_{IJ} \bar X^I \bar X^J
		- 3 \bar F)	
	\eol & \quad
	+ S^{ij} \bar S_{ij} K
	+ 2 G^{a \, ij} G_{a\, ij} K
	- g^2 g^{IJ} D_{I} D_J~,
	\label{eq:VM_Lpot} \displaybreak[1] \\[0.5ex]
\cL_{\text{ferm}} &=
	g_{IJ} \,\l_i^I \sigma^a \bar \l^{j J} \Big(
		\frac{1}{2} \delta^i_j G_a + i G_a{}^i{}_j\Big)
	+ \frac{i}{4} (\l_i^I \l_j^J) S^{ij} F_{IJK} X^K
	- \frac{i}{4} (\bar \l^{iI} \bar \l^{jJ}) \bar S_{ij} \bar F_{IJK} \bar X^K
	\eol & \quad
	- \frac{i}{8} (\l^{k I} \s^{ab} \l_k^J) (\bar \cZ_{ab} \bar X^K +\cZ_{ab} X^K)\, F_{IJK} 
	+ \frac{i}{8} (\bar \l^{k I} \bsigma^{ab} \bar \l_k^J)
		(Z_{ab} X^K + \bar\cZ_{ab} \bar X^K)\,\bar F_{IJK} 
	\eol & \quad
	- \frac{i}{48} F_{IJKL} (\l^{i I} \l^{j J})(\l_i^{K} \l_j^L)
	+ \frac{i}{48} \bar F_{IJKL} (\bar \l^{iI} \bar \l^{j J})(\bar \l_i^{K} \bar \l_j^L)
	\eol & \quad
	- \frac{1}{2} g \l^{k I} \l_k^J \,g_{IK} \bar J_{J}{}^K
	- \frac{1}{2} g \bar \l^{k I} \bar \l_k^J \,g_{IK} J_J{}^K~.
\end{align}
We have utilized the Killing potentials (or moment maps) for $J_{\hat I}{}^J$ given by
\begin{align}
D_{\hat I} = f_{\hat I \hat J}{}^{K} (X^{\hat J} \bar F_K + \bar X^{\hat J} F_K)~.
\end{align}
The moment maps $D_I$ associated with the dynamical vector multiplets
may also be written
\begin{align}\label{eq:DINice}
D_I = -i g_{IJ} X^{\hat K} \bar X^{\hat L} f_{\hat K \hat L}{}^J~.
\end{align}

Suppose now that the prepotential $F$ is not gauge invariant but transforms as
\begin{align}
\delta F = \frac{1}{2} \L^{\hat K} C_{\hat K, \hat I \hat J} X^{\hat I} X^{\hat J}~,
\end{align}
with real $C_{\hat K, \hat I \hat J}$.
This transformation lies in a subgroup of \eqref{eq:VM_ISPnew} with infinitesimal
$w_{I \hat J} = \L^{\hat K} C_{\hat K, I \hat J}$.
Provided we write the $B F$ coupling as in \eqref{eq:VM_BF}, the
Lagrangian fails to be gauge invariant only due to the generalized
$\theta$ term involving $F_{IJ} + \bar F_{IJ}$ in \eqref{eq:VM_F2},
the moment couplings involving $\eps^{abcd} \cZ_{cd}$ in \eqref{eq:VM_Oab},
and the potential terms involving products like
$\cZ_{ab} \cZ_{cd} \eps^{abcd}$ in \eqref{eq:VM_Lpot}. These have a natural interpretation
as generalized $\theta$ terms involving the background vector
multiplets.
As in the Minkowski case, gauge invariance can be restored by adding
the Chern-Simons like term \cite{dWLvP, dWHR}
\begin{align}
\cL_{\rm CS-like} = -\frac{2}{3} \,g \,\eps^{mnpq} \,C_{\hat K, \hat I \hat J}\,
	A_m{}^{\hat K}\, A_n{}^{\hat I}
	(\pa_p A_q{}^{\hat J} - \frac{3}{8} g f_{\hat M \hat N}{}^{\hat J} A_p{}^{\hat M} A_q{}^{\hat N})~,
\end{align}
which involves both the physical connections $A_m{}^I$ and the background
connections $A_m{}^{(i)}$. In addition, one must modify the Killing potentials
in the various expressions above to
\begin{align}
D_{\hat I} = f_{\hat I \hat J}{}^{K} (X^{\hat J} \bar F_K + \bar X^{\hat J} F_K)
	- C_{\hat I, \hat J \hat K} X^{\hat J} \bar X^{\hat K}~,
\end{align}
with \eqref{eq:DINice} still holding.
These modifications also restore supersymmetry.

\subsubsection*{BPS conditions for the vector multiplet}
From the supersymmetry transformations \eqref{eq:VM_SUSYNAb}, we may
characterize the moduli space of
supersymmetric configurations for a vector multiplet in a generic rigid background.
Recall that in Minkowski spacetime a supersymmetric configuration for a
vector multiplet is given by constant scalar $X^I$ and vanishing fermions,
field strengths, and auxiliary fields.
In a generic rigid $\cN=2$ background characterized by the background fields
$S^{ij}$, $\cZ_{ab}$, $G_a$, and $G_a{}^i{}_j$, the situation differs.

Let us first assume that $G_a{}^{ij} = 0$ and that we have eliminated the
spacetime ${\rm U}(1)_R$ connection. Requiring $\delta X^I = 0$ for eight linearly
independent supercharges implies that the fermions $\l_{\alpha i}{}^I$ vanish.
Requiring $\delta \l_{\alpha i}{}^I = 0$ leads to the additional constraints 
\begin{alignat}{2}\label{eq:VM_BPS}
Y_{ij}{}^I &= -2 S_{ij} X^I = -2 \bar S_{ij} \bar X^I~, &\qquad
\cD_a X^I &= 0~, \eol
F_{ab}{}^I &= -\cZ_{ab} X^I - \bar \cZ_{ab} \bar X^I~, &\qquad
X^{\hat J} \bar X^{\hat K} f_{\hat K \hat J}{}^I &= 0~.
\end{alignat}
The first condition fixes the auxiliary field and relates the phase of $X^I$
to that of $S^{ij}$ (provided $S^{ij}$ is non-vanishing).
The condition on $F_{ab}{}^I$ is a BPS attractor equation in a fully supersymmetric
background; equivalently, given a field strength
(which must be related to $\cZ_{ab}$) it fixes the values of
the scalars $X^I$.\footnote{This result
generalizes the attractor equation found in the near horizon limit
of BPS black holes
\cite{FerraraKalloshStrominger, FK:Attractor1, FK:Attractor2, CdWM, CdWKM},
where the supersymmetry algebra is given by the
supergroup $\rm D(2, 1; -1)$ with
$\cZ_{ab} = -W_{ab}^+ = -\frac{1}{4} T_{ab}^+$.}
The conditions on the right force $X^I$ to be covariantly constant
and constrain the VEVs of the scalars when non-abelian couplings are present.
Note that the usual result
$[X^I T_I,\bar X^J T_J] = X^I \bar X^J f_{IJ}{}^K T_K = 0$
is deformed by the background vector multiplets.

If instead we have a background with $G_a{}^{ij} \neq 0$, the situation is drastically simpler.
We find that $X^I$ must vanish and so the entire multiplet vanishes.
This is a consequence of the non-trivial $R$-symmetry appearing in the supersymmetry
algebra: since the superfield $\cX^I$ carries $R$-charge, it must completely vanish.

\subsection{Hypermultiplets}
Hypermultiplets are on-shell representations of the supersymmetry algebra consisting
of $4n$ real scalars $\phi^\mu$ with $\mu=1, \cdots, 4n$ and $2n$
chiral fermions $\z_\alpha{}^\ra$, with $\ra = 1, \cdots, 2n$, obeying
$(\z_\alpha{}^\ra)^* = \bar \z_\dalpha{}^{\bar \ra}$. We follow the conventions
of \cite{Butter:HKP}. As in a Minkowski background,
the scalars parametrize the target space of a hyperk\"ahler manifold with metric
$g_{\mu \nu}$ and three covariantly constant complex structures
$(\cJ_A)^\mu{}_\nu$, obeying the quaternion algebra
\begin{align}\label{eq:HM_JQ1}
\cJ_A \cJ_B = -\delta_{AB} + \eps_{ABC} \cJ_C~.
\end{align}
Introducing $\cJ{}^i{}_j = \frac{i}{2} (\sigma_A)^i{}_j \cJ_A$ where $\sigma_A$ are the
three Pauli matrices, we can construct three hyperk\"ahler two-forms
$(\Omega_{ij})_{\mu\nu} = \eps_{ik} g_{\mu\rho} (\cJ^k{}_j)^\rho{}_\nu$.

There are three important classes of isometries.
The first are the triholomorphic isometries $J^\mu$, which obey
$\cL_J g_{\mu\nu} = \cL_J (\Omega_{ij})_{\mu\nu} = 0$.
These may be gauged by vector multiplets (including the background ones), in which
case we denote them by $J_{\hat I}{}^\mu$. Associated with each $J^\mu$ is
a moment map (or Killing potential) $D^{ij}$ obeying
\begin{align}
\nabla_\mu D_{\hat I}{}^{ij} = - (\Omega^{ij})_{\mu\nu} J_{\hat I}{}^\nu~.
\end{align}
This defines the moment map up to a constant Fayet-Iliopoulos term.

The second class of isometry, denoted by $V^\mu$, rotates the complex structures,
\begin{align}
\cL_V g_{\mu\nu} = 0~, \qquad
\cL_{V} (\Omega_{ij})_{\mu\nu} = -2 v^k{}_{(i} (\Omega_{j) k})_{\mu\nu}~, \qquad
\cL_{V} \cJ_A = -2 \eps_{ABC} \,v_B \cJ_C~,
\end{align}
where $v^{ij}$ is a symmetric pseudoreal normalized SU(2) triplet, $v^{ij} v_{jk} = \delta^i_k$,
equivalent to a normalized real SO(3) vector, $v^A = -\frac{i}{2} (\sigma_A)^i{}_j v^j{}_i$.
Because $V^\mu$ is holomorphic with respect to the complex structure $v^A \cJ_A$,
there is a corresponding moment map, which we denote $\cK$, for this specific complex
structure. We normalize it so that
\begin{align}
\nabla_\mu \cK = v_{ij} (\Omega^{ij})_{\mu\nu} V^\nu = v^A (\cJ_A)_\mu{}^\nu V_\nu~.
\end{align}
$\cK$ is also a K\"ahler potential for the metric $g_{\mu \nu}$ with respect to any complex structure
perpendicular to $v^A \cJ_A$. That is for $w^A \cJ_A$ with $w^A v_A = 0$, one can show that
\begin{align}
g_{\mu\nu} = \frac{1}{2} (\delta_\mu{}^\rho \delta_\nu{}^\sigma
	+ w^A w^B \cJ_A{}_\mu{}^\rho \cJ_B{}_\nu{}^\sigma)
	\nabla_\rho \nabla_\sigma \cK~.
\end{align}
This isometry is relevant whenever $S^{ij} = \mu \,v^{ij}$ is non-vanishing, which implies
the presence of an $\rm SO(2)$ subgroup of ${\rm SU}(2)_R$ in the supersymmetry
algebra; this isometry is manifested on the target space as $V^\mu$.\footnote{Even though
$G_a{}^{ij}$ admits a similar decomposition, it does not generate an $R$-symmetry
transformation in the SUSY algebra and has no effect on the target
space geometry.} This requirement was observed in \cite{BK:AdSSigma}.

The final case of interest is when the target space is
a hyperk\"ahler cone. Then there is a homothetic conformal Killing vector
$\chi^\mu$, obeying $\nabla_\mu \chi^\nu = \delta_\mu{}^\nu$,
and the target space has a globally defined hyperk\"ahler potential
$\chi = \frac{1}{2} \chi^\mu \chi_\mu$.
One can construct a family of isometries that rotate the complex structures
in any direction: they are given by $V_A^\mu = -(\cJ_A)^\mu{}_\nu \chi^\nu$.
The isometries generated by $\chi^\mu$ and $V_A^\mu$
are the target space realization of the superconformal dilatation and
${\rm SU(2)}_R$ generators, and so hyperk\"ahler cones are precisely those
target spaces that may be coupled directly to conformal supergravity.
(The hypermultiplets are inert under ${\rm U}(1)_R$.)
In fact, the presence of the ${\rm SU(2)}_R$
isometries is sufficient to deduce the presence of the dilatation isometry
on the target space. For the cases with $\cZ_{ab}^- = Y_{ab}^-$ and/or
$G_a$ nonzero, the supersymmetry algebra generates an arbitrary ${\rm SU(2)}_R$
element, and so these superalgebras require a hyperk\"ahler cone for
the hypermultiplets.

\subsubsection*{Supersymmetry transformations}
There is one additional feature necessary to
describe the hypermultiplet supersymmetry transformations: the structure group is $\Sp(n) \times \Sp(1)$. That is, one can introduce
a target space vielbein $f_\mu{}_i{}^\ra$ and its inverse $f_{\ra}{}^{i\,\mu}$
(see \cite{SierraTownsend, BaWi:QK, dWKV, dWKV:Rigid}) where $\ra = 1, \cdots, 2n$,
obeying
\begin{align}
f_\mu{}_i{}^{\ra} f_{\ra}{}^{i\,\nu} = \delta_\mu{}^\nu~, \qquad
f_{\ra}{}^{i \,\mu} f_\mu{}_j{}^{\rb} = \delta_\ra{}^\rb \delta^i{}_j~, \qquad
f_\mu{}_i{}^\ra = -\eps_{ij} \,\omega^{\ra\rb} g_{\mu \nu}\, f_\rb{}^{j\,\nu }~, \eol
g_{\mu\nu} = \eps^{ij} \omega_{\ra\rb}\, f_\mu{}_i{}^{\ra} f_{\nu}{}_j{}^{\rb} ~, \quad
(\cJ_A)^\mu{}_\nu = i  f_{\nu}{}_i{}^{\ra} (\sigma_A)^i{}_j f_{\ra}{}^{j \,\mu}~, \quad
(\Omega^{ij})_{\mu\nu} = f_\mu{}^{\ra (i} f_\nu{}^{\rb j)} \,\omega_{\ra\rb}~,
\end{align}
where $\omega_{\ra\rb}$ is an antisymmetric matrix with
$\omega^{\ra \rb}$ obeying $\omega^{\ra\rb} \omega_{\rb \rc} = -\delta^\ra_\rc$.
One can introduce the complex conjugate of $f_\mu{}_i{}^\ra$, given by
$\Big(f_\mu{}_i{}^{\ra}\Big)^* = f_\mu{}^{i \bar \ra}$, so that
\begin{align}
g_{\mu\nu} = f_\mu{}_i{}^\ra f_\nu{}^i{}^{\bar \rb} g_{\ra \bar \rb}
\end{align}
in terms of an \Sp(n) metric $g_{\ra \bar \rb}$. This implies that
$f_\mu{}^{i \bar \ra} = \eps^{ij} g^{\bar \ra \ra} \omega_{\ra \rb} f_\mu{}_j{}^\rb$.
If the \Sp(n) indices are chosen to be flat tangent space indices, then one can
choose $g_{\ra \bar\rb} = \delta_{\ra \bar\rb}$ and take
$\omega_{\ra\rb}$ to be the canonical antisymmetric tensor of \Sp(n).
Following \cite{dWKV, dWKV:Rigid},
we will instead keep a non-trivial \Sp(n) metric and a covariantly constant
$\omega_{\ra\rb}$.
Any vector $V^\mu$ can be related to an $\Sp(n) \times \Sp(1)$ vector
$V_i{}^{\ra} = V^\mu f_\mu{}_i{}^{\ra}$, and similarly for tensors.
The hyperk\"ahler Riemann tensor is valued in \Sp(n) alone,
\begin{align}
R_\ra{}^i{}_\rb{}^j{}_\rc{}^k{}_\rd{}^l := f_\ra{}^{i\, \mu} f_\rb{}^{j\, \nu} f_\rc{}^{k\, \rho} f_\rd{}^{l\, \sigma} R_{\mu \nu \rho \sigma}
	= R_{\ra\rb\rc\rd} \,\eps^{ij} \eps^{kl}
\end{align}
where $R_{\ra\rb\rc\rd}$ is totally symmetric.
One can always take the $\Sp(1)$ connection to vanish, and then the
\Sp(n) connection $\Gamma_{\nu\rb}{}^\ra$ is determined by
requiring  $f_\mu{}_i{}^{\ra}$ to be covariantly constant.

The supersymmetry transformations of the hypermultiplet fields are
\begin{align}
\delta \phi^\mu &= \xi_i \z^\rb \, f_\rb{}^i{}^\mu + \bar \xi^{i} \bar \z^{\bar \rb}\, f_{\bar \rb i}{}^\mu~, \displaybreak[1] \eol
\delta \z_\alpha^\ra
	&=
	\Big(2 i \,\cD_{\alpha\dbeta} \phi^\mu
		- 4 \,G_{\alpha \dbeta} \chi^\mu \Big) f_\mu{}_i{}^\ra \,\bar\xi^{\dbeta i} 
	+ \Big(2 \mu V^\mu + 4 g \bar X^{\hat I} J_{\hat I}{}^\mu \Big) f_\mu{}_i{}^\ra \eps^{ij} \xi_{\alpha j}
	\eol & \quad
	+ 2 \,Y_\alpha{}^\beta \chi^\mu f_\mu{}_i{}^{\ra} \eps^{ij} \xi_{\beta j} 
	- \Gamma_{\mu \rb}{}^\ra \delta \phi^\mu \, \z_\alpha^\rb ~,\displaybreak[1] \eol	
\delta \bar\z^{\dalpha \bar\ra} &=
	\Big(2 i \,\cD^{\dalpha\beta} \phi^\mu
		+ 4 \,G^{\dalpha \beta} \chi^\mu \Big) f_\mu{}^{i \bar \ra} \xi_{\beta i} 
	- \Big(2\bar\mu V^{\mu} + 4 g X^{\hat I} J_{\hat I}{}^{\mu} \Big) f_\mu{}^{i \bar \ra} \eps_{ij} \bar \xi^{\dalpha j} 
	\eol & \quad
	- 2 \,\bar Y^\dalpha{}_\dbeta \chi^\mu f_\mu{}^{i \bar \ra} \eps_{ij} \bar \xi^{\dbeta j} 
	- \Gamma_{\mu \bar\rb}{}^{\bar\ra} \delta \phi^{\mu}
		\, \bar\z^{\dalpha \bar\rb}~.
\label{eq:SUSYHyper}
\end{align}
This is an on-shell supersymmetry algebra only. The component Lagrangian is
\begin{align}
\cL &=
	-\frac{1}{2} \cD^m \phi^\mu\, \cD_m \phi^{\nu}\, g_{\mu\nu}
	- \frac{i}{4} \, g_{\rb \bar \rb}\,
		(\z^{\alpha \rb} \overleftrightarrow{{\widehat\cD}}_{\alpha\dalpha} \bar\zeta^{\dalpha \bar \rb})
	+ \frac{1}{16} \z^\ra \z^\rb \bar \z^{\bar \ra} \bar\z^{\bar \rb}\, R_{\ra \bar \ra\, \rb \bar \rb}
	\eol & \quad
	+ g\, \z^{\ra} \l_{i}{}^I J_{I \ra}{}^i
	+ g\, \bar\z^{\bar \ra} {\bar\l}^{i\, I} J_{I \bar \ra i}
	- 2 g^2\, X^{\hat I} \bar X^{\hat J} J_{\hat I}{}^\mu J_{\hat J}{}^\nu g_{\mu\nu}
	+ g\, Y_{ij}{}^{I} \, D_{I}{}^{ij}
	\eol & \quad
	- \frac{g}{4} \z^\ra \z^\rb X^{\hat I} \, \nabla_{\mu} J_{\hat I}{}_\nu
		f_{\ra j}{}^\mu f_{\rb}{}^j{}^\nu
	- \frac{g}{4} \bar \z^{\bar\ra} \z^{\bar\rb} \bar X^{\hat I} \, \nabla_{\mu} J_{\hat I}{}_\nu
		f_{\bar \ra j}{}^\mu f_{\bar \rb}{}^j{}^\nu
	\eol & \quad	
	+ \cL_S + \cL_{G_{ij}} + \cL_{ZG}~.
\end{align}
We have written the Riemann curvature term using
$R_{\ra \bar \rb \rc \bar \rd} := \omega_{\bar \rb}{}^\rb \omega_{\bar \rd}{}^\rd R_{\ra\rb\rc\rd}$.
This and other terms in the first three lines are straightforward covariantizations of the
general gauged hyperk\"ahler sigma model in a Minkowski background. The covariant
derivatives are
\begin{align}
\cD_m \phi^\mu &= \pa_m \phi^\mu - A_m{}^{\hat I} J_{\hat I}{}^\mu~,\eol
\hat \cD_m \z_\alpha{}^\rb &= \pa_m \z_\alpha{}^\rb
	+ \frac{1}{2} \omega_m{}^{ab} (\sigma_{ab})_\alpha{}^\beta \z_\beta{}^\rb
	+ i A_m \z_\alpha{}^\rb
	\eol & \quad
	- \frac{1}{2} A_m{}^{\hat I} \z_\alpha{}^\rc f_\rc{}^{j \mu} \nabla_\mu J_{\hat I}{}^\nu f_\nu{}_j{}^\rb 
	+ \cD_m \phi^\mu \,\z_\alpha{}^\rc\,\Gamma_\mu{}_\rc{}^\rb ~.
\end{align}

The remaining terms have been set apart corresponding to their dependence
on the background fields.
The terms in $\cL_S$ are present only when $S^{ij}$ is non-vanishing:
\begin{align}
\cL_S &=
	- \frac{1}{8} \bar \mu\, \z^\ra \z^\rb\,
		f_{\ra j}{}^\mu f_{\rb}{}^j{}^\nu\, \nabla_{\mu} V_\nu
	- \frac{1}{8} \mu\, \bar\z^{\bar\ra} \z^{\bar\rb}\,
		f_{\bar\ra j}{}^\mu f_{\bar\rb}{}^j{}^\nu\, \nabla_{\mu} V_\nu
	\eol & \quad
	- \frac{1}{2} |\mu|^2 \,V^\mu V_\mu
	+ 3 |\mu|^2 \,\cK
	- g \mu X^{\hat I} J_{\hat I}{}^\mu V_\mu
	- g \bar \mu \bar X^{\hat I} J_{\hat I}{}^\mu V_\mu
	\eol & \quad
	- 2 g \mu X^I D_I{}^{ij} v_{ij}
	- 2 g \bar \mu \bar X^I D_I{}^{ij} v_{ij}
	- 6 g \, \mu X^{(k)} D_{(k)}{}^{ij} v_{ij}~.
\end{align}
Here we left $S^{ij}$ complex, i.e. $S^{ij} = \mu v^{ij}$ and $\bar S^{ij} = \bar \mu v^{ij}$,
to distinguish which terms arise from $\bar S^{ij}$.
This distinction will be important in the Euclidean case.
Recall that when $S^{ij} \neq 0$, the target space must possess an isometry
$V^\mu$ that rotates the complex structures.
In this case, the triholomorphic isometries associated with
frozen vector multiplets may be absorbed into a redefinition of $V^\mu$ \cite{BKLT-M:AdS4}.

The terms involving $G_a{}^{ij}$
are simplest when written in terms of its dual two-form,
\begin{align}
\cL_{G_{ij}} &=
	\eps^{mnpq} B_{mn\, ij} \,\Big(\cD_p \phi^\mu \, \cD_q \phi^\nu \Omega_{\mu\nu}{}^{ij}
		+ g\, F_{pq}{}^{\hat I} D_{\hat I}{}^{ij} \Big) \eol
	&= \eps^{mnpq} B_{mn\, ij} \,\pa_p \phi^\mu \pa_q \phi^\nu \Omega_{\mu\nu}{}^{ij}
		-\frac{2}{3} g\,\eps^{mnpq} H_{mnp}{}^{ij}  A_q{}^{\hat I} D_{\hat I\, ij}~,
\end{align}
with equality holding up to total derivatives.
The first term is just the spacetime pullback of the hyperk\"ahler two-forms.
This expression may be interpreted as the
hyperk\"ahler analogue of the special K\"ahler couplings involving
$G^a$ \eqref{eq:VM_BF}. It should be emphasized that when $G_a{}^{ij}$ is present,
no constraint is placed on the hyperk\"ahler target space.

Finally, we give the remaining terms involving $\cZ_{ab}$ or $G_a$. We give
them both in terms of $\cZ_{ab}$ and in terms of $Y_{ab}^- = \cZ_{ab}^-$
and $W_{ab}^+ = -\cZ_{ab}^+$:
\begin{align}
\cL_{ZG} &=
	\frac{1}{8} (W^-_{ab} Y^{ab-} + W_{ab}^+ Y^{ab+})\, \chi
	+ 4 G^a G_a \chi
	- \frac{1}{4} \z^{\alpha \ra} \z^{\beta \rb}\, W_{\alpha\beta} \,\omega_{\ra\rb}
	- \frac{1}{4} \bar\z^{\dalpha \bar\ra} \bar \z^{\dbeta \bar \rb}\,
		\bar W_{\dalpha\dbeta} \,\omega_{\bar\ra\bar\rb}\eol
	&= - \frac{1}{8} \cZ_{ab} \bar \cZ^{ab}\, \chi
	+ 4 G^a G_a \chi
	+ \frac{1}{4} \z^{\alpha \ra} \z^{\beta \rb}\, \bar \cZ_{\alpha\beta} \,\omega_{\ra\rb}
	+ \frac{1}{4} \bar\z^{\dalpha \bar\ra} \bar \z^{\dbeta \bar \rb}\,
		\cZ_{\dalpha\dbeta} \,\omega_{\bar\ra\bar\rb}~.
\end{align}
The hyperk\"ahler potential $\chi$ appears only when the target space must be
superconformal, that is when either $G^a$ or $Y_{ab}^- = \cZ_{ab}^-$ is nonzero.

\subsubsection*{BPS conditions for hypermultiplets}
As with the vector multiplets, it is easy to find the conditions for full
supersymmetry:
\begin{align}\label{eq:HM_BPS}
\cD_a \phi^\mu = 0~, \qquad 
\mu V^\mu  = -2 g \bar X^{\hat I} J_{\hat I}{}^\mu~, \qquad
G_a \chi^\mu = Y_{ab}^\pm \chi^\mu = 0~.
\end{align}
The first condition is simple enough to understand: covariant constancy of the hypermultiplet scalars.
The second condition leads to a complicated alignment criterion between the hyperscalars
and the vector scalars that does not always admit a solution.\footnote{For instance, take Minkowski spacetime ($\mu=0$) and target space $\mathbb R^4$
and gauge a constant shift symmetry with a central charge. Then
$X^{(i)} J_{(i)}{}^\mu$ is everywhere non-vanishing.}
The last condition implies that when $G_a$ or $Y_{ab}^\pm$ are nonzero, in which
case the target space is a cone, the scalars must lie at the origin of the cone where
$\chi^\mu$ vanishes.

\subsection*{The off-shell origin of on-shell hypermultiplets}
We have given the on-shell hypermultiplet SUSY transformations
and action without derivation. It turns out they can be derived directly from the off-shell
formulation for hypermultiplets given in curved projective
superspace \cite{KLRT-M1, KLRT-M2, KT-M:DiffReps}. Let us briefly sketch this
topic using the conventions of \cite{Butter:CSG4d.Proj}.

An off-shell hypermultiplet is described by a complex arctic superfield $\Upsilon^+$ living
on $\cM^{4|8} \times \SU(2)$ where $\cM^{4|8}$ is the original $\cN=2$
superspace and $\SU(2)$ is an auxiliary manifold parametrizing the infinite number
of auxiliary fields, with coordinates $v^{i+}$ and $v_i^- = (v^{i+})^*$ obeying
$v^{i+} v_i^- = 1$. Actually, only the space $\mathbb CP^1 = \SU(2) / \U(1)$ plays
a role, and the charge on the superfield $\Upsilon^+$ denotes its weight under $\U(1)$.
A general hypermultiplet Lagrangian is described in flat superspace by an action
\begin{align}
S = -\frac{1}{2\pi} \oint_\cC v_i^+ \rd v^{i+} \,\int \rd^4x\, \rd^4\q^+
	\cF^{++} (\Upsilon^+, \breve \Upsilon^+, v^{i+})
\end{align}
where $\cF^{++}$ is an arbitrary charge-two function of its arguments:
the arctic superfield $\Upsilon^+$, its antarctic conjugate $\breve \Upsilon^+$, and
the auxiliary coordinates $v^{i+}$. The auxiliary integral is over a contour
$\cC$ in $\mathbb CP^1$. When no further restriction is imposed on
$\cF^{++}$, the target space is a generic hyperk\"ahler manifold
upon eliminating the auxiliary fields.

When coupled to conformal supergravity, the action generalizes to
\begin{align}\label{eq:HKconeAction}
S = -\frac{1}{2\pi} \oint_\cC \rd \tau \,\int \rd^4x\, \rd^4\q^+\, \cE^{--}\,
	\cF^{++} (\Upsilon^+, \breve \Upsilon^+)
\end{align}
where $\cE^{--}$ is an appropriate superspace measure, including the
generalization of $v_i^+ \pa_\tau v^{i+}$, where $\tau$ parametrizes the contour $\cC$.
The Lagrangian $\cF^{++}$
is now superconformal, possessing no explicit dependence on $v^{i+}$.
The component reduction of this class was discussed in \cite{Butter:HKP},
where it was shown how to recover the Lagrangian of a
hyperk\"ahler cone coupled to conformal supergravity \cite{dWKV}.

For the rigid supergeometries of interest in this paper, there are three
cases to consider:
\begin{enumerate}[(1)]
\item The most restrictive case is when $Y_{ab}^\pm$ and/or $G_a$
are turned on; then the SUSY algebra generates the full $\SU(2)_R$
group. This requires that the superspace Lagrangian $\cF^{++}$
is covariant under $\SU(2)$ diffeomorphisms, so it cannot
depend explicitly on the coordinates $v^{i+}$.
It takes the same form as \eqref{eq:HKconeAction} and upon reduction to components leads to a
hyperk\"ahler cone.

\item The next case is when $S^{ij} = \mu \,v^{ij}$, corresponding 
to $\rm AdS_4$ in Lorentzian signature. The most general
action is of the form \cite{KT-M:CFlat}
\begin{align}\label{eq:HKU1action}
S = -\frac{1}{2\pi} \oint_\cC \rd \tau \,\int \rd^4x\, \rd^4\q^+\, \cE^{--}\,
	\cF^{++} (\Upsilon^+, \breve \Upsilon^+, v^{++})~, \qquad v^{++} = v^{ij} v_i^+ v_j^+~.
\end{align}
Here only an $\SO(2)_R$ subgroup of $\SU(2)_R$ must be preserved. 
At the component level, this leads to a hyperk\"ahler target space with an $\SO(2)_R$ isometry that
rotates the complex structures. These actions have already been discussed extensively in
four dimensions \cite{BK:AdSSigma, BKLT-M:AdS4}; similar results hold in five
\cite{KT-M:5DCFlat, BX} and three dimensions \cite{BKT-M:AdS3D}.

\item The final case involves background field configurations with only
$W_{ab}^\pm$ or $G_a{}^{ij}$. Because no
$\SU(2)_R$ symmetry survives in the SUSY algebra, the Lagrangian need
not respect $\SU(2)$ diffeomorphisms and may depend arbitrary on the coordinates
$v^{i+}$,
\begin{align}\label{eq:HKaction}
S = -\frac{1}{2\pi} \oint_\cC \rd \tau \,\int \rd^4x\, \rd^4\q^+\, \cE^{--}\,
	\cF^{++} (\Upsilon^+, \breve \Upsilon^+, v^{i+})~.
\end{align}
This leads, as in a flat background, to an unconstrained hyperk\"ahler target space.
\end{enumerate}

The procedure of reducing the various cases above to the explicit on-shell
actions is a straightforward application of the covariant techniques of \cite{Butter:HKP}.
Because it is rather cumbersome, we do not give the derivation explicitly.
A more roundabout derivation will be sketched for cases (1) and (2) in the next section.

\subsection{Conformal supergravity and the origin of rigid actions}
In the preceding sections, we have emphasized the origin of the
vector multiplet and hypermultiplet actions within a purely rigid supersymmetric
framework, sketching their derivation from rigid superspace.
It is instructive to briefly discuss how to reproduce the above results
within the context of conformal supergravity and existing component
actions.\footnote{For a complete and pedagogical review of conformal
supergravity-matter systems, we refer the reader to the recent textbook \cite{FvP}.}

This approach is easiest to understand when applied to the vector multiplets.
When frozen vector multiplets $X^{(i)}$ are included, the prepotentials $F(X^I)$
may be lifted (albeit non-uniquely) to homogeneous conformal prepotentials
$F(X^I, X^{(i)})$. The general action coupling vector multiplets to
conformal supergravity was given in \cite{dWLvP}.
Moving from the conformal framework to the rigid supersymmetric
framework involves freezing the vector multiplets $X^{(i)}$ to constant values
(a Weyl-U(1) gauge-fixing), turning off all background fermions including the gauginos
$\l^{(i)}$ (a choice of $S$-gauge), and setting to zero
the dilatation connection (a conformal gauge-fixing). The auxiliary field $D$,
the Ricci scalar, and the auxiliary fields $T_{ab}^\pm$ are fixed as
\begin{align}
D &= \frac{1}{12} \cZ_{ab} \bar \cZ^{ab}~, \qquad
\frac{1}{4} T_{ab}^- \equiv W_{ab}^- = - \bar \cZ_{ab}^-~, \qquad
\frac{1}{4} T_{ab}^+ \equiv W_{ab}^+ = - \cZ_{ab}^+~, \eol
\cR &= -\cZ_{ab} \bar \cZ^{ab} + 6 S_{ij} \bar S^{ij} + 24 G^2 + 12 G^{a ij} G_{a ij}~,
\label{eq:CSG_BFields}
\end{align}
and one must redefine the $R$-symmetry connections, as discussed in footnote
\ref{foot:RConns}.
The field strengths and auxiliary fields of the frozen vector multiplets are
\begin{align}
F_{ab}{}^{(i)} = -\cZ_{ab} X^{(i)} - \bar \cZ_{ab} \bar X^{(i)}~, \qquad
Y_{ij}{}^{(k)} = -2 S_{ij} X^{(k)} = -2 \bar S_{ij} \bar X^{(k)}~.
\end{align}

For the hypermultiplets, the problem is more subtle. If the hyperk\"ahler target space
we seek is a cone, we may directly couple it to conformal supergravity, apply the
identifications \eqref{eq:CSG_BFields} for the Weyl multiplet,
and redefine the $\U(1)_R$ and $\SU(2)_R$ connections.
However, in the case that the target space is
not a cone, one must introduce additional hypermultiplet compensators and identify
the appropriate rigid limit.

For the class (2) given in \eqref{eq:HKU1action},
the object $v^{++} = v^{ij} v_i^+ v_j^+$ may be identified as a frozen tensor multiplet
$L^{++} = L^{ij} v_i^+ v_j^+$. In the component setting, one can take a hyperk\"ahler
cone with an abelian isometry, dualize the hypermultiplets associated with the
isometry into a tensor multiplet (and reintroduce its auxiliary fields),
and then freeze the multiplet to a rigid configuration.
A residual $\SO(2)_R$ isometry will survive, as required.

For the class (3) given in \eqref{eq:HKaction}, one may
identify $v^{i+}$ as frozen values of an arctic multiplet $\Upsilon_0^+$
and its conjugate $\breve\Upsilon_0^+$. However, these multiplets must be frozen
\emph{prior to the elimination of the auxiliary fields} -- they are \emph{not} on-shell 
multiplets -- and so it is unclear to us how the superspace procedure is related to taking the
rigid $4n$-dimensional limit of a $4n+4$-dimensional hyperk\"ahler cone.
Thankfully, there are no terms in the action unique to this class -- all
expressions in the rigid hypermultiplet action can be compared
with the rigid Minkowski case, the hyperk\"ahler cone action coupled to
conformal supergravity \cite{dWKV}, or the class (2) sketched above.

A related question is whether the actions and their corresponding supersymmetry
algebras can be derived \emph{dynamically} from some supergravity-matter action.
We will return to this issue in the final section.

\subsection{A simple example: The $\cN=2^*$ action}
As a simple application, we will give the $\cN=2^*$ Lagrangian
in a rigid background. The matter content consists of $n$ vector multiplets $X^I$
transforming in some compact non-Abelian gauge group of dimension $n$ with metric
$g_{IJ} = \delta_{IJ}$, coupled to $4n$ hypermultiplets transforming
in the adjoint. We set the $\theta$ term to zero for simplicity.
In a Minkowski background, the on-shell field content is
an $\cN=4$ multiplet; the $\cN=2^*$ theory arises after giving the
hypermultiplet a mass by coupling it to a background vector multiplet.

Let us sketch here the relevant geometric data for the hyperk\"ahler manifold. We
choose a complex basis for the bosons so that the third complex structure is diagonal,
\begin{align}
\phi^\mu = (A^I, B_I, \bar A_I, \bar B^I)~, \qquad
(\cJ_3)^\mu{}_\nu = \text{diag}(i \delta^I{}_J, i\delta_I{}^J, - i \delta_I{}^J, - i\delta^I{}_J)~.
\end{align}
Because $g_{IJ} = \delta_{IJ}$, the complex fields $A^I$ and $B_I$ transform in the same
(adjoint) representation. The three hyperk\"ahler two-forms are
\begin{align}
\Omega_{\1\1} = \rd A^I \wedge \rd B_I~, \quad
\Omega_{\2\2} = \rd \bar A_I \wedge \rd \bar B^I~, \quad
\Omega_{\1\2} = \frac{1}{2} \rd A^I \wedge \rd \bar A_I
	+ \frac{1}{2} \rd B_I \wedge \rd \bar B^I~.
\end{align}
The target space is a cone, so
$\chi = A^I \bar A_I + B_I \bar B^I$ is the hyperk\"ahler potential for all complex structures.
The fermions are
\begin{align}
\z^\ra = (\psi^I, \rho_I)~, \qquad \bar \z^{\bar \ra} = (\bar \psi_I, \bar \rho^I)
\end{align}
and the target space vielbein $f_\mu{}_i{}^\ra$ can be identified from
$\phi^\mu f_\mu{}_\1{}^\ra = (A^I, B_I)$ and
$\phi^\mu f_\mu{}_\2{}^\ra = (-\bar B^I, \bar A_I)$.
We charge the hypermultiplets under a $\U(1)$ associated with the background
vector multiplet $X^{(1)} = 1$ so that $A$ and $\psi$ have charge $+e$ while
$B$ and $\rho$ have charge $-e$.

In a Minkowski background, the Lagrangian is 
\begin{align}
\cL &= - \cD_m \bar A_I \cD^m A^I
	- \cD_m \bar B^I \cD^m B_I
	- \cD_m \bar X^I \cD^m X^I
	\eol & \quad
	- \frac{i}{4} \psi^I \overleftrightarrow{\slashed{\cD}} \bar \psi_I
	- \frac{i}{4} \rho_I \overleftrightarrow{\slashed{\cD}} \bar \rho^I
	- \frac{i}{4} \l_j^I \overleftrightarrow{\slashed{\cD}} \bar \l^{j I}
	- \frac{1}{8} F_{ab}{}^I F^{ab I}
	\eol & \quad
	+ \frac{1}{2} g X^{\textbf{ij}\, I} \l_{\textbf i}{}^J \l_{\textbf j}{}^K f_{IJK}
	+ \frac{1}{2} g X_{\textbf {ij}}{}^I \bar \l^{\textbf i}{}^J \bar \l^{\textbf j}{}^K f_{IJK}
	- \frac{1}{2} g^2 \textrm{Tr} \Big([X_{\textbf{ij}}, X_{\textbf{kl}}]
		[X^{\textbf{ij}}, X^\textbf{kl}]\Big)
	\eol & \quad
	- 2 e^2 (A^I \bar A_I + B_I \bar B^I)
	- 2 i e (\psi^I \rho_I)
	+ 2 i e (\bar\psi_I \bar \rho^I)~,
\end{align}
after integrating out the auxiliary field $Y_{ij}{}^I$.
The $\U(1)$ charge corresponds to a hypermultiplet mass $e \sqrt {2}$,
and in the massless limit, we recover $\cN=4$ SYM.
For that reason, we have grouped some terms
together into an $\SU(4)$ covariant form,
\begin{align}
X^{\textbf{ij}} = (X_{\textbf{ij}})^* =
\begin{pmatrix}
0 & \bar X & -\bar A & \bar B \\
-\bar X & 0 & B & A \\
\bar A & -B & 0 & X \\
-\bar B & -A & -X & 0
\end{pmatrix}~, \qquad
\l_{\textbf{i}} = (\l_i, \psi, \rho) = (\bar \l^{\textbf i})^*~,
\end{align}
with the index $\textbf{i}$ labelling the $\mathbf 4$ of $\SU(4)$.
We normalize the trace so that
\begin{align}
\textrm{Tr} \Big([X_{\textbf{ij}}, X_{\textbf{kl}}]
		[X^{\textbf{ij}}, X^\textbf{kl}]\Big)
	= X_{\textbf{ij}}{}^I  X_{\textbf{kl}}{}^J f_{IJ}{}^K
	X^{\textbf{ij}\, L} X^{\textbf{kl}\, M} f_{LM}{}^K~.
\end{align}

When $S^{ij} = \mu\, v^{ij}$ is non-vanishing, one additional piece of information is
required: the form of the Killing vector $V^\mu$.
We choose $v^{ij} = -\delta^{ij}$ so that $V^\mu = (\bar B^I, -\bar A_I, B_I, -A^I)$;
the identification is simple because the target space is a cone.
Introducing all of the required couplings leads to
\begin{align}
\cL &= - \cD_m \bar A_I \cD^m A^I
	- \cD_m \bar B^I \cD^m B_I
	- \cD_m \bar X^I \cD^m X^I
	\eol & \quad
	- \frac{i}{4} \psi^I \overleftrightarrow{\slashed{\cD}} \bar \psi_I
	- \frac{i}{4} \rho_I \overleftrightarrow{\slashed{\cD}} \bar \rho^I
	- \frac{i}{4} \l_j^I \overleftrightarrow{\slashed{\cD}} \bar \l^{j I}
	- \frac{1}{8} F_{ab}{}^I F^{ab I}
	\eol & \quad
	+ \frac{1}{2} F_{ab}{}^{I} (W^{ab+} X^I+ W^{ab-} \bar X^I)
	+ \cL_{BF}
	+ \cL_{\rm pot}
	+ \cL_{\rm ferm}~.
\end{align}
The potential terms are given by
\begin{align}
\cL_{\rm pot} &= 2 (|\mu|^2 - e^2) (A^I \bar A_I + B_I \bar B^I)
	+ 2 |\mu|^2 X^I \bar X^I
	+ 2 i \,\mu \,e\, (A^I B_I - \bar A_I \bar B^I)
	\eol & \quad
	- \frac{1}{4} Z_{ab} \bar Z^{ab}
		\Big(X^I \bar X^I + \frac{1}{2} A^I \bar A_I + \frac{1}{2} B_I \bar B^I\Big)
	- \frac{1}{4} (W_{ab}^+)^2 X^I X^I
	- \frac{1}{4} (W_{ab}^-)^2 \bar X^I \bar X^I
	\eol & \quad
	+ 2 G_{a \, ij} G^{a\, ij} X^I \bar X^I
	+ 4 G^2 (A^I \bar A_I + B_I \bar B^I)
	\eol & \quad
	- \frac{1}{2} g^2 \textrm{Tr} \Big([X_{\textbf{ij}}, X_{\textbf{kl}}]
		[X^{\textbf{ij}}, X^\textbf{kl}]\Big)~.
\end{align}
Because we chose $X^{(1)} = 1$, $\mu$ must be real.
The fermionic couplings are
\begin{align}
\cL_{\rm ferm} &=
	(\l_i^I \sigma^a \bar \l^{j I}) (\frac{1}{2} \delta^i_j G_a + i G_a{}^i{}_j)
	- 2 i e (\psi^I \rho_I) + 2 i e (\bar \psi_I \bar \rho^I)
	\eol & \quad
	+ \frac{1}{2} \psi^{\alpha I} \rho^{\beta}{}_I \bar Z_{\alpha\beta}
	+ \frac{1}{2} \bar \psi_{\dalpha I} \bar\rho_{\dbeta}{}^I Z^{\dalpha\dbeta}
	\eol & \quad
	+ \frac{1}{2} g X^{\textbf{ij}\, I} \l_{\textbf i}{}^J \l_{\textbf j}{}^K f_{IJK}
	+ \frac{1}{2} g X_{\textbf {ij}}{}^I \bar \l^{\textbf i}{}^J \bar \l^{\textbf j}{}^K f_{IJK}~.
\end{align}
Finally, the generalized $BF$ terms are
\begin{align}
\cL_{BF} &= 2 i \, \eps^{mnpq} B_{mn} \pa_p X^I \pa_q \bar X^I
	+ \eps^{mnpq} B_{mn}{}^{ij} \pa_p \phi^\mu \pa_q \phi^\nu \Omega_{\mu\nu\, ij}
	\eol & \quad
	+ \frac{2}{3} g \eps^{mnpq} H_{mnp} A_q{}^I D_I
	- \frac{2}{3} g \eps^{mnpq} H_{mnp}{}^{ij} A_q{}^I D_{I\, ij}~,
\end{align}
where the special K\"ahler and hyperk\"ahler moment maps are
\begin{alignat}{2}
D_I &= -i X^J \bar X^K f_{JKI}~, &\qquad
D_{I\,\1\2} &= \frac{1}{2} (A^J \bar A^K + B^J \bar B^K) f_{JKI}~, \eol
D_{I\,\1\1} &= A^J B^K f_{JKI}~, &\qquad
D_{I\,\2\2} &= \bar A^J \bar B^K f_{JKI}~.
\end{alignat}

\section{The general Euclidean supersymmetry algebra}\label{sec:E_SUSY}

Now we turn to the case of Euclidean SUSY. As in the Lorentzian case,
we seek first a general rigid Euclidean superalgebra admitting eight rigid
supersymmetries. A straightforward way to construct such an algebra is via
analytic continuation from the Lorentzian case; the approach we follow
resembles that chosen by \cite{KZ:ExtSUSY}.

This can briefly be described as follows. First complexify the Lorentzian
superalgebra \eqref{eq:LSusyAlg},
relaxing the reality conditions on all of the operators and fields. The algebra still
closes because the Bianchi identity is holomorphic in these quantities.
Next, one makes a Wick rotation on all vector quantities $V$, taking
$V_0 \rightarrow i V_4$ and $V^0 \rightarrow -i V^4$, and similarly
for any tensors. The algebra retains the same form provided we take
$\sigma^0 = -i \sigma^4$ and $\bsigma^0 = -i \bsigma^4$ and
replace $\eta_{ab}$ with the Euclidean $\delta_{ab}$. We
also exchange the Lorentzian $\eps_{abcd}$ (with $\eps_{0123} = 1$)
for $i \,\veps_{abcd}$ (with $\veps_{1234} = 1$).
Now one must impose a reality condition so that the
momentum generators are Hermitian, but how exactly (if at all) to impose this condition
on the supercharges is an interesting question.

This is an old topic in the literature and one can identify two schools of
thought. A real Lorentzian supersymmetry algebra maps
naturally under Wick rotation to a \emph{reflection positive} Euclidean supersymmetry
algebra. This means that a Euclidean SUSY algebra (or action)
arising directly from analytic continuation does not need to be real in the conventional
sense, but rather real in the sense of Osterwalder and Schrader.
This agrees with the approach taken by Nicolai \cite{Nicolai:ESusy}.
The second approach, originally proposed by Zumino \cite{Zumino:ESusy},
involves maintaining reality of the SUSY algebra and associated actions
by choosing Majorana supercharges, but this is possible only for $\cN\geq 2$.
Real Euclidean actions and algebras naturally continue to $CT$-even
(but potentially complex) Lorentzian actions and algebras.
As noted in \cite{CortesMohaupt1}, the first case automatically
gives the correct Green's functions under Wick rotation, while the
latter case arises via timelike dimensional reduction from 5D.
Because the first possibility is just a Wick rotation of the
Lorentzian case (and so should offer no new features),
we will focus on the second possibility exclusively.

A real Euclidean superalgebra requires a real Euclidean superspace.\footnote{Euclidean
superspaces for both real and holomorphic SUSY were introduced in \cite{Lukierski:1982hr}.}
We take
\begin{align}\label{eq:DBarEuc}
(\cD_a)^* = \cD_a~, \qquad (\cD_\alpha{}^i)^* = -\cD^\alpha{}_i~, \qquad
(\bar\cD^{\dalpha i})^* = -\bar\cD_{\dalpha i}~,
\end{align}
so that the Killing spinors are symplectic Majorana-Weyl,
\begin{align}
(\xi_{\alpha}{}^i)^* = \xi^{\alpha}{}_i~, \qquad
(\bar\xi^{\dalpha i})^* = \bar\xi_{\dalpha i}~.
\end{align}
Consistency with the flat space Euclidean supersymmetry algebra
implies that the $R$-symmetry group $\U(1)$ must map to
the non-compact group $\SO(1,1)$ corresponding to chiral
dilatations \cite{Zumino:ESusy}.
We account for this by exchanging the generator $\mathbb A$ for
$i \mathbb U$ where
\begin{align}
[\mathbb U, \cD_\alpha{}^i] = -\cD_\alpha{}^i~, \qquad
[\mathbb U, \bar\cD^\dalpha{}_i] = +\bar\cD^\dalpha{}_i~.
\end{align}
We correspondingly continue the $\U(1)_R$ connection to an $\SO(1,1)_R$ connection.
The $\SU(2)_R$ generator and connection are unchanged.
Note that we keep the notation $\bar \xi^\dalpha{}_i$ and
$\bar \cD^{\dalpha i}$ even though these are not the complex conjugates
of the unbarred quantities.

After these modifications, the rigid Euclidean superspace algebra is 
\begin{align}
\{\cD_\alpha{}^i, \cD_\beta{}^j\}
	&= 4 S^{ij} M_{\alpha\beta}
	+ \eps^{ij} \eps_{\alpha\beta} \,\newG^{cd} M_{cd}
	+ 2 \eps^{ij} \eps_{\alpha\beta} \,S^k{}_l I^l{}_k
	- 4 \,\newG_{\alpha\beta} I^{ij}~, \eol{}
\{\bar \cD^\dalpha{}_i, \bar \cD^\dbeta{}_j\}
	&= 4 \bar S_{ij} \bar M^{\dalpha \dbeta}
	- \eps_{ij} \eps^{\dalpha \dbeta} \,\bar \newG^{cd} M_{cd}
	- 2 \eps_{ij} \eps^{\dalpha\dbeta} \bar S^k{}_l I^l{}_k
	- 4 \bar \cZ^{\dalpha \dbeta} I_{ij}~,\eol{}
\{\cD_\alpha{}^i, \bar \cD_{\dbeta j}\}
	&= -2i \,\delta^i_j (\sigma_a)_{\alpha\dbeta} \cD_a
	+ 2 (\sigma_a)_{\alpha\dbeta} \veps^{abcd} (\delta^i_j G_b + i G_b{}^i{}_j) M_{cd}
	- 8 G_{\alpha\dbeta} I^i{}_j
	+ 2i\, G_{\alpha \dbeta}{}^i{}_j \mathbb U~, \eol{}
[\cD_a, \cD_\beta{}^j]
	&=
	\frac{i}{2} (\sigma_a)_{\beta \dgamma} S^{jk} \bar \cD^\dgamma{}_k
	- \frac{i}{2} \newG_{ab} (\sigma^b)_{\beta \dgamma} \bar \cD^{\dgamma j}
	-2 i G^b  (\sigma_{ba})_\beta{}^{\gamma} \cD_\gamma{}^j
	- G_b{}^j{}_k (\sigma_a \bsigma^b)_\beta{}^\gamma \cD_\gamma{}^k~, \eol {}
[\cD_a, \bar\cD^\dbeta{}_j]
	&=
	\frac{i}{2} (\bsigma_a)^{\dbeta \gamma} \bar S_{jk} \cD_\gamma{}^k
	+ \frac{i}{2} \bar\newG_{ab} (\bsigma^b)^{\dbeta \gamma} \cD_{\gamma j}
	+ 2 i G^b (\bsigma_{ba})^\dbeta{}_{\dgamma} \bar\cD^\dgamma{}_j
	+G_b{}^k{}_j (\bsigma_a \sigma^b)^\dbeta{}_\dgamma \bar\cD^\dgamma{}_k~,\eol {}
[\cD_a, \cD_b] &= -\frac{1}{2} R_{ab}{}^{cd} M_{cd}~.
\label{eq:ESusyAlg}
\end{align}
The Riemann tensor is explicitly determined to be
\begin{align}
R_{ab}{}^{cd} &=
	- \frac{1}{2} (\newG_{ab} \bar \newG^{cd} + \bar \newG_{ab} \newG^{cd})
	+ 8 \,G^2 \delta_a{}^{[c} \delta_b{}^{d]}
	- 16 \,G_{[a} G^{[c} \delta_{b]}{}^{d]}
	\eol & \quad
	+ 4  G^f_{ij} G_{f}^{ij} \delta_a{}^{[c} \delta_b{}^{d]}
	- 8 G_{[a}^{ij} G^{[c}_{ij} \delta_{b]}{}^{d]}
	+ S^{ij} \bar S_{ij} \delta_a{}^{[c} \delta_b{}^{d]}~.
\end{align}
It will be useful to retain the same decomposition \eqref{eq:DefZ} for $\cZ_{ab}$.
The space is conformally flat when both $\cZ_{ab}^- \bar \cZ_{cd}^-$ and
$\cZ_{ab}^+ \bar \cZ_{cd}^+$ vanish and superconformally flat when
$\cZ_{ab}^+ = \bar\cZ_{ab}^-= 0$.

The Killing spinor equations are
\begin{equation}
\begin{aligned}
\label{killing spinor equations euclidean}
\cD_b \xi_{\alpha}{}^i &=
	- 2i G^c (\sigma_{cb} \xi^i)_\alpha
	- G^c{}^i{}_j (\sigma_c \bsigma_b \xi^j)_\alpha
	- \frac{i}{2} \bar S^{ij} (\sigma_b \bar \xi_j)_\alpha
	- \frac{i}{2} \bar \newG_{bc} (\sigma^c \bar\xi^i)_\alpha~, \\
\cD_b \bar \xi^{\dalpha i} &=
	+ 2i G^c (\bsigma_{cb} \bar \xi^i)^\dalpha
	- G^c{}^i{}_j (\bsigma_c \sigma_b \bar\xi^j)^\dalpha
	+ \frac{i}{2} S^{ij} (\bsigma_b \xi_j)^\dalpha
	+ \frac{i}{2} \newG_{bc} (\bsigma^c \xi^i)^\dalpha~.
\end{aligned}
\end{equation}
Because $\xi_\alpha{}^i$ and $\bar \xi^{\dalpha i}$ are symplectic Majorana-Weyl,
these two equations are independent.
The reality conditions on the curvature fields are
\begin{align}\label{eq:EReality}
(\cZ_{ab})^* = -\cZ_{ab}~, \qquad
(S^{ij})^* = - S_{ij}~, \qquad
(G_a)^* = -G_a~, \qquad
(G_a{}^{ij})^* = G_a{}_{ij}~,
\end{align}
and similarly for $\bar\cZ$ and $\bar S_{ij}$. We emphasize that $\cZ_{ab}$,
$S^{ij}$ and $G_a$ are (pseudo)imaginary in Euclidean signature. Note that
the $\SO(1,1)$ weights of the various fields are given by 
\begin{align}
w(S^{ij}) = w (Z_{ab}) = -2~, \qquad
w(G_a) = w (G_a{}^i{}_j) = 0~.
\end{align}
An important feature of Euclidean signature is that the barred and
unbarred fields, e.g. $\cZ_{ab}$ and $\bar \cZ_{ab}$, are
\emph{completely independent}.

\section{Euclidean backgrounds} \label{sec:EBacks}

We may again introduce torsion by redefining the spin connection as
\begin{align}
\widetilde \cD_a := \cD_a + i \veps_a{}^{bcd} G_b M_{cd}~.
\end{align}
Because $G_a$ is imaginary in Euclidean signature, this modification
leaves the spin connection real. The modified superspace algebra is then
\begin{align}
\{\cD_\alpha{}^i, \cD_\beta{}^j\}
	&= 4 S^{ij} M_{\alpha\beta}
	+ \eps^{ij} \eps_{\alpha\beta} \,\newG^{cd} M_{cd}
	+ 2 \eps^{ij} \eps_{\alpha\beta} \,S^k{}_l I^l{}_k
	- 4 \,\newG_{\alpha\beta} I^{ij}~, \displaybreak[1]\eol{}
\{\bar \cD^\dalpha{}_i, \bar \cD^\dbeta{}_j\}
	&= 4 \bar S_{ij} \bar M^{\dalpha \dbeta}
	- \eps_{ij} \eps^{\dalpha \dbeta} \,\bar \newG^{cd} M_{cd}
	- 2 \eps_{ij} \eps^{\dalpha\dbeta} \bar S^k{}_l I^l{}_k
	- 4 \bar \newG^{\dalpha \dbeta} I_{ij}~,\displaybreak[1]\eol{}
\{\cD_\alpha{}^i, \bar \cD_{\dbeta j}\}
	&= -2i \,\delta^i_j (\sigma_a)_{\alpha\dbeta} \widetilde\cD_a
	+ 2 i (\sigma_a)_{\alpha\dbeta} \veps^{abcd} G_b{}^i{}_j M_{cd}
	- 8 G_{\alpha\dbeta} I^i{}_j
	+ 2 i \,G_{\alpha \dbeta}{}^i{}_j \mathbb U~, \displaybreak[1]\eol{}
[\widetilde \cD_a, \cD_\beta{}^j]
	&=
	\frac{i}{2} (\sigma_a)_{\beta \dgamma} S^{jk} \bar \cD^\dgamma{}_k
	- \frac{i}{2} \newG_{ab} (\sigma^b)_{\beta \dgamma} \bar \cD^{\dgamma j}
	- 4 i G^b  (\sigma_{ba})_\beta{}^{\gamma} \cD_\gamma{}^j
	- G_b{}^j{}_k (\sigma_a \bsigma^b)_\beta{}^\gamma \cD_\gamma{}^k~, \displaybreak[1]\eol {}
[\widetilde \cD_a, \bar\cD^\dbeta{}_j]
	&=
	\frac{i}{2} (\bsigma_a)^{\dbeta \gamma} \bar S_{jk} \cD_\gamma{}^k
	+ \frac{i}{2} \bar\newG_{ab} (\bsigma^b)^{\dbeta \gamma} \cD_{\gamma j}
	+ 4 i G^b (\bsigma_{ba})^\dbeta{}_{\dgamma} \bar\cD^\dgamma{}_j
	+G_b{}^k{}_j (\bsigma_a \sigma^b)^\dbeta{}_\dgamma \bar\cD^\dgamma{}_k~,\eol {}
[\widetilde \cD_a, \widetilde \cD_b] &= - \widetilde T_{ab}{}^c \widetilde\cD_c
	-\frac{1}{2} \widetilde R_{ab}{}^{cd} M_{cd}~,
\end{align}
where the torsion and Lorentz curvature tensors are given by
\begin{align}
\widetilde T_{ab}{}^c &= -4 i\,\veps_{ab}{}^{cd} G_d~, \eol
\widetilde R_{ab}{}^{cd} &=
	- \frac{1}{2} (\newG_{ab} \bar \newG^{cd} + \bar \newG_{ab} \newG^{cd})
	+ 4 \,G^f_{ij} G_{f}^{ij} \delta_a{}^{[c} \delta_b{}^{d]}
	- 8 \,G_{[a}^{ij} G^{[c}_{ij} \delta_{b]}{}^{d]}
	+ S^{ij} \bar S_{ij} \delta_a{}^{[c} \delta_b{}^{d]}~.
\end{align}

Now we can analyze the integrability conditions for all of these
background fields exactly as in Lorentzian signature, leading again to
\eqref{eq:LProductConditions}. These now lead to \emph{five} possibilities:
\begin{enumerate}[(I)]
\item $S^{ij}$ and/or $\bar S_{ij}$ nonzero, all other fields vanishing;
\item $G_a{}^i{}_j \neq 0$, all other fields vanishing;
\item $G_a \neq 0$, perhaps with some of $\cZ_{ab}^\pm, \bar\cZ_{ab}^\pm$ nonzero;
\item $\cZ_{ab}^\pm$ and/or $\bar \cZ_{ab}^\pm$ nonzero, but all other fields vanishing;
\item $S^{ij}$ and $\cZ_{ab}^+ = -W_{ab}^+$ nonzero, but all other fields vanishing.
\end{enumerate}
The fifth case was not possible in Lorentzian signature.
In the Euclidean case, both $S^{ij}$ and $\cZ_{ab}^+ = -W_{ab}^+$ are pseudoimaginary and
not the complex conjugates of $\bar S_{ij}$ and $\bar\cZ_{ab}^- = - W_{ab}^-$, and so it is possible to
keep one set while discarding the other. This leads to a SUSY algebra where
only the left or right-handed generators are deformed.
Actually, the fact that the right-handed and left-handed SUSY generators are no longer
related by complex conjugation leads to somewhat different possibilities in the other
cases as well. Such variants correspond to full supersymmetric relatives of the
$\Omega$ background.
The analysis of the $R$-symmetry connections is analogous to the Lorentzian case.

Now the additional integrability conditions are identical to the Lorentzian case,
\begin{gather}
S^{ij} \propto \bar S^{ij}~, \qquad
G_{[a}{}^{ij} G_{b]}{}^{kl} = 0~, \qquad 
G^a \cZ_{ab} = 0~, \qquad
\veps^{abcd} \cZ_{ab} \bar \cZ_{cd} = 0~, \eol{}
\cZ_{ab}^\pm \propto \bar \cZ_{ab}^\pm~, \qquad
\widetilde\cD_a G_b = \widetilde\cD_a G_b{}^{ij} = 0~, \qquad
\widetilde \cD_a \cZ_{bc} = \widetilde \cD_a\bar \cZ_{bc} = 0~.
\end{gather}
As before, $\cZ_{ab}$ is a closed complex two-form,
but its dual is not closed unless $G_a = 0$. 
We summarize in Table \ref{tab:EBacks} the resulting consistent Euclidean backgrounds.

\begin{table}[th!]
\def\twocol{\multicolumn{2}{l}}
\tablespacings
\centering
\begin{tabular}{ll@{\hskip 6em}l}
\toprule
\twocol{Background fields} & Geometry\\
\otoprule
\twocol{$S^{ij}\neq0$ and $\bar S^{ij}\neq0$}              & $S^4$ and $H^4$                                          \\[.5ex]
\twocol{$G_a{}^i{}_j\neq0$}                                & Two-sheeted $H^3\times\bbR$                              \\[.5ex]
\twocol{$G_a\neq0$}                                        & $S^3\times\bbR$                                          \\
&$\cZ\cdot\bar\cZ > 32 \,G^2$                              & Warped $S^3\times\bbR$                                   \\
&$\cZ\cdot\bar\cZ = 32 \,G^2$                              & Heis$_3\times\bbR$                                               \\
&$\cZ\cdot\bar\cZ < 32 \,G^2$                              & Warped one-sheeted $H^3\times\bbR$                      \\[.5ex]
\twocol{$\cZ_{a}{}^{b}\neq 0$  but $G_a = 0$}              & $H^2\times S^2$, $\bbR^2\times S^2$ and  $H^2\times \bbR^2$ \\[.5ex]
\twocol{$S^{ij}\neq0$, $\cZ\neq0$, $S^{ij}\cZ^-=0$}        & Flat space (deformed susy)
\cbottomrule
\end{tabular}
\caption{Consistent Euclidean backgrounds.}
\label{tab:EBacks}
\end{table}

\subsection{$S^4$ and $H^4$}

The first cases we consider are associated to non-vanishing $S^{ij}$ and $\bar S^{ij}$.
For definiteness, let us gauge-fix them to $S^{ij} = \ii \mu \delta^{ij}$ and $\bar S^{ij} = \ii \bar\mu \delta^{ij}$ for real $\mu,\ \bar\mu$.
Depending on the relative sign of $\mu$ and $\bar\mu$, two different superalgebras arise with the associated supercoset spaces:
\begin{equation}
\frac{\OSp(2|4)}{\SO(4)\times\SO(2)}
\qquad\text{and}\qquad
\frac{\OSp(2|2,2)}{\SO(4)\times\SO(2)},
\end{equation}
with the latter obtained for $\bar\mu=-\mu$ and corresponding to the Wick rotation of the AdS$_4$ spacetime and gives the hyperboloid $\rm SO(4,1)/SO(4)$, noting the isomorphism of the algebras $\Sp(2,2)\simeq\SO(4,1)$.
The $\OSp(2|4)$ case corresponds to a round $S^4$ geometry.
As usual, further details and explicit expressions for the Killing spinors are given in Appendix~\ref{app:EBack}.

\subsection{Squashed and stretched $S^3\times \bbR$ and $S^3\times S^1$}

The (squashed) $\bbR\times S^3$ geometry generated by $G_a$ and $\cZ_{ab}$ in Lorentzian signature can be Wick rotated to Euclidean and the background fields satisfy appropriate reality conditions.
In fact, in Euclidean signature we have more freedom in deforming the geometry of the three-sphere and the associated supersymmetry algebra, basically because of the independence of $\cZ$ and $\bar\cZ$.

In analogy with the Lorentzian case, non-vanishing $G$ gives rise to an $S^3\times \bbR$ geometry with a (centrally extended) $\SU(2|2)\times\SU(2)$ supersymmetry algebra.
The flat direction is generated by the central charge, hence there is no obstruction to compactifying it to $S^1$.
Turning on $\cZ$ and $\bar\cZ$ fluxes along $S^3$ and defining $\cZ^2\equiv-8\l^2$, $\bar\cZ^2\equiv-8\tilde\l^2$ and $G^2=-g^2$, we obtain a warped $S^3$ geometry with metric 
\begin{align}
\label{metric warped S3 euclidean signature}
\rd s^2 = \frac{\upsilon}{16g^2}
	\big[ \rd\theta^2+\sin^2\theta\,\rd\omega^2 + \upsilon(\rd\phi+\cos\theta\,\rd\omega)^2\big]
	+\rd z^2
\end{align}
and warping parameter $\u\equiv1/\big(1+\frac{\l\tilde\l}{4g^2}\big)$, provided $\l\tilde\l>-4g^2$.
Notice that independence of $\cZ$ and $\bar\cZ$ permits to source both squashing and stretching of the $S^3$.
The superspace is again \eqref{squashed S3 supercoset} with the substitution $P_0\to -\ii G\cdot P$.

When both $\cZ$ and $\bar\cZ$ are non-vanishing, we can set $|\l|=|\tilde\l|$ by an $\SO(1,1)_R$ gauge choice.
An interesting possibility is to set $\bar\cZ =0$ and keep $\cZ\neq0$ (or vice versa).
In this case there is no squashing of the $S^3$ geometry, but the supersymmetry algebra is deformed in the sense that the isometry SU(2) group of the sphere does not preserve chirality of the supercharges.
This fact also breaks the residual Lorentz symmetry to the U(1) stabilizing $\cZ$, and the superalgebra is SU($2|2$) with two central charges just as in the squashed case.

In full analogy with the discussion in Lorentzian signature, the Killing spinors we find do not depend on the Euclidean time specified by $G\cdot P$, and the $R$-symmetry connections are vanishing.%
\footnote{We point out again that these connections have been redefined in footnote~\ref{foot:RConns}. 
Our statement is unambiguous in the sense that a theory coupled to this background need not have (full) $R$-symmetry, even when we compactify to $S^3\times S^1$.}

The requirement for real supersymmetry in Euclidean signature excludes the possibility of an $\SU(2|1)^2$ realization of $S^3\times\bbR$.
This is reflected in the different reality conditions \eqref{eq:EReality} for $G_a$ and $G_a{}^i{}_j$.%
\footnote{Of course, breaking the reality condition on the fermions one could consider the Wick rotation of the Lorentzian background.
Compactification of the flat direction would then always break half the supersymmetries.}

\subsection{Warped one-sheeted $H^3\times \bbR$}

In the previous section we imposed $\l \tilde\l >-4g^2$ in order to obtain an $S^3$ geometry.
If we consider now $\l \tilde\l <-4g^2$, the resulting manifold is AdS$_3\times\bbR$ with a warped Euclidean signature metric.
This is basically the converse of what happens for the $S^3$ with Lorentzian metric in Section~\ref{sec:Lorentzian S3}.
The choice of Euclidean signature automatically breaks the isometry group from the natural $\SU(1,1)^2$ to $\SU(1,1)\times\U(1)$.
In fact, depending on $\l ,\ \tilde\l $ the geometry is warped along the circular fiber of the Hopf fibration $S^1 \hookrightarrow H^3 \to H^2$ (this is what we would call timelike warped AdS$_3$ if the metric were Lorentzian).
The supercoset structure of this space is the same as for timelike stretched AdS$_3$, cfr. \eqref{supercoset timelike tretched AdS3}, the difference being the choice of signature and the fact that we are now allowed to both stretch and squash along the circular fiber.
In the same coordinates we use for AdS$_3$, we have the metric
\begin{equation}
\label{metric one sheeted H3}
\rd s^2 = \frac{\u}{16g^2}\left[
\u(\rd\tau+\cosh\rho \,\rd\phi)^2 + \rd\rho^2 +\sinh^2\rho \,\rd\phi^2
\right] +\rd z^2,\quad
\u\equiv-\Big(1+\frac{\l\tilde\l}{4g^2}\Big)^{-1}\!>0.
\end{equation}

\subsection{The Heis$_3\times\bbR$ limit}

We shall now consider the threshold case between $S^3$ and $H^3$, obtained by setting $\l \tilde\l =-4g^2$.
In complete analogy with Section~\ref{ovestretched AdS3}, the supersymmetry algebra is a non-semisimple contraction of the (centrally extended) SU$(2|2)$ superalgebra of $S^3\times \bbR$ containing Heis$_3\rtimes \U(1)\times \bbR$ as spacetime isometries.
The independent factor is generated by $G\cdot P$ as usual, and the central charge of $\rm Heis_3$ is a linear combination of a translation and Lorentz generator.
We arrive at the metric
\begin{equation}
\label{metric Heis3xR}
\rd s^2 = (\rd w + 2g\, y\,\rd x- 2g\,x\,\rd y)^2 +\rd x^2+\rd y^2+\rd z^2.
\end{equation}

\subsection{The two-sheeted $H^3\times \bbR$}

Another hyperbolic background arises from $G_a{}^i{}_j$.
Euclidean reality conditions do not allow for $\rm SU(2|1)^2$ or $\rm SU(1,1|1)^2$ superalgebras.
What turns out to be allowed instead is the superalgebra $\rm SL(2|1,\bbC)$.
Its bosonic part is isomorphic to $\SO(3,1)\times \U(1)\times \SO(1,1)_R$.
We have therefore the supercoset space
\begin{equation}
\frac{\SL(2|1,\bbC)}{\SU(2)\times\SO(1,1)_R},
\end{equation}
which covers one sheet of the hyperboloid $X^2-Y^2-V^2-W^2=1$.
We may take
\begin{align}
X &= \cosh\rho                         \,,  \quad
Y = \sinh\rho \cos\theta              \,,  \quad
V = \sinh\rho \sin\theta \cos\phi     \,,  \quad
W = \sinh\rho \sin\theta \sin\phi     \,. 
\end{align}
These coordinates have range and periodicity $\rho \in [0,+\infty)$, $\theta\in[0,\pi]$ and $\phi\simeq \phi+2\pi$.
The metric is the standard
\begin{equation}
\label{metric two sheeted H3}
\rd s^2 = \frac1{4g^2}\left(
\rd\rho^2+\sinh^2\!\rho\, ( \rd\theta^2 +\sin^2\!\theta\, \rd\phi^2 )
\right) + \rd z^2\,.
\end{equation}

Let us comment on the properties of the Killing spinors:
the supercharges $Q_\a{}^\1$ and $Q_\a{}^\2$ have opposite charges under $G\cdot P$, and the same holds for the antichiral ones.
This is reflected in the Killing spinors depending on $z$ with factors $e^{\pm\ii g z}$ (see \eqref{killing spinors two sheeted H3}).
If we take the universal cover $\bbR$ for the $z$ direction, these Killing spinors are globally defined.
However, compactifying to $H^3\times S^1$ we find periodic Killing spinors only for $z\simeq z+\frac{2\pi k}{g}$, $k\in\bbZ$.

\subsection{$H^2\times S^2$ and D$(2,1;\alpha)$}

The appropriate Wick rotation of AdS$_2\times S^2$ gives an $H^2\times S^2$ space, where $H^2$ is (one sheet of) the two-sheeted hyperboloid.
The discussion follows again lines similar to the Lorentzian case.
The background is determined by $\cZ$ and $\bar\cZ$, breaking the Lorentz symmetry to $\U(1)\times \U(1)$.
One obtains the same real form of $\rm D(2,1;\alpha)$, and the supercoset space is
\begin{equation}
\frac{\rm D(2,1;\alpha)}{{\rm SU}(2)_R\times \rm U(1)^2}\ ,
\end{equation}
with $\alpha\equiv \l_+/\l_-$.
The radii of curvature of $H^2$ and $S^2$ are $1/|\l_-|$ and $1/|\l_+|$ respectively.
Let us specify $\cZ_{ab}=2\ii(\l_+\delta^{12}_{ab}-\l_-\delta^{34}_{ab})$ and 
$\bar\cZ_{ab}=-2\ii(\l_+\delta^{12}_{ab}+\l_-\delta^{34}_{ab})$.
The only differences with the AdS$_2\times S^2$ case are that now the \SU(1,1) generators read
\begin{align}
T_{\tilde a\tilde b}&\equiv 
\begin{pmatrix}
\frac{1}{\l_-} (\ii P_3-P_4)   & -\ii M_{34}                        \\
-\ii M_{34}                       & \frac{1}{\l_-}(\ii P_3+P_4)  
\end{pmatrix}\,,
\end{align}
and the reality conditions now are
\begin{align}
(Q_{\tilde a\,\tilde\alpha\,i})^* &= (\sigma_1)_{\tilde a}{}^{\tilde b} \e^{\tilde\a\,\tilde\b} Q_{\tilde b\,\tilde\b}{}^i\,,\quad
(T_{\tilde a\tilde b})^* = -(\sigma_1)_{\tilde a}{}^{\tilde c} (\sigma_1)_{\tilde b}{}^{\tilde d} T_{\tilde c\tilde d}\,,\quad
(T_{\tilde\alpha\tilde\beta})^* = \e^{\tilde\a\,\tilde\g}\e^{\tilde\b\,\tilde\d}  T_{\tilde\g\tilde\d}\,.
\end{align}

The analysis of the superalgebra isomorphisms for special values of $\alpha$ is identical to AdS$_2\times S^2$ up to an obvious difference for $\alpha=\infty$, which now corresponds to $\rm SU(2|2)\rtimes ISO(2)$ associated with an $\bbR^2\times S^2$ geometry.
Choosing polar coordinates $\omega,\ \rho$ on $H^2$, we obtain the metric 
\begin{equation}
\label{metric H2xS2}
\rd s^2 = \frac1{\l_-^2}(\rd\rho^2+\sinh^2\!\rho\,\rd\omega^2) 
+ \frac1{\l_+^2}(\rd\theta^2+\sin^2\!\theta\,\rd\phi^2).
\end{equation}

\subsection{Deformed supersymmetry in flat space}

The facts that in Euclidean signature the reality conditions on spinors do not mix chiralities and the self- and anti self-dual parts of a field strength are independent imply the possibility of deforming only the chiral or the anti-chiral part of the supersymmetry algebra.
In our notation, this corresponds to turning on $S^{ij}$ and $\cZ$ only.
As the Riemann tensor vanishes, these backgrounds correspond to flat Euclidean space with a deformed supersymmetry algebra.

Closure of the supersymmetry algebra imposes the constraint $S^{ij} \cZ^- = 0$, so that there are essentially two possibilities: either we turn on a generic $\cZ$, or we turn on a self-dual $\cZ$ and $S^{ij}$.
Because this constraint does not arise as an integrability condition for the Killing spinor equations \eqref{killing spinor equations euclidean}, we can give a common solution:
\begin{align}
\xi_\alpha{}^i=\epsilon_\alpha{}^i,\qquad
\bar\xi^\dalpha{}_i = \bar\epsilon^\dalpha{}_i 
-\frac\ii2 x^a (S_{ij}\delta_{ab} -\epsilon_{ij}\cZ_{ab}) \bar\sigma^{b\,\dalpha\alpha} \epsilon_\alpha{}^j\qquad(S^{ij}\cZ^-_{ab}=0).
\end{align}

The self-dual part of $\cZ$ is proportional to the $T^+$ tensor of the Weyl multiplet.
We identify the background obtained by $\cZ^+\neq 0$ alone as the $\epsilon_1 = -\epsilon_2$ limit of the $\Omega$-background in flat space (see e.g. \cite{Hama.Hosomichi:SWTheory,KZ:ExtSUSY}).
A generic $\Omega$-deformation on flat space breaks half of the supersymmetry, so it cannot arise in our analysis.

\section{Rigid Euclidean supersymmetric actions} \label{sec:E_Actions}

\subsection{Vector multiplets}

The structure of $\cN=2$ Euclidean vector multiplets has been discussed elsewhere
in a number of places
(see e.g. \cite{Hama.Hosomichi:SWTheory, KZ:ExtSUSY, GM:AllSols, BBRT:S2xS2, Sinamuli, R-GS:ToricK}).
Of particular relevance is the work of Cortes et al.
which constructs vector multiplets in Euclidean signature via a timelike reduction from
5D \cite{CortesMohaupt1}. Because their construction naturally
leads to a real Euclidean superalgebra,
the approach we sketch below will naturally match theirs up to
conventions.

It is simplest to motivate the reality conditions of the vector multiplet from superspace.
If we straightforwardly analytically continue all of the $\cF_{AB}{}^I$ field strengths,
requiring that the new ones be real in Euclidean signature, we discover
that $\cX^I$ and $\bar \cX^I$ are imaginary superfields. It will be convenient to make
the formal replacements
\begin{align}\label{eq:SuperVectorWick}
\cX^I = i \cX_+^I~, \qquad
\bar\cX^I = -i \cX_-^I~,
\end{align}
where $\cX_+^I$ and $\cX_-^I$ are real chiral and antichiral
superfields obeying the Bianchi identity
\begin{align}
(\cD^{ij} + 4 S^{ij}) \cX_+^I = -(\bar \cD^{ij} + 4 \bar S^{ij}) \cX_-^I~.
\end{align}
We define now the covariant components of $\cX_\pm{}^I$ as
\begin{align}
\l_\alpha{}^i = -i \cD_\alpha{}^i \cX_+^I \loco~, \qquad
\bar\l^{\dalpha i} = -i \bar \cD^{\dalpha i} \cX_-^I \loco~, \qquad
Y^{ij I} := -\frac{i}{2} (\cD^{ij} + 4 S^{ij}) \cX_+^I \loco
\end{align}
so that they formally do not change upon continuation to Euclidean signature.
The fermions are now symplectic Majorana-Weyl
but the auxiliary field is still pseudoreal,
\begin{align}
(\l_\alpha{}^i)^* = \l^\alpha{}_i~, \qquad
(\bar\l^{\dalpha i})^* = \l_{\dalpha i}~, \qquad
(Y^{ij I})^* = Y_{ij}{}^I~.
\end{align}
The component two-form field strength is
\begin{align}
F_{ab}{}^I = \frac{i}{4} (\sigma_{ab})^{\beta \alpha} \cD_{\beta\alpha} \cX_+^I\loco
	- \frac{i}{4} (\bsigma_{ab})^{\dbeta \dalpha} \bar\cD_{\dbeta \dalpha} \cX_-^I \loco
	- i \cZ_{ab} \cX_+^I \loco 
	+ i \bar \cZ_{ab} \cX_-^I \loco~.
\end{align}
We follow the notation $\cX_\pm^I$ of \cite{CortesMohaupt1},
which denotes the chirality of the associated gaugino.
Note that $\cX_+^I$ contains the anti self-dual field strength $F^-_{ab}{}^I$.

The supersymmetry transformations are easy to find: one simply makes the replacements
$X \rightarrow i X_+$ and $\bar X \rightarrow -i X_-$ everywhere, leading to
\begin{subequations}
\begin{align}
\delta X_+^I &= i\xi^\alpha_i \l_\alpha^i{}^I~, \qquad \qquad
\delta X_-^I = i\bar \xi_\dalpha^i \bar \l^\dalpha_i{}^I~, \displaybreak[1]\\[0.5ex]
\delta \l_{\alpha i}{}^I &=
	(F_{ab}{}^I + i \cZ_{ab} X_+^I - i \bar \cZ_{ab} X_-^I) (\sigma^{ab} \xi_i)_\alpha
	+ (Y_{ij}{}^I + 2 i S_{ij} X_+^I) \xi_\alpha{}^j
	\eol & \quad
	+ 2\,\cD_{a} X_+^I \, (\sigma^a \bar \xi_i)_\alpha
	- 4 \,G_{a\, ij} X_+^I \, (\sigma^a \bar\xi^{j})_\alpha
	- 2 g \,X_-^{\hat J} J_{+ \hat J}{}^I\,\xi_\alpha{}_i 
	~, \displaybreak[1] \\[0.5ex]
\delta \bar \l^{\dalpha i}{}^I &=
	(F_{ab}{}^I + i \cZ_{ab} X_+^I - i \bar \cZ_{ab}\bar X_-^I) (\bsigma^{ab} \bar \xi^i)^\dalpha
	- (Y^{ij}{}^I - 2 i \bar S^{ij} \bar X_-^I) \bar \xi^\dalpha{}_j
	\eol & \quad
	+ 2 \,\cD_a \bar X_-^I \, (\bsigma^a \xi^i)^\dalpha
	+ 4 \,G_{a}{}^{ij} \bar X_-^I \,(\bsigma^a \xi_{j})^\dalpha
	- 2 g \,X_+^{\hat J} \bar J_{-\hat J}{}^I \, \bar \xi^\dalpha{}^i 
	~, \displaybreak[1] \\[0.5ex]
\delta Y_{ij}{}^I &=
	2i \,\xi_{(i} {\slashed{\cD}} \bar \l_{j)}{}^I
	- 4i G_a{}_{k (i} \,\xi_{j)} \sigma^a \bar\l{}^{k I}
	- 2 G_a \, \xi_{(i} \sigma^a \bar \l_{j)}{}^I
	\eol & \quad
	- 2i \,\bar\xi_{(i} {\slashed{\cD}} \l_{j)}{}^I
	+ 4i G_a{}_{k (i} \,\bar \xi_{j)} \bsigma^a \l{}^{k I}
	- 2 G_a \, \bar \xi_{(i} \bsigma^a \l_{j)}{}^I
	\eol & \quad
	+ 4 i g\, \xi_{(i} \l_{j)}{}^J\, \bar J_{-J}{}^I
	- 4 i g\, \bar\xi_{(i} \bar \l_{j)}{}^J\, J_{+J}{}^I
	~, \displaybreak[1]\\[0.5ex]
\delta A_m{}^I &= i (\xi_j \sigma_m \bar \l^j{}^I) + i (\bar \xi^j \bsigma_m \l_j{}^I)~.
\end{align}
\end{subequations}

We normalize the vector multiplet action in superspace as
\begin{align}
- \int \rd^4x\, \rd^4\q\, \cE\, F(X_+)  + \int \rd^4x\, \rd^4\bar \q\, \bar \cE\, \bar F(X_-)
\end{align}
where $F = F(X_+)$ and $\bar F = \bar F(X_-)$ are both real functions, i.e.
$F(X_+)^* = F(X_+)$.
These can derived from the Lorentzian case by formally replacing $F \rightarrow -i F$ and
$\bar F \rightarrow -i \bar F$ and simultaneously replacing their arguments by \eqref{eq:SuperVectorWick}.
This amounts to
\begin{gather}\label{eq:VectorWick}
X^I \rightarrow i X_+^I~, \qquad \bar X^I \rightarrow -i X_-^I~, \qquad
F \rightarrow -i F~, \qquad \bar F \rightarrow -i \bar F~, \eol
F_I \rightarrow -F_{I}~, \qquad \bar F_I \rightarrow \bar F_{I}~, \qquad
F_{IJ} \rightarrow i F_{IJ}~, \qquad \bar F_{IJ} \rightarrow i \bar F_{IJ}~.
\end{gather}
Now the special K\"ahler potential and metric are given by
$K = X_-^I F_I - X_+^I \bar F_I$ and $g_{IJ} = F_{IJ} - \bar F_{IJ}$.
The target space geometry is actually a special para-K\"ahler
manifold given in terms of adapted coordinates. Our conventions here
differ somewhat from \cite{CortesMohaupt1}.

Consistent with the modifications \eqref{eq:VectorWick}, we should take
\begin{align}
J_I{}^J \rightarrow i J_{+ I}{}^J~, \qquad
\bar J_I{}^J \rightarrow -i J_{- I}{}^J~, \qquad
D_I \rightarrow i D_I^{(E)}~, \qquad
C_{\hat I, \hat J \hat K} \rightarrow i C_{\hat I, \hat J \hat K}
\end{align}
where the real Euclidean moment map $D_I^{(E)}$ is given by
\begin{align}
D_{\hat I}^{(E)} = f_{\hat I \hat J}{}^{K} (X^{\hat J} \bar F_K + \bar X^{\hat J} F_K)
	- C_{\hat I, \hat J \hat K} X^{\hat J} \bar X^{\hat K}~.
\end{align}

The symplectic vectors are not just the straightforward Wick rotations of the
Lorentzian case. An additional factor of $i$ is needed for the dual field strengths
to account for exchanging
the Lorentzian $\eps_{abcd}$ for the Euclidean $\veps_{abcd}$. Our
choice of conventions above account for this so that the symplectic vectors take
the same form, 
$(X_+^I, F_{I})$
and
$(X_-^I, \bar F_{I})$.
Duality transformations are still described by $\textrm{ISp}(2n, \mathbb R)$
but now the inhomogeneous terms are real numbers $R_\pm^I$ and $R_{I}^\pm$.
This reflects the
presence of two real background vector multiplets $X_\pm^{(i)}$ that can be
introduced in each case, corresponding to the \emph{two real central charges} possible
in the Euclidean supersymmetry algebra. A conventional choice for
the background vector multiplets is
\begin{align}
X_+^{(1)} = X_-^{(1)} = 1~, \qquad
X_+^{(2)} = - X_-^{(2)} = 1~,
\end{align}
but sometimes a different choice is necessary.
As before, the presence of various background fields affects the geometry and the
possibilities for the frozen multiplets. If $G_a{}^{ij}$ is present so that an $\SO(1,1)_R$
symmetry is maintained, the frozen multiplets must be absent and the action must
be superconformal, with the prepotentials both homogeneous of degree two. If $S^{ij}$
and/or $\bar S^{ij}$ is present, the background multiplets must obey
$S^{ij} X_+^{(i)} = -\bar S^{ij} X_-^{(i)}$
so only one choice of background vector multiplets is possible. Generally if both $S^{ij}$ and $\bar S^{ij}$ are non-vanishing,
we can choose either $S^{ij} = \bar S^{ij}$ or $S^{ij} = -\bar S^{ij}$ via an
$\SO(1,1)_R$ gauge choice; this allows either $X_\pm^{(2)}$ or $X_\pm^{(1)}$ given above
but not both. In the degenerate case
$S^{ij} \neq 0$ but $\bar S^{ij} = 0$, we must have
$X_+^{(1)} = X_+^{(2)} = 0$, and only one of $X_-^{(1)}$ or $X_-^{(2)}$ is needed.
This situation has no analogue in the Lorentzian theory.

We will not explicitly give the Euclidean action, but it is straightforward to
write down by applying the necessary changes to the Lorentzian action.

\subsection{Hypermultiplets}

Unlike the vector multiplets, hypermultiplets in Euclidean signature can be defined
without any alteration of the target space. This can be understood in the
framework of \cite{CortesMohaupt1} by dimensionally reducing the 5D hypermultiplet
action along a timelike circle. The reality properties of the hyperk\"ahler
metric and its associated vielbeins $f_\mu{}_i{}^\ra$ are unchanged.
Requiring the same supersymmetry conditions for $\delta \phi^\mu$ as in the first line of
\eqref{eq:SUSYHyper} leads to the following reality conditions for the fermions:
\begin{align}
(\z_\alpha{}^\rb)^* = \omega^{\bar\rb}{}_{\rc} \z^{\alpha \rc}
	= g^{\bar \rb \rb} \omega_{\rb \rc} \z^{\alpha \rc}~, \qquad
(\bar\z^\dalpha{}^{\bar\rb})^* = g^{\rb \bar \rb} \omega_{\bar\rb \bar\rc} \bar \z_\dalpha{}^{\bar \rc}
	= \omega^{\rb \rc} g_{\rc \bar \rc} \bar \z_\dalpha{}^{\bar \rc}~.
\end{align}
It would be reasonable
to denote the antichiral spinor as $\bar \z^{\dalpha \rb}$ by contracting with
a factor involving the $\Sp(n)$ metric and symplectic two-form
but we will avoid doing so to keep
our formulae as similar to the Lorentzian case as possible.

Now it is quite easy to convert the Lorentzian SUSY rules and action to
Euclidean signature. The only issue to keep in mind is that we must now take
\begin{align}
S^{ij} = i \mu v^{ij}~, \qquad \bar S^{ij} = i \tilde\mu v^{ij}~,
\end{align}
with $\mu$ and $\tilde\mu$ in principle different. This means we must
analytically continue $\mu \rightarrow i \mu$ and $\bar\mu \rightarrow i \tilde \mu$
from the Lorentzian formulae. The SUSY transformations become
\begin{align}
\delta \phi^\mu &= \xi_i \z^\rb \, f_\rb{}^i{}^\mu + \bar \xi^{i} \bar \z^{\bar \rb}\, f_{\bar \rb i}{}^\mu~, \eol
\delta \z_\alpha^\ra
	&=
	\Big(2 i \,\cD_{\alpha\dbeta} \phi^\mu
		- 4 \,G_{\alpha \dbeta} \chi^\mu \Big) f_\mu{}_i{}^\ra \,\bar\xi^{\dbeta i} 
	+ \Big(2 i \mu V^\mu - 4 i g X_-^{\hat I} J_{\hat I}{}^\mu \Big) f_\mu{}_i{}^\ra \eps^{ij} \xi_{\alpha j}
	\eol & \quad
	+ 2 \,Y_\alpha{}^\beta \chi^\mu f_\mu{}_i{}^{\ra} \eps^{ij} \xi_{\beta j} 
	- \Gamma_{\mu \rb}{}^\ra \delta \phi^\mu \, \z_\alpha^\rb~, \eol	
\delta \bar\z^{\dalpha \bar\ra} &=
	\Big(2 i \,\cD^{\dalpha\beta} \phi^\mu
		+ 4 G^{\dalpha \beta} \chi^\mu \Big) f_\mu{}^{i \bar \ra} \xi_{\beta i} 
	- \Big(2i \tilde \mu V^{\mu} + 4 i g X_+^{\hat I} J_{\hat I}{}^{\mu} \Big) f_\mu{}^{i \bar \ra} \eps_{ij} \bar \xi^{\dalpha j} 
	\eol & \quad
	- 2 \,\bar Y^\dalpha{}_\dbeta \chi^\mu f_\mu{}^{i \bar \ra} \eps_{ij} \bar \xi^{\dbeta j} 
	- \Gamma_{\mu \bar\rb}{}^{\bar\ra} \delta \phi^{\mu}
		\, \bar\z^{\dalpha \bar\rb}~.
\end{align}
We do not explicitly give the action in Euclidean signature, but it is easy to work
out by making the appropriate replacements in the Lorentzian action.

\section{Discussion and conclusions} \label{sec:D&C}

In this paper we classified the rigid backgrounds and actions
that admit full (real) $\cN=2$ supersymmetry in Lorentzian and Euclidean signatures.
The principle guiding this classification has been the identification of the
supercoset spaces which arise in curved superspace when the requirement of full
supersymmetry is imposed.
It is worth noting that similar results as those in Section~\ref{sec:L_SUSY}
were found for full $\cN=1$ Lorentzian supersymmetry in 5D~\cite{KNT-M:5DSCS}.
Our analysis regards the supercoset spaces as global manifolds.
Quotents by discrete isometries are allowed if they preserve the supercharges.
For example, in an appropriate Lorentz gauge it is straightforward to see that the \SU(2|2) realizations of $S^3\times\bbR$ admit quotienting by discrete right isometries.
This includes in particular lens spaces $S^3/\bbZ_p$, where the quotient acts freely on the Hopf fiber, and warping is also allowed.
We leave to future work a full analysis of all discrete quotients for general backgrounds.

One interesting issue that we have not addressed is the dynamical origin of
the rigid supersymmetric backgrounds. In particular, which of them solve Einstein's equations
of some 4D supergravity? Since any such theory is on-shell
equivalent to conformal supergravity coupled to two compensators -- a
vector multiplet and hypermultiplet -- one can look for simultaneous solutions of
the BPS conditions \eqref{eq:VM_BPS} and \eqref{eq:HM_BPS}. The equation of motion of the Weyl multiplet
auxiliary $D$ relates the hyperk\"ahler potential and the special K\"ahler potential
to the Planck scale;
because both potentials require non-vanishing VEVs for the vector and hyperk\"ahler
scalars, one must choose
$G_a = Y_{ab}^\pm = G_a{}^{ij} = 0$ in either signature. This reproduces the well known
fact \cite{Hristov:2009uj} that the only fully supersymmetric dynamical vacua arising in 4D Lorentzian
supergravity are $\rm AdS_4$ with $S^{ij} \neq 0$ and
${\rm AdS}_2 \times S^2$ (or its Penrose limit) with
$W_{ab}^\pm = \frac{1}{4} T_{ab}^\pm \neq 0$. Analogous statements hold for
(real) Euclidean supergravity where we find only $S^4$, $H^4$, or $H^2 \times S^2$.
However, this does not mean the other backgrounds are unphysical:
they might arise from a higher derivative theory similarly to what
happens in three dimensions \cite{Kuzenko:SCSReview},
or they could arise via dimensional reduction from higher dimensions.
For example, the general class of ${\rm AdS}_2 \times S^2$ with unequal radii can arise
from a higher dimensional supergravity theory with an ${\rm AdS}_2 \times S^2 \times S^2$
factor (see e.g. \cite{Zarembo:2010sg, Wulff:2014kja}).\footnote{We thank Dmitri Sorokin for this observation and for bringing these references to our attention.}
The latter possesses the supersymmetry algebra $\mathrm D(2,1;\alpha)$; upon reduction,
the isometry group of the internal $S^2$ becomes the 4D $R$-symmetry
group. It would be interesting to understand better possible uplifts of the other cases to higher dimensions.

Another interesting feature of many of the backgrounds is the presence of a single
timelike or spacelike $\bbR$ or $S^1$ factor. When the Killing spinors are independent
of this dimension, such as with the squashed $S^3$, one recovers $\cN=4$ Killing spinors
after a timelike or spacelike dimensional reduction to three dimensions.

We should emphasize again that we have
restricted the Euclidean backgrounds to those admitting real supercharges.
This excluded such cases as $S^2 \times S^2$ recently discussed in \cite{BBRT:S2xS2,Sinamuli,R-GS:ToricK} and reflects the well-known fact that the
$\mathrm D(2,1;\alpha)$ superalgebra possesses no real form with bosonic part
$\SU(2) \times \SU(2) \times \SU(2)$ \cite{Parker:1980af,Frappat:1996pb}.\footnote{Note that one can obtain four real supercharges on $S^2 \times S^2$ via an equivariant twist \cite{BBRT:S2xS2,R-GS:ToricK}.}
Nevertheless, one may still exploit our
results to investigate such cases by relaxing the requirement of
real supercharges.

For instance, one might allow a superisometry algebra
where supercharges and bosonic isometries appear with complex coefficients.
A nice example is offered by $S^2 \times S^2$; this arises by analytically continuing $\lambda_- \rightarrow i \lambda_-$ and $\rho \rightarrow i \rho$
in the $H^2 \times S^2$ supercoset space, leading to a complex form of $\mathrm D(2,1;\alpha)$ possessing imaginary $\alpha$.
One finds eight complex Killing spinors corresponding
to the analytic continuation of \eqref{killing spinors H2xS2}.
These correspond to the (untwisted) Killing spinors
discussed in \cite{BBRT:S2xS2, Sinamuli}.
The actions we have found in Section~\ref{sec:E_Actions} hold with all of the matter fields now understood as complex quantities.

An interesting intermediate case between pure reality and pure complexity
in Euclidean signature
would be to demand that some bosonic fields, such as the vector multiplet connection
$A_\mu{}^I$, remain real under repeated application of supersymmetry.
This would impose only that the bosonic isometries generated in the superalgebra
possess real coefficients. In the interesting case of $S^2 \times S^2$, one can
show that this is impossible. In particular, there is no non-trivial subalgebra
where the supercharges generate real non-vanishing bosonic isometries. In other words,
any choice of supercharges either generates bosonic isometries with complex coefficients,
or the supercharges are purely nilpotent.\footnote{Alternatively, one might consider $S^2 \times S^2$ with split
signature, $\eta_{ab} = \textrm(-1,-1,+1,+1)$.
Now the supergravity $R$-symmetry group becomes $\SL(2) \times \SO(1,1)$.
The superalgebra on $S^2 \times S^2$ becomes a real form of $\mathrm D(2,1;\alpha)$ with
bosonic group $\SU(2) \times \SU(2) \times \SL(2)$.}

Finally, in both Lorentzian and Euclidean cases, we have found a modified set of
full supersymmetry conditions for the vector and hypermultiplets. It would be interesting
to understand what role these may play in the analysis of quantum field theories
on these curved manifolds, especially in light of the results of \cite{Hama.Hosomichi:SWTheory}
on ellipsoids.

\section*{Acknowledgements}
We are grateful to Bernard de Wit, Kiril Hristov,
Sergei Kuzenko, and Dmitri Sorokin for insightful discussions over
the lifetime of this project.
This work was supported in part by the ERC Advanced Grant no. 246974,
{\it ``Supersymmetry: a window to non-perturbative physics''}
and by the European Commission
Marie Curie IIF grant no.
PIIF-GA-2012-627976.

\appendix

\section{Conventions}\label{App:Conv}
\addtocontents{toc}{\setcounter{tocdepth}{1}}

\subsection{Lorentzian signature}
We employ conventions similar to \cite{BGG} and \cite{WB}, with minor modifications.
Undotted Greek indices $\alpha, \beta, \cdots$ denote left-handed spinors
and dotted Greek indices $\dalpha, \dbeta, \cdots$ denote right-handed spinors.
These are raised and lowered using $\eps_{\alpha\beta}$ and $\eps_{\dalpha\dbeta}$,
obeying $\eps^{12} = \eps_{\dot 2 \dot 1} = 1$.
We denote ${\rm SU}(2)_R$ indices by $i, j, k, \cdots$ with $i=\1,\2$ and raise
and lower them with $\eps_{ij}$ and $\eps^{ij}$ as with spinor indices.

The sigma matrices $(\sigma^a)_{\alpha \dalpha}$ are defined as
\begin{equation}
{\sigma}^0=
\begin{pmatrix}
1 & 0 \\
0 & 1
\end{pmatrix}~, \qquad
{\sigma}^1=
\begin{pmatrix}
0 & 1 \\ 1 & 0
\end{pmatrix}~, \qquad
{\sigma}^2 =
\begin{pmatrix}
0 & -i \\ i & 0
\end{pmatrix}~, \qquad
{\sigma}^3 =
\begin{pmatrix}
1 & 0 \\ 0 & -1
\end{pmatrix}~,
\end{equation}
and the antisymmetric symbol $\eps_{abcd}$ is normalized as
$\eps_{0123} \,=\,+1$ and $\eps^{0123} \,=\,-1$.
The conjugate sigma matrix $(\bar\sigma^a)^{\dalpha\alpha}$ is given by
\begin{align}
({\bsigma}^a){}^{\dalpha \alpha} &=\,{\eps}^{\dalpha \dbeta} {\eps}^{\alpha \beta} ({\sigma}^{a})_{\beta \dbeta} ~, \qquad
{\bsigma}^0 =\,{\sigma}^0, \quad {\bsigma}^{1,2,3} \,=\,-{\sigma}^{1,2,3}.
\end{align}
The product of $\sigma^a$ with $\bar\sigma^b$ is
\begin{align}
 {\sigma}^a {\bsigma}^b = - {\eta}^{ab} + 2{\sigma}^{ab}~, \qquad
\bsigma^a \sigma^b = - \eta^{ab} + 2\bsigma^{ab}~.
\end{align}
The tensors $(\sigma^{ab})_\alpha{}^\beta$ and $(\bsigma^{ab})^\dalpha{}_\dbeta$
are anti-selfdual and selfdual, respectively,
\begin{align}
\frac{1}{2} {\eps}_{abcd}\, {\sigma}^{cd} \,=\,-i{\sigma}_{ab}, \quad
\frac{1}{2} {\eps}_{abcd} \,{\bsigma}^{cd} \,=\,+i{\bsigma}_{ab}~.
\end{align}
The product of three sigma matrices is
\begin{align}
 {\sigma}^{a} {\bsigma}^{b} {\sigma}^c
	&= -{\eta}^{ab}{\sigma}^c
	+{\eta}^{ca} \sigma^b
	- {\eta}^{bc}{\sigma}^{a}
	+ i{\eps}^{abcd} {\sigma}_{d}~, \eol
{\bsigma}^{a} {\sigma}^{b} {\bsigma}^c
	&= -{\eta}^{ab}{\bsigma}^c
	+{\eta}^{ca} \bsigma^b
	- {\eta}^{bc}{\bsigma}^{a}
	- i{\eps}^{abcd} {\bsigma}_{d}~.
\end{align}
We will also use the four-component gamma matrices and charge conjugation matrix:
\begin{align}\label{eq:GammaConventions}
\gamma^a =
\begin{pmatrix}
0 & i (\sigma^a)_{\alpha\dbeta} \\
i (\bsigma^a)^{\dalpha \beta} & 0
\end{pmatrix}~,\quad
(\gamma^a)^\dag = \gamma_a~, \quad \{\gamma^a, \gamma^b\} = 2 \,\eta^{ab}~, \quad
C =\begin{pmatrix}
\eps^{\alpha \beta} & 0 \\
0 & \eps_{\dalpha \dbeta}
\end{pmatrix}~.
\end{align}

Associated with any vector $V^a$ is the bispinor
$V_{\alpha\dbeta} = V^a (\sigma_a)_{\alpha\dbeta}$.
Given a tensor $F_{ab}$, we define the dual $\tilde F_{ab}$ and
selfdual (antiselfdual) components $F_{ab}^\pm$ by
\begin{gather}
F_{ab} = F_{ab}^- + F_{ab}^+~, \qquad
\tilde F_{ab} = \frac{1}{2} \eps_{abcd} F^{cd} = -i F_{ab}^- + i F_{ab}^+~, \qquad
F_{ab}^\pm = \frac{1}{2} (F_{ab} \mp i \tilde F_{ab})~.
\end{gather}
The selfdual (anti-selfdual) components are related to symmetric
bispinors $F_{\dalpha\dbeta}$ ($F_{\alpha\beta}$) by
\begin{align}
F_{ab}^- = -(\sigma_{ab})^{\alpha \beta} F_{\alpha\beta}
	= (\sigma_{ab})_\alpha{}^\beta F_\beta{}^\alpha~, \qquad
F_{ab}^+ = -(\bsigma_{ab})^{\dalpha \dbeta} F_{\dalpha\dbeta}
	= (\bsigma_{ab})^\dalpha{}_\dbeta F^\dbeta{}_\dalpha~.
\end{align}

Complex conjugation exchanges dotted for undotted spinors but does not change their
positions. That is, $(\psi_\alpha)^* = \bar\psi_\dalpha$, $(\psi^\alpha)^* = \bar\psi^\dalpha$
and $\big((\sigma^a)_{\alpha \dbeta}\big)^* = (\sigma^a)_{\beta \dalpha}$.
As usual, complex conjugation transposes Grassmann quantities, so that
$(\psi_\alpha \rho_\beta)^* = \bar \rho_\dbeta\, \bar\psi_\dalpha 
= - \bar\psi_\dalpha \,\bar \rho_\dbeta$. If $\psi$ and $\rho$ are operators
with a non-trivial anticommutator, one must interpret this statement as
$(\psi_\alpha \rho_\beta)^* = -\bar\psi_\dalpha \bar \rho_\dbeta$.
If $V_a$ is a real vector, then
$(V_{\alpha\dbeta})^* = V_{\beta \dalpha}$,
and if $F_{ab}$ is a real tensor, then 
$(F_{\alpha\beta})^* = - F_{\dalpha\dbeta}$.
Killing spinors $\xi_{\alpha i}$ and $\bar \xi^{\dalpha i}$ obey
\begin{align}
(\xi_{\alpha i})^* = \bar \xi_{\dalpha}{}^i~, \qquad
(\bar \xi^{\dalpha i})^* = \xi^\alpha{}_i~.
\end{align}
Because $(\eps^{ij})^* = -\eps_{ij}$ one has
$(\xi_{\alpha}{}^i)^* = -\bar \xi_{\dalpha i}$ and
$(\bar \xi^\dalpha{}_i)^* = -\xi^\alpha{}^i$.
For the covariant derivatives of superspace, one finds
\begin{align}
(\cD_{\alpha \dbeta})^* = \cD_{\beta \dalpha}~, \qquad
(\cD_\alpha{}^i)^* = \bar \cD_{\dalpha i}~, \qquad
(\bar \cD^\dalpha{}_i)^* = \cD^{\alpha i}~.
\end{align}

\subsection{Euclidean signature}
Our Euclidean conventions amount to taking $V^4 = i V^0$ for all vector and
tensor-valued objects. In particular,
\begin{align}
(\sigma^4)_{\alpha\dbeta} = i (\sigma^0)_{\alpha\dbeta} =
\begin{pmatrix}
i & 0 \\
0 & i 
\end{pmatrix}~, \qquad
(\bsigma^4)^{\dalpha\beta} = i (\bsigma^0)^{\dalpha\beta} =
\begin{pmatrix}
i & 0 \\
0 & i 
\end{pmatrix}~.
\end{align}
We keep all the other $\sigma$ matrices and $\eps_{\alpha\beta}$
unchanged. The Euclidean metric is $\eta_{ab} = \delta_{ab}$.
We trade the antisymmetric symbol $\eps_{0123} = +1$
for the Euclidean $\veps_{1234} = +1$, which amounts to exchanging
$\eps_{abcd}$ for $i \veps_{abcd}$.
Now one finds
\begin{align}
 {\sigma}^a {\bsigma}^b &= - {\delta}^{ab} + 2{\sigma}^{ab}~, \qquad
\bsigma^a \sigma^b = - \delta^{ab} + 2\bsigma^{ab}~, \eol
\frac{1}{2} {\veps}_{abcd}\, {\sigma}^{cd} \,&=\,-{\sigma}_{ab}, \qquad
\frac{1}{2} {\veps}_{abcd} \,{\bsigma}^{cd} \,=\,+{\bsigma}_{ab}~, \eol
 {\sigma}^{a} {\bsigma}^{b} {\sigma}^c
	&= -{\eta}^{ab}{\sigma}^c
	+{\eta}^{ca} \sigma^b
	- {\eta}^{bc}{\sigma}^{a}
	- {\veps}^{abcd} {\sigma}_{d}~, \eol
{\bsigma}^{a} {\sigma}^{b} {\bsigma}^c
	&= -{\eta}^{ab}{\bsigma}^c
	+{\eta}^{ca} \bsigma^b
	- {\eta}^{bc}{\bsigma}^{a}
	+ {\veps}^{abcd} {\bsigma}_{d}~.
\end{align}
The four-component $\gamma^a$ and charge conjugation matrix are
\begin{align}
\gamma^a =
\begin{pmatrix}
0 & i (\sigma^a)_{\alpha\dbeta} \\
i (\bsigma^a)^{\dalpha \beta} & 0
\end{pmatrix}~,\quad
(\gamma^a)^\dag = \gamma^a~, \quad \{\gamma^a, \gamma^b\} = 2 \,\delta^{ab}~, \quad
C =\begin{pmatrix}
\eps^{\alpha \beta} & 0 \\
0 & \eps_{\dalpha \dbeta}
\end{pmatrix}~.
\end{align}
The self-dual and anti-self-dual components of $F_{ab}$ are
\begin{gather}
F_{ab} = F_{ab}^- + F_{ab}^+~, \qquad
\tilde F_{ab} = \frac{1}{2} \veps_{abcd} F^{cd} = -F_{ab}^- + F_{ab}^+~, \qquad
F_{ab}^\pm = \frac{1}{2} (F_{ab} \pm \tilde F_{ab})~.
\end{gather}Under complex conjugation undotted indices remain undotted but are raised or
lowered,
\begin{align}
((\sigma^a)_{\alpha\dbeta})^* = -(\bsigma^a)^{\dbeta \alpha}~, \quad
((\sigma^{ab})_\alpha{}^\beta)^* = -(\sigma^{ab})_\beta{}^\alpha~, \quad
((\sigma^{ab})_{\alpha\beta})^* = (\sigma^{ab})^{\alpha\beta}~,
\end{align}
so if $V_a$ and $F_{ab}$ are real, then
\begin{align}
(V_{\alpha \dbeta})^* = -V^{\dbeta \alpha}~, \qquad
(F_{\alpha\beta})^* = F^{\alpha\beta}~.
\end{align}
The Killing spinors are chosen to be symplectic Majorana-Weyl,
\begin{gather}
(\xi_\alpha{}^i)^* = \xi^\alpha{}_i~, \qquad
(\bar\xi^\dalpha{}^i)^* = \bar\xi_\dalpha{}_i~.
\end{gather}
Keeping in mind that
$(\eps^{\alpha\beta})^* = -\eps_{\alpha\beta}$, one can see that
$(\xi^{\alpha i})^* = -\xi_{\alpha i}$, so the positions of the indices must be
observed. These conditions imply that
\begin{align}
(\cD_\alpha{}^i)^* = - \cD^\alpha{}_i~, \qquad
(\bar\cD^{\dalpha i})^* = -\bar\cD_{\dalpha i}~.
\end{align}

\section{General action principle in rigid superspace}\label{App:Actions}
For the rigid superspace geometry discussed in this paper, there are a few general
formulae that will be relevant for relating superspace actions to component ones.
We emphasize that these expressions are valid only for the rigid supergeometries discussed.
In particular, they assume the covariant constancy of the various
superfields $S^{ij}$, $Y_{\alpha\beta}$, $W_{\alpha\beta}$, etc.

First, let us relate a chiral superspace action
to a component one. We take
\begin{align}\label{eq:ChiralAction}
S = \int \rd^4x\, \rd^4\q\, \cE\, \mathscr{L}_c
\end{align}
where $\mathscr{L}_c$ is a covariantly chiral superfield,
$\bar \cD^\dalpha_i \mathscr{L}_c = 0$, and $\cE$ is the appropriate superspace
measure constructed from the superdeterminant (Berezinian) of the chiral part of the superspace
vielbein. The component Lagrangian constructed from \eqref{eq:ChiralAction} is
\begin{align}
\cL &= \frac{1}{96} \cD^{ij} \cD_{ij} \mathscr{L}_c
	- \frac{1}{96} \cD^{\alpha\beta} \cD_{\alpha\beta} \mathscr{L}_c
	+ \frac{2}{3} S_{ij} \cD^{ij} \mathscr{L}_c
	- \frac{1}{3} Y^{\alpha\beta} \cD_{\alpha\beta} \mathscr{L}_c
	\eol & \quad
	+ \Big(3 S^{ij} S_{ij} 
	- Y^{\alpha\beta} Y_{\alpha\beta}
	+ \bar W_{\dalpha\dbeta} \bar W^{\dalpha\dbeta} \Big) \mathscr{L}_c
\end{align}
where we have defined $\cD^{ij} := \cD^{\gamma (i} \cD_\gamma^{j)}$
and $\cD_{\alpha\beta} := \cD_{(\alpha}^k \cD_{\beta) k}$ and the projection to $\theta=0$
is assumed.
This is a special case of the chiral action presented in \cite{GKT-M:ChiralActions}.

Next is the action relating projective superspace actions to component actions.
We use the projective superspace action principle 
adapted to the rigid superspace geometries discussed in this paper,
\begin{align}
S &= -\frac{1}{2\pi} \oint_\cC \rd \tau \int \rd^4x\, \rd^4\q^+\, \cE^{--}\, \mathscr{L}^{++}
	= -\frac{1}{2\pi} \oint_\cC v_i^+ \rd v^{i+} \int \rd^4x\, e\, \cL^{--}~,
\end{align}
with the $\rm SL(2, \mathbb C)$ representation of projective superspace
\cite{KLRT-M1, KLRT-M2, KT-M:DiffReps}
so that only a single contour integral is needed
(see the discussion in \cite{Butter:CSG4d.Proj}).
The component Lagrangian is
\begin{align}
\cL^{--}
	&= \frac{1}{16} (\cD^-)^2 (\bar \cD^-)^2  \mathscr{L}^{++}
	+ \frac{i}{2} G^{\dalpha \alpha --} [\cD_\alpha^-, \bar \cD_\dalpha^-] \mathscr{L}^{++}
	\eol & \quad
	+ \frac{3}{4} S^{--} (\bar \cD^-)^2 \mathscr{L}^{++}
	+ \frac{3}{4} \bar S^{--} (\cD^-)^2 \mathscr{L}^{++}
	+ 9 \bar S^{--} S^{--} \mathscr{L}^{++}
	\eol & \quad
	- \frac{i}{4} \cV_m^{--} (\bsigma^m)^{\dalpha \alpha} [\cD_\alpha^-, \bar\cD_\dalpha^-]\mathscr{L}^{++}
	- 12 \cV_a^{--} G^{a --} \mathscr{L}^{++}~.
\end{align}
This expression is real under the modified complex conjugation of projective superspace.
One must still perform the contour integral to arrive at a standard Lagrangian
in $x$-space.

For completeness, we include also the relation between full superspace and chiral superspace
actions, although the former generically involve higher derivative interactions and play no
role in this paper:
\begin{align}
\int \rd^4x\, \rd^4\q\, \rd^4\bar\q\, E\,\mathscr{L}
	= \int \rd^4x\, \rd^4\q\, \cE \Big(
	\frac{1}{48} \bar\cD^{ij} \bar \cD_{ij} \mathscr{L}
	+ \frac{1}{12} \bar S_{ij} \bar \cD^{ij} \mathscr{L}
	+ \frac{1}{4} \bar Y_{\dalpha\dbeta} \bar \cD^{\dalpha\dbeta} \mathscr{L}
	\Big)~.
\end{align}

\section{Details of Lorentzian backgrounds} \label{app:LBack}

\subsection{Warped AdS$_3$ spaces ($\mathrm{wAdS}_3\times\bbR$)}
\label{App:AdS3}

We gauge fix $G^a=(0,0,0,g)$ for constant $g$.
First consider the case without warping, where no other background fields are turned on.
Then the bosonic part of the superalgebra
involves only the generators $P_a = (P_I, P_3)$ with $I=0,1,2$,
\begin{align}
[ P_I, P_3 ] = 0~, \qquad 
[P_I, P_J] = 4 g\, \veps_{IJK} \eta^{KL} P_L~,\qquad
\eta=\text{diag}(-1,1,1)~.
\end{align} 
The dimensionless generators used to construct the group manifold are just $T_I \equiv \frac{1}{4g} P_I$.
We choose the explicit parameterization which leads to a global set of coordinates for AdS$_3$ with $\tau\in[0,4\pi)$:
\begin{equation}
L = e^{\phi T_2} e^{\rho T_1} e^{\tau T_0} e^{z P_3},
\end{equation}
Both choices of orthonormal frame specified in \eqref{SU(1,1) invariant vielbein} can be used to construct the physical vielbein and arrive at the metric \eqref{metric AdS3}.

In the spinor representation \eqref{eq:spinorial representation for G and Z case} we can decompose $L$ in terms of the embedding coordinates \eqref{coset repr for two sheeted H3}:
\begin{align}
L      = X + 2(Y\,T_0+V\,T_1+W\,T_2),\quad
L^{-1} = X - 2(Y\,T_0+V\,T_1+W\,T_2)\,.
\end{align}
In absence of warping, using $e^I=E^I/4g$ as vielbein, the Killing spinors read 
\begin{align}
\label{round AdS3 killing spinors}
\xi_{\alpha\,i} &= [X
-2 (Y \sigma^{12} +W\sigma^{01} +V\sigma^{20}) ]_\alpha{}^\beta \epsilon_\beta{}_i,
\end{align}
while if we choose $e^I=E'^I/4g$ we must flip the sign of $Y,\ W$ and $V$ (i.e. we exchange $L$ with $L^{-1}$).

\paragraph{Timelike stretched AdS$_3\times \bbR$}

Keeping $G^a=(0,0,0,g)$, we turn on $\cZ_{ab}=4\lambda\delta^{12}_{ab}$ and  require $|\lambda|<2|g|$.
The appropriate choice of dimensionless generators is now
\begin{align}
T_0&\equiv \upsilon\left( \frac1{4g}P_0 - \frac{\lambda^2}{4g^2} M_{12} \right),\qquad
T_{1,2} \equiv \sqrt\upsilon \frac1{4g}P_{1,2},\qquad
U \equiv \frac1{4g} P_0 - M_{12},
\end{align}
with $\upsilon = \left(1 - \frac{\lambda^2}{4g^2}\right)^{-1} \ge 1$.
We keep the same choice of coset representatives. The left-invariant vielbein reads
\begin{align}
e^0 &= \upsilon\frac1{4g}E^0,
\qquad e^{1,2} = \frac{\sqrt\upsilon}{4g}E^{1,2},
\qquad e^3 = dz.
\end{align}
The final metric is provided in equation \eqref{metric timelike stretched ads3}.

The Killing spinors are still given by the spinorial representation \eqref{eq:spinorial representation for G and Z case} of $L$, which now gives
\begin{align}
\xi_\alpha{}_i &=
[X -2Y\sigma^{12}-2\sqrt\u (W\sigma^{01}+V\sigma^{20})]_\alpha{}^\beta \epsilon_\beta{}_i 
+\ii\sqrt\upsilon\frac{\lambda}{2g} (W\sigma^1-V\sigma^2)_{\alpha\dbeta} \bar\epsilon^\dbeta{}_i\,.
\end{align}

Finally, the potentials for $\cZ$ and $*G$ are
\begin{align}
C_{(1)} = \frac{\lambda}{2g} e^0,\qquad
B_{(2)} &= \frac{\u^2}{64g^2} \sinh\rho\,\rd\phi\,\rd\tau~.
\end{align}

\paragraph{Spacelike squashed AdS$_3\times \bbR$}
This time we take $\cZ_{ab}=4\lambda\delta^{01}_{ab}$.
We define $\upsilon=\left(1+\frac{\lambda^2}{4g^2}\right)^{-1},\ 0<\upsilon\le1$ and
\begin{align}
T_{0,1}&\equiv \frac{\sqrt\upsilon}{4g}P_{0,1},\qquad
T_2\equiv \upsilon\left( \frac1{4g}P_2 - \frac{\lambda^2}{4g^2} M_{01} \right),\qquad
U \equiv \frac1{4g} P_2 + M_{01}.
\end{align}
The residual Lorentz generator as well as $U$ are non-compact.
It proves convenient to choose the right-invariant forms $E'$ \eqref{SU(1,1) invariant vielbein} to obtain the physical vielbein
\begin{equation}
 e^{0,1} = \frac{\sqrt\upsilon}{4g}E'^{0,1},\qquad
 e^2 = \frac\upsilon{4g}E'^2~.
\end{equation}
The metric takes the form \eqref{metric spacelike squashed ads3}.

The Killing spinors are computed as usual:
\begin{align}
\xi_\alpha{}_i &=
[X +2( W\sigma^{01}+\sqrt\upsilon Y\sigma^{12}+\sqrt\upsilon V\sigma^{20})]_\alpha{}^\beta \epsilon_\beta{}_i 
-\ii\sqrt\upsilon\frac{\lambda}{2g} (V\sigma^0-Y\sigma^1)_{\alpha\dbeta} \bar\epsilon^\dbeta{}_i\ .
\end{align}

The background potentials for $\cZ$ and $*G$ are now respectively
\begin{align}
\label{potential B2 timelike warped AdS3}
C_{(1)} = -\frac{\l}{2g} e^2,\qquad
B_{(2)} = - \frac{\u^2}{64g^2} \sinh\rho\,\rd\phi\,\rd\tau\,.
\end{align}

\paragraph{Null warped AdS$_3\times\bbR$}
We take $\cZ$ to be null and fix it to $\cZ_{ab}=4\lambda(\delta^{02}_{ab}+\delta^{12}_{ab})$ with $\lambda>0$.
The usual procedure yields the \SU(1,1) generators
\begin{align}
\nonumber
T_0 &= \left(1+\frac\upsilon2\right)\frac1{4g}P_0 - \frac\upsilon{8g}P_1 +\upsilon (M_{02}-M_{12}),
&
U   &= M_{12}-M_{02}+\frac1{4g}(P_1-P_0)~,
\\
T_1 &= \left(1-\frac\upsilon2\right)\frac1{4g}P_1 + \frac\upsilon{8g}P_0 +\upsilon (M_{02}-M_{12}),
&
T_2 &= \frac1{4g}P_2~.
\end{align}
where this time $\upsilon=\frac{\lambda^2}{4g^2}\ge0$.
Using again the one forms $E'^I$ for convenience, we construct the physical vielbein giving rise to the metric \eqref{metric null warped ads3}
\begin{align}
e^0 &= \frac1{4g}\left( \big(1+\frac\upsilon2\big)E'^0+\frac\upsilon2 E'^1 \right),\quad
e^1 = \frac1{4g}\left( \big(1-\frac\upsilon2\big)E'^1-\frac\upsilon2 E'^0 \right),\quad
e^2 = \frac1{4g}E'^2.
\end{align}
We also find the Killing spinors:
\begin{align}
\begin{split}
\xi_\alpha{}_i =\ &
[X +2( W\sigma^{01}+Y\sigma^{12}+V\sigma^{20})-\upsilon(Y+V)(\sigma^{12}-\sigma^{20})]_\alpha{}^\beta \epsilon_\beta{}_i \\
&-\ii\,\frac{\l}{2g}[W(\sigma^0+\sigma^1)-(Y+V)\sigma^2]_{\alpha\dbeta} \bar\epsilon^\dbeta{}_i\ .
\end{split}
\end{align}

As usual we provide the potentials for $\cZ$ and $*G$:
\begin{align}
C_{(1)} = \frac{\l}{8 g^2} (E'^0+E'^1)\,,\qquad
B_{(2)} = -\frac{1}{64g^2}\sinh\rho\,\rd\phi\,\rd\tau.
\end{align}

{If we introduce Poincar\'e coordinates as in equation \eqref{metric null warped ads3 Poincarè}, the appropriate coset representative, dimensionless vielbein and embedding coordinates read}
\begin{equation}
\begin{aligned}
L&=e^{\sqrt2x_-T_-}r^{-2T_2}e^{\sqrt2x_+T_+},\quad T_\pm=\frac1{\sqrt2}(T_1\pm T_0)~,\\[1ex]
E'^0 &=  \rd x_- +\frac{2x_-}{r}\rd r - \frac{x_-^2+1}{r^2}\rd x_+ ,\quad
E'^1 = -\rd x_- +\frac{2x_-}{r}\rd r + \frac{x_-^2-1}{r^2}\rd x_+ ,\\
E'^2 &= \frac{2\rd r}{r}+\frac{2x_-\rd x_+}{r^2}\,,\\[1ex]
X &= \frac{r^2+1+x_-x_+}{2r},  \quad
Y = \frac{x_+-x_-}{2r},  \quad
V = \frac{x_++x_-}{2r},    \quad
W = \frac{r^2-1+x_-x_+}{2r}   \,.
\end{aligned}
\end{equation}

\paragraph{AdS$_3\times\bbR$ and $\rm SU(1,1|1)^2$ supersymmetry}
The second realization of $\cN=2$ supersymmetry on a `round' AdS$_3\times\bbR$ is supported by $G_a{}^i{}_j$, which we gauge-fix to $G_a{}^i{}_j=g_a(\ii \sigma_3)^i{}_j$, $g_a=(0,0,0,g)$.
Analogously to the $S^3$ case we define
\begin{equation}
\Delta_a^{(i)}\equiv P_a+(-)^ig(\delta_a^3\mathbb A+\epsilon_{a3}{}^{cd}M_{cd})
\end{equation}
which gives us the non-vanishing (anti)commutators of $\SU(1,1|1)^2$
\begin{formula}
\{Q_\alpha{}^i,\bar Q_\dbeta{}_j\}&=-2\ii\delta^i_j\Delta_{\alpha\dbeta}^{(i)},\\
[\Delta^{(i)}_a,Q_\alpha{}^j] &= (-)^{i+1}\delta^{ij}\,2g\left[\ii\delta^3_a\delta_\alpha^\beta + \epsilon_{a3cd}(\sigma^{cd})_\alpha{}^\beta\right]Q_\beta^i,\\
[\Delta_a^{(i)},\Delta_b^{(j)}] &= (-)^i\delta^{ij} 4g\epsilon_{ab}{}^c{}_3\Delta_c^{(i)}.
\end{formula}
We will choose $T_I\equiv\frac1{4g}\Delta_I^{(2)}$ and generate the flat direction using $\Delta^{(2)}_3$.
We can take coordinates and (left invariant) vielbein as in Section~\ref{sec:AdS3}, so that the U(1)$_R$ connection turns out to be $A=g\, e^3 = g\,\rd z$ and we arrive at the Killing spinors $\hat\xi$
\begin{align}
\hat\xi_{\alpha \1}  = \eps_{\alpha \1}~, \qquad
\hat\xi_{\alpha\,\2} = e^{2 i g z} \xi_{\alpha\,\2},
\end{align}
where $\xi_{\alpha\,\2}$ is defined in \eqref{round AdS3 killing spinors}.
If the spatial direction generated by $P_3$ is compact, the choice of $\mathrm U(1)_R$ connection is not necessarily pure gauge and might correspond to a Wilson line along the circle.
For the Killing spinors $\hat\xi_\2$ to be well-defined the radius of the $z$ circle must be a multiple of $n\pi/g$, and the U$(1)_R$ connection is non-trivial for odd $n$.
The potential for $*G^i{}_j$ is $\ii \sigma_3$ times the potential $B_{(2)}$ of \eqref{potential B2 timelike warped AdS3}.

\subsection{AdS$_2\times S^2$ spacetimes and D$(2,1;\alpha)$}

Let us gauge-fix $\cZ_{ab}=-2\lambda_-\delta^{03}_{ab}+2\ii\lambda_+\delta^{12}_{ab}$, with $\lambda_\pm$ real.
We define coset representatives using dimensionless coordinates $\tau,\rho$ on $\rm AdS_2$ and $\phi,\ \theta$ on $S^2$
\begin{equation}
L\equiv e^{\tau\frac1{\l_-} P_0} e^{\rho\frac1{\l_-} P_3}
e^{\phi\, M_{12}} e^{\theta \frac{1}{\l_+}P_2} e^{-\phi\, M_{12}} ,
\end{equation}
from which we obtain a vielbein associated with the metric \eqref{metric AdS2xS2}
\begin{align}
e^0 &= \frac1{\l_-}\cosh\rho\,\rd\tau, & 
e^1 &= -\frac1{\l_+}(\sin\phi\,\rd\theta+\cos\phi\,\sin\theta\,\rd\phi) \eol
e^3 &= \frac1{\l_-}\rd\rho, &
e^2 &= \frac1{\l_+}(\cos\phi\,\rd\theta-\sin\phi\,\sin\theta\,\rd\phi).
\end{align}
The choice of Lorentz gauge in $L$ is convenient to guarantee that it is well-behaved at the north pole of $S^2$.
\begin{equation}
\label{change of gauge S2}
L(\tau,\rho,\phi,\theta) \to
L^{\rm(south)}(\tau,\rho,\phi,\theta) \equiv
L(\tau,\rho,\phi,\theta-\pi)e^{\pi\frac1{\l_+} P_2}
=L(\tau,\rho,\phi,\theta)e^{2\phi M_{12}}\,.
\end{equation}
As evidenced by the last equality, this is not a change of coordinates, but rather just a change in local Lorentz gauge.%
\footnote{The isotropy group of the south pole is the same of the north pole and
the transformation $\exp{\frac{\pi}{\l_+}P_2}$ induces the appropriate automorphism that gives rise to a single-valued gauge at the south pole.
This way to induce the change of gauge easily generalizes to higher dimensional spheres.}
The associated vielbein along the $S^2$ is 
\begin{equation}
e^1_{\rm(south)} =\frac1{\l_+}(\sin\phi\,\rd\theta-\cos\phi\,\sin\theta\,\rd\phi),
\qquad
e^2_{\rm(south)} = \frac1{\l_+}(\cos\phi\,\rd\theta+\sin\phi\,\sin\theta\,\rd\phi)~.
\end{equation}

In the northern gauge we can write Killing spinors in terms of the matrices
\begin{align}
A=\ &\cosh\frac\rho2 \left(\cos\frac\theta2 \cos\frac\tau2 
   +2\sin\frac\theta2 \sin\frac\tau2 
   (\sin\phi\,\sigma^{01} -\cos\phi\,\sigma^{02})\right) \eol
   &+2\ii \sinh\frac\rho2 \left(\cos\frac\theta2 \sin\frac\tau2 \sigma^{12}
   -\sin\frac\theta2 \cos\frac\tau2 (\cos\phi\,\sigma^{01}+\sin\phi\,\sigma^{02}) 
   \right)\,,\eol[1ex]
B=\ &-\ii\sinh\frac\rho2 \left(\cos\frac\theta2
   \cos\frac\tau2+\sin\frac\theta2 \sin\frac\tau2 (\sin\phi\,\sigma^1  - \cos\phi\,\sigma^2)\right) \eol
   &+\cosh\frac\rho2 \left(\ii \cos\frac\theta2 \sin\frac\tau2 \sigma^3
   + \sin\frac\theta2 \cos\frac\tau2 (\cos\phi\,\sigma^1+\sin\phi\,\sigma^2)\right)\,,
\end{align}
so that the Killing spinors of AdS$_2\times S^2$ are simply
\begin{align}
\xi_{\alpha\,i} = 
A_\alpha{}^\beta \epsilon_{\beta\,i} 
+ B_{\alpha\dbeta}\bar\epsilon^\dbeta{}_i\ .
\end{align}

Notice that these Killing spinors differ from \cite{Lu:1998nu}, because our choice of Lorentz gauge makes them periodic in $\phi$ and well-behaved at the north pole of $S^2$.
For each Killing spinor we can compute another expression that differs by a local Lorentz gauge transformation and is single-valued at the south pole.
This is the transformation described in \eqref{south pole substitution S4} and on the spinors it corresponds to the substitution
\begin{equation}
\xi_{(i)\hat\a}(\tau,\rho,\phi,\theta) \longrightarrow \xi^{\rm(south)}_{(i)\hat\a} \equiv (\ii \g_1)_{\hat\a}{}^{\hat\b}\xi_{(i)\hat\b}(\tau,\rho,\phi,\theta-\pi).
\end{equation}
For the spacetimes $\bbR^{1,1}\times S^2$ and AdS$_2\times \bbR^2$, the isometry generators associated with the flat directions become trivially represented in the spinorial representation and
the Killing spinors are obtained setting to zero the corresponding coordinates ($\tau,\, \rho$ and $\phi,\,\theta$ respectively) in the expression for AdS$_2\times S^2$.

The background complex two-form $\cZ$ has potential 
\begin{equation}
C_{(1)} = \frac1{\l_-} \sinh\rho\,\rd\tau-\,\frac\ii{\l_+}(\cos\theta\pm1)\,\rd\phi\,,
\end{equation}
the plus/minus sign corresponding to the northern and southern patches of $S^2$.

\subsection{Lorentzian $S^3\times\bbR$}

We take $G^a=(0,0,0,g)$ and $\cZ_{ab}=4\lambda\delta^{12}_{ab}$, imposing now $\lambda^2>4g^2$.
We introduce generators $T_0,\, T_1,\, T_2,$ satisfying the SU(2) algebra, defining $\upsilon = (\frac{\l^2}{4g^2}-1)^{-1}>0$ and
\begin{align}
T_0&\equiv \upsilon\left( -\frac1{4g}P_0 + \frac{\lambda^2}{4g^2} M_{12} \right),\qquad
T_{1,2} \equiv \sqrt\upsilon \frac1{4g}P_{1,2},\qquad
U \equiv \frac1{4g} P_0 - M_{23},
\end{align}
with $U$ commuting with everything.
We can choose the same group representative $L$ as for the standard $S^3$, trade $T_3$ there for $T_0$ here and write in the spinorial representation
\begin{align}
L = e^{\phi T_0} e^{\theta T_2} e^{\omega T_0}
  = X +2(Y T_0 + V T_1 + W T_2),
\end{align}
where we have also introduced the embedding coordinates \eqref{S3 embedding coordinates}.
We can also recycle the left-invariant dimensionless vielbein in \eqref{eq:S3RoundV} and identify the physical vielbein
\begin{align}
 e^0 = \frac\upsilon{4g} E^3,\quad
 e^{1,2} = \frac{\sqrt\upsilon}{4g}E^{1,2},\quad
 e^3 = \rd z.
\end{align}
The metric is given in equation \eqref{metric Lorentzian S3}.

The Killing spinors are
\begin{align}
\xi_{\alpha\,i} &= 
[X\mathbb I_2 -2(W\sigma^{12}+\sqrt\upsilon V\sigma^{20}+\sqrt\upsilon Y\sigma^{01})]_\alpha{}^\beta \epsilon_{\beta\,i}
+\ii\sqrt\upsilon\frac{\lambda}{2g}(Y\sigma^1-V\sigma^2)_{\alpha\dbeta}\bar\epsilon^{\dbeta}{}_i\ .
\end{align}

The background potentials are $C_{(1)}=-\frac{\l}{2g}e^0$ and $B_{(2)}$ given in \eqref{two form potential S3}.

\subsection{Lightlike $S^3\times\bbR$}

We choose $G^a = \frac{1}{\sqrt2}(g,0,0,g)$ and $\cZ_{ab}=4\lambda\delta^{12}_{ab}$. 
The absolute value of $g$ has no physical relevance as it can be rescaled by a Lorentz boost.
The appropriate choice of dimensionless generators turns out to be
\begin{align}
T_{1,2}&\equiv\frac1{2\lambda}P_{1,2},\quad
T_3 \equiv M_{12}-\frac{g}{\lambda^2}P_+,\quad
U \equiv\frac1{4g}P_- +M_{12}~,
\end{align}
with $P_\pm =\frac1{\sqrt2}(P_3\pm P_0)$.
Together with $P_+$, these generators form $\mathrm{SU(2)\times U(1)}_{U}\times \mathrm U(1)_{P_+}$.
We can choose the usual coordinates and expressions \eqref{eq:S3RoundV} for the SU(2) manifold, generating the fourth direction by $\exp({u\frac1{4g}P_-})$.
We then read off the physical vielbein giving rise to the metric \eqref{metric lightlike S3}
\begin{align}
e^- = \frac{1}{4g}\rd u,\quad e^+=\frac{g}{\lambda^2}E^3,\quad e^{1,2}=\frac1{2\lambda}E^{1,2}.
\end{align}

Killing spinors are computed in terms of the embedding coordinates \eqref{S3 embedding coordinates}:
\begin{align}
\xi_{\alpha\,i} &= 
\big[X\mathbb I_2 
-2\big(W\sigma^{12}
+\sqrt2\tfrac{g}{\lambda}(V+\ii Y)(\sigma^{20}-\ii \sigma^{01})\big)\big]_\alpha{}^\beta \epsilon_{\beta\,i}
+\ii(Y\sigma^1-V\sigma^2)_{\alpha\dbeta}\bar\epsilon^{\dbeta}{}_i\ .
\end{align}

The background forms are analogous to the other spheres: 
\begin{align}
C_{(1)} = -\frac{\l}{2g} e^+,\qquad
B_{(2)} &= \frac{1}{16\l^2}(\cos\theta\pm1)\,\rd\phi\,\rd u.
\end{align}

\subsection{`Overstretched' AdS$_3$}

There is a threshold case between timelike stretched AdS$_3\times\bbR$ and the Lorentzian sphere.
We take $G^a=(0,0,0,g)$ and $\cZ =4\lambda \delta^{12}_{ab}$, and choose the specific value $\lambda = 2g$.
The commutation relations read
\begin{align}
\begin{split}
[P_0,\,P_1] &= 4g P_2,              \\
[P_0,\,P_2] &= -4g P_1,             \\
[P_1,\,P_2] &= -4g P_0+4gM_{12} \equiv -4g H,    
\end{split}
&
\begin{split}
[M_{12},\,P_1] &= P_2,              \\[.5ex]
[M_{12},\,P_2] &= -P_1.
\end{split}
\end{align}
The isometry generators $P_1$ and $P_2$ close on a central charge, and we obtain the algebra of ${\rm Heis_3\rtimes U(1)}_M\times \mathrm U(1)_{P_3}$ or, alternatively, $\mathrm{ISO}(2)_{(H)}\times \mathrm U(1)_{P_3}$, $H$ being a central extension of ISO(2).
We are left with a group manifold 
${\rm Heis}_3\times \mathrm U(1)_{P_3}$, of which we parameterize a generic element as
$
L = e^{t H} e^{x P_1 + y P_2} e^{z P_3} 
$
and easily arrive at the vielbein 
\begin{equation}
e^0 = \rd t+2g(x\rd y-y\rd x),\quad
e^1=\rd x,\quad
e^2=\rd y,\quad
e^3=\rd z
\end{equation}
and the metric \eqref{metric relativistic fluid}.
The Killing spinors are:
\begin{align}
\xi_{\alpha\,i} &= 
[\mathbb I_2 -4g x \sigma^{20} -4g y \sigma^{01} ]_\alpha{}^\beta \epsilon_{\beta\,i}
-2\ii g( x \sigma^2-y\sigma^1 )_{\alpha\dbeta}\bar\epsilon^{\dbeta}{}_i\ .
\end{align}

The background potentials are
$C_{(1)}=2g(x\rd y-y\rd x)$ and 
$B_{(2)}=\frac14 C_{(1)}\wedge\rd t$.

\subsection{Plane waves}

We have $G^a=\frac{1}{\sqrt2}(g,0,0,g)=g\d^a_+$ and $\cZ_{ab}=2\sqrt2\lambda_+\delta_{ab}^{-1} -2\sqrt2\ii \lambda_- \delta_{ab}^{-2}$ as in the main text.
The generic isometry algebra is $\bbR_{P_-}\ltimes\mathrm{Heis_5}$.
The generator $P_-$ is an elliptic generator of the superalgebra, despite the fact that it corresponds to a null direction in spacetime.
Its orbit on $\rm Heis_5$ is not necessarily closed, depending on the values of $\l_\pm$.
We pick the coset representative
\begin{equation}
L\equiv e^{vP_+}e^{u P_-}e^{x P_1 + yP_2} e^{4g yM_{+1}-4gxM_{+2}}.
\end{equation}
This choice allows us to take advantage of the solvability of the Heisenberg algebra to compute the Cartan--Maurer form explicitly: we obtain the vielbein giving rise to the metric \eqref{metric plane wave}
\begin{align}
\label{vielbein and metric null cases}
e^+&= \rd v+2g(y\,\rd x-x\,\rd y)-(\lambda_-^2x^2+\lambda_+^2y^2)\rd u\,,\quad
e^- =\rd u,\quad 
e^1 = \rd x,\quad 
e^2=\rd y.
\end{align}
Note that we can also switch to Brinkmann coordinates as shown in equation \eqref{metric plane wave brinkmann coords}.

Expressions for the Killing spinors can be derived as usual computing $R[L]$ from \eqref{eq:spinorial representation for G and Z case}.
We write them as
\begin{gather}
\label{null case killing spinors}
\xi_{\alpha\,i} = 
A_\alpha{}^\beta \epsilon_{\beta\,i} 
+ B_{\alpha\dbeta}\bar\epsilon^\dbeta{}_i\ ,
\\[2ex]
A=\begin{pmatrix}
 \cos\kp\!u -\frac{2ig}{\kp} \sin\kp\!u &
 0\\[2ex]
 -\frac{\l_++\l_-}{\sqrt2\kp}(\l_+ x+ i\l_- y)\sin\kp\!u \quad   &
 \cos\km\!u +\frac{2ig}{\km} \sin\km\!u 
\end{pmatrix}
,\\[1ex]
B=\begin{pmatrix}
 0 &
 -i \frac{\l_++\l_-}{\sqrt2\kp} \sin\kp\!u  \\[2ex]
 i \frac{\l_--\l_+}{\sqrt2\km}\sin\km\!u    \ \  &
 (\l_- y-\ii \l_+ x)\big(\cos\kp\!u +\frac{2ig}{\kp} \sin\kp\!u\big)
\end{pmatrix}.
\end{gather}
We have also defined $k_\pm = \left( 4g^2+(\l_+\pm\l_-)^2/2 \right)^{1/2}$.

The background potentials  for $\cZ$ and $*G$ are
$C_{(1)} = ( \l_+ e^2 + \ii\l_- e^1 )/2\sqrt2g$
and 
$B_{(2)} = \frac14 e^-\wedge e^+$.

The above geometry for $\lambda_+=\lambda_-=0$ admits a second realization of supersymmetry obtained by trading $G_a$ for $G_a{}^i{}_j$. Using the same light-cone coordinates as above,
we can choose $G_a{}^i{}_j = g\delta^-_a\,\ii (\sigma_3)^i{}_j$, and introduce the two commuting sets of isometry generators $\Delta_a^{(i)}$:
\begin{equation}
\Delta_a^{(i)} \equiv P_a 
+ (-)^i g(\delta_a^-\mathbb A - \epsilon_{+a}{}^{cd}M_{cd})\,.
\end{equation}
The superalgebra is easily computed and corresponds to two copies of the contraction of $\rm SU(2|1)$, containing $\U(1)_{P_-}\!\ltimes\rm Heis_3$ as bosonic subalgebra.
The central charges $\Delta^{(i)}_+$ of $\rm Heis_3^2$ extend to central charges of the full superalgebra.
Following the analogy with the previous cases based on $G_a{}^i{}_j$, it is not surprising that generating the coset space using $T_a\equiv\frac1{4g}\Delta_a^{(2)}$ we induce a choice of spin and U(1)$_R$ connections such that half of the Killing spinors are entirely constant.
The other half is
$
\hat\xi_{\alpha\,\2} = e^{2\ii g u} \xi_{\alpha\,\2}
$,
with $\xi_{\alpha\,\2}$ defined in \eqref{null case killing spinors} for $\cZ=\l_\pm=0$.

\addtocontents{toc}{\protect\enlargethispage{3\baselineskip}}
\section{Details of Euclidean backgrounds} \label{app:EBack}

\subsection{$S^4$ and $H^4$}

We can set $S^{ij} = \ii \mu \delta^{ij}$ and $\bar S^{ij} = \ii \bar\mu \delta^{ij}$ for real $\mu,\ \bar\mu$.
Whenever they are both non-vanishing we can use the SO(1,1) $R$-symmetry to set $|\mu|=|\bar\mu|$.

For $\mu=\bar\mu$ the geometry is $S^4$.
The standard sphere line element can be obtained by the coset representative
\begin{align}
\label{coset repres S4}
L \equiv e^{\omega M_{12}}  e^{\phi M_{23}}  e^{\rho M_{34}} e^{\theta \frac1\mu P_4}
         e^{-\rho M_{34}} e^{-\phi M_{23}} e^{-\omega M_{12}},
\end{align}
where the specific choice of local Lorentz gauge renders $L$ well-behaved at the north pole.
In order to define single-valued objects at the south pole we perform a change of Lorentz gauge analogous to that discussed around \eqref{change of gauge S2}:
\begin{equation}
\label{south pole substitution S4}
L(\omega,\phi,\rho,\theta) 
\longrightarrow 
L^{\rm (south)}\equiv L(\omega,\phi,\rho,\theta-\pi)e^{\frac\pi\mu P_4}\,.
\end{equation}
In practice the effect is to flip the sign of $\rho$ in the rightmost exponential of \eqref{coset repres S4}.
In the northern gauge \eqref{coset repres S4} can be rewritten as $L=\exp\left(\frac\theta\mu \tilde x^a P_a\right)$ with 
\begin{align}
\tilde x_1=-\sin\rho\,\sin\phi\,\sin\omega,\ \ 
\tilde x_2=\sin\rho\,\sin\phi\,\cos\omega,\ \ 
\tilde x_3=-\sin\rho\,\cos\phi,\ \ 
\tilde x_4=\cos\rho\,.
\end{align}
The vielbein then can be written in the compact form 
\begin{align}
e^a &=\frac1{\mu}( \tilde x^a \rd\theta +\sin\theta\, \rd\tilde x^a),&&\text{northern patch}\eol
e^{a\neq4} &= -\frac1{\mu}( \tilde x^a \rd\theta -\sin\theta\, \rd\tilde x^a),\quad
e^4 = \frac1{\mu}( \tilde x^4 \rd\theta -\sin\theta\, \rd\tilde x^4),&&\text{southern patch.}
\end{align}
The metric is given by the standard line element
\begin{equation}
\rd s^2 = \frac1{\mu^2}\Big[
\rd\theta^2 + \sin^2\!\theta \,\big(
\rd\rho^2 + \sin^2\!\rho \,(
\rd\phi^2 + \sin^2\!\phi \, \rd \omega^2)\big)\Big]\,.
\end{equation}

In this case we use four-component Killing spinors $\xi_{(i)\,\hat\a}=(\xi_{\alpha\,i},\ \bar\xi^{\dalpha\,i})$ and a similar form for a constant spinor $\epsilon_{(i)\,\hat\a}$.
We then compute
\begin{align}
\xi_{(i)\hat\a} =
\Big(
\cos\frac\theta2+\sin\frac\theta2\, \tilde x^a\g_a\gamma_5
\Big)_{\hat\a}{}^{\!\!\hat\b} 
\epsilon_{(i)\hat\b}\,.
\end{align}
These spinors are periodic in $\phi$ and well-behaved at the north
pole.\footnote{They coincide with those of \cite{Pestun} up to a coordinate transformation and
those of \cite{Lu:1998nu} up to a local Lorentz transformation.}
The above change of Lorentz gauge for the southern patch corresponds to the substitution
\begin{equation}
\xi_{(i)\hat\a}(\omega,\phi,\rho,\theta) \longrightarrow \xi^{\rm(south)}_{(i)\hat\a} \equiv -(\g_4\g_5)_{\hat\a}{}^{\hat\b}\xi_{(i)\hat\b}(\omega,\phi,\rho,\theta-\pi)~.
\end{equation}

Setting now $\bar\mu=-\mu$ we find the surface $-X^2+Y^2+Z^2+V^2+W^2=-1$.
We content ourselves with the description of one connected component.
We can use the same expression for the coset representative as above, though now $P_4$ is a noncompact generator and the vielbein and metric read 
\begin{align}
e^a &= \frac1{\mu}(\tilde x^a \rd\theta +\sinh\theta\,\rd\tilde x^a)\,,\eol
\rd s^2 &= 
\frac1{\mu^2}\Big[
\rd\theta^2 + \sinh^2\!\theta \,\big(
\rd\rho^2 + \sin^2\!\rho \,(
\rd\phi^2 + \sin^2\!\phi \, \rd \omega^2)\big)\Big]\,.
\end{align}
with the same $\tilde x_a$.
The Killing spinors are
\begin{align}
\xi_{(i)\hat\a} =
\Big(
\cosh\frac\theta2+\sinh\frac\theta2\, \tilde x^a\g_a
\Big)_{\hat\a}{}^{\!\!\hat\b} 
\epsilon_{(i)\hat\b}\,.
\end{align}

\subsection{Squashed $S^3\times \bbR$ and $S^3\times S^1$}

Let us fix $G^a =(0,0,0,-\ii g)$, $g\in\bbR$ and 
$
\cZ_{ab} = 4\ii \l  \delta^{12}_{ab},\ 
\bar\cZ_{ab} = -4\ii \tilde\l  \delta^{12}_{ab}
$ with real $\l ,\ \tilde\l $.
Whenever they are both non-vanishing, we are free to set $|\l |=|\tilde\l |$.
Keeping independent $\l ,\ \tilde\l $ for the time being, we still find an $S^3$ geometry if $\l \tilde\l  >-4g^2$.
Defining the squashing parameter $\upsilon = \left(1+\frac{\l \tilde\l }{4g^2}\right)^{-1}$,
we can use the coset representative 
\begin{equation}
L=e^{x_4 P_4} e^{\phi T_3} e^{\theta T_2} e^{\omega T_3}
\end{equation}
analogous to Section~\ref{sec:round S3 lorentzian signature} and reuse the expressions (\ref{eq:S3RoundV}, \ref{S3 T generators}, \ref{squashed S3 vielbein}) for the generators and vielbein.
In terms of the embedding coordinates \eqref{S3 embedding coordinates}, the Killing spinors are
\begin{align}
\xi_{\alpha\,i} =\ & 
[X\mathbb I_2 -2(W\sigma^{12}+\sqrt\upsilon V\sigma^{23}+\sqrt\upsilon Y\sigma^{31})]_\alpha{}^\beta \epsilon_{\beta\,i}
+\frac{\sqrt\u}{2g}(\l Y\sigma^1-\tilde\l V\sigma^2)_{\alpha\dbeta}\bar\epsilon^{\dbeta}{}_i, \eol
\bar\xi^{\dalpha}{}_i =\ & 
[X\mathbb I_2 -2(W\bar\sigma^{12}+\sqrt\upsilon V\bar\sigma^{23}+\sqrt\upsilon Y\bar\sigma^{31})]^\dalpha{}_\dbeta\bar\epsilon^{\dbeta}{}_i 
+\frac{\sqrt\u}{2g}(\l Y\bar\sigma^1-\tilde\l V\bar\sigma^2)^{\dalpha\beta}\epsilon_{\beta\,i}\ .
\end{align}

The potentials for $\cZ,\ \bar\cZ$ and $*G$ are respectively
\begin{equation}
C_{(1)} = -\ii\frac\l{2g} e^3,\qquad
\bar C_{(1)} = \ii\frac{\tilde\l}{2g} e^3,\qquad
B_{(2)} = \ii\frac{\u^2}{64g^2}(\cos\theta\pm1)\rd\phi\,\rd\omega~.
\end{equation}
The two signs in $B_{(2)}$ are associated with the northern and southern patches.

\subsection{The one-sheeted $H^3\times \bbR$}

We consider now $\l \tilde\l <-4g^2$, define $\u=-\big(1+\frac{\l \tilde\l }{4g^2}\big)^{-1}>0$ and choose
\begin{align}
T_0 = -\u\left( \frac1{4g}P_3+ \frac{\l \tilde\l }{4g^2} M_{12}  \right),\qquad
T_{1,2} = \sqrt\u\frac1{4g}P_{1,2},\qquad 
U \equiv \frac1{4g} P_3 - M_{12},
\end{align}
so that the algebra is formally the same as Section~\ref{sec:AdS3}.
Borrowing results from there, we arrive at the physical vielbein
$
e^{1,2}=\frac{\sqrt\u}{4g} E^{1,2},\ 
e^3 = -\frac{\u}{4g} E^{0},\ 
e^4=\rd x_4
$ and the metric \eqref{metric one sheeted H3}.
The Killing spinors are (in terms of \eqref{eq:AdS3 embedding coordinates})
\begin{align}
\xi_\alpha{}_i &=
[X -2Y\sigma^{12}-2\sqrt\u (W\sigma^{23}+V\sigma^{31})]_\alpha{}^\beta \epsilon_\beta{}_i 
+\sqrt\upsilon\frac{\tilde\l }{2g} (W\sigma^1-V\sigma^2)_{\alpha\dbeta} \bar\epsilon^\dbeta{}_i\,,\eol
\bar\xi^\dalpha{}_i &=
[X -2Y\bar\sigma^{12}-2\sqrt\u (W\bar\sigma^{23}+V\bar\sigma^{31})]^\dalpha{}_\dbeta \bar\epsilon^\dbeta{}_i 
+\sqrt\upsilon\frac{\l }{2g} (W\bar\sigma^1-V\bar\sigma^2)^{\dalpha\beta} \epsilon_\beta{}_i\,.
\end{align}

The potentials for $\cZ,\ \bar\cZ$ and $*G$ are respectively
\begin{equation}
C_{(1)} = -\ii\frac\l{2g} e^3,\qquad
\bar C_{(1)} = \ii\frac{\tilde\l}{2g} e^3,\qquad
B_{(2)} = \ii\frac{\u^2}{64g^2}\sinh\rho\,\rd\phi\,\rd\tau\,.
\end{equation}

\subsection{The Heis$_3\times\bbR$ limit}

We shall now set $\l \tilde\l =-4g^2$.
A flat direction is generated by $P_4$ as usual, while $[P_1,\ P_2] = \frac1{4g}P_3-M_{12}\equiv H$ identifies the central charge $H$ of Heis$_3$.
Using $L=e^{x P_1+ y P_2} e^{w H} e^{z P_4}$ we arrive at the vielbein 
\begin{equation}
e^1=\rd x,\quad e^2= \rd y,\quad e^3 = \rd w + 2g(y\rd x-x\rd y),\quad e^4 = \rd z
\end{equation}
 and the metric \eqref{metric Heis3xR}.
Gauge-fixing $\l =-\tilde\l =2g$, the Killing spinors are
\begin{align}
\xi_{\alpha\,i} &= 
[\mathbb I_2 -4g\,x \sigma^{23} -4g\,y \sigma^{31} ]_\alpha{}^\beta \epsilon_{\beta\,i}
+2g( x \sigma^2-y\sigma^1 )_{\alpha\dbeta}\bar\epsilon^{\dbeta}{}_i\,, \eol[.5ex]
\bar\xi^{\dalpha}{}_{i} &= 
[\mathbb I_2 -4g\,x \bar\sigma^{23} -4g\,y \bar\sigma^{31} ]^\dalpha{}_\dbeta \bar\epsilon^{\beta}{}_{i}
-2g( x \bar\sigma^2-y\bar\sigma^1 )^{\dalpha\beta}\epsilon_{\beta}{}_i\ .
\end{align}
The background forms $\cZ,\ \bar\cZ$ and $*G$ are easily integrated to potentials
\begin{equation}
C_{(1)} = \ii\l(x\rd y-y\rd x),\quad
\bar C_{(1)} = -\ii\tilde\l(x\rd y-y\rd x),\quad
B_{(2)} = -\frac{\ii g}2(x\rd y-y\rd x)\rd w\,.
\end{equation}

\subsection{The two-sheeted $H^3\times \bbR$}

We turn on and gauge-fix $G_a{}^i{}_j = \ii g \delta^4_a (\sigma_3)^i{}_j$.
The structure of the superalgebra is then most evident if we define
$
\Delta_a^{(i)} \equiv P_a + (-)^{i+1} \ii g \big(\epsilon_{abc4} M^{bc} + \delta^4_a{\mathbb U}\big)
$,
which together with their complex conjugates generate $\rm SL(2,\bbC)\times GL(1,\bbC)$.
We choose as coset representatives
\begin{align}
\label{coset repr for two sheeted H3}
L \equiv e^{z P_4} e^{\phi M_{12}} e^{\theta M_{31}} e^{\frac\rho{2g} P_3} e^{-\theta M_{31}} e^{-\phi M_{12}},
\end{align}
which represents polar coordinates along one sheet of the hyperboloid in a convenient gauge.
We can use embedding coordinates satisfying $X^2-Y^2-V^2-W^2=1$:
\begin{align}
X = \cosh\rho                         \,,  \quad
Y = \sinh\rho \cos\theta              \,,  \quad
V = \sinh\rho \sin\theta \cos\phi     \,,  \quad
W = \sinh\rho \sin\theta \sin\phi     \,. 
\end{align}
The ranges and periodicities are $\rho \in [0,+\infty)$, $\theta\in[0,\pi]$ and $\phi\simeq \phi+2\pi$.
The vielbein is
\begin{equation}
\begin{aligned}
e^1&=\frac1{2g}\big(\cos\phi\sin\theta\,\rd\rho+\sinh\rho\,\rd(\cos\phi\sin\theta)\big),\\
e^2&=\frac1{2g}\big(\sin\phi\sin\theta\,\rd\rho+\sinh\rho\,\rd(\sin\phi\sin\theta)\big),\\
e^3&=\frac1{2g}(\cos\theta\,\rd\rho-\sinh\rho\sin\theta\,\rd\theta),\quad
e^4=\rd z
\end{aligned}
\end{equation}
and the metric is \eqref{metric two sheeted H3}.

The supercharges $Q_{\a,1}$ and $Q_{\a,2}$ have opposite charges under $P_a$, and the same holds for the antichiral ones.
Hence, we find:%
\begin{equation}
\begin{aligned}
\label{killing spinors two sheeted H3}
\xi_{\alpha\,1} =\ & 
e^{-\ii g z} \tfrac{1}{\sqrt{1+X}}
[1+X +2(Y\ii\sigma^{12}+ V\ii\sigma^{23}+ W\ii\sigma^{31})]_\alpha{}^\beta \epsilon_{\beta\,1}  \,,\\
\xi_{\alpha\,2} =\ & 
e^{+\ii g z} \tfrac{1}{\sqrt{1+X}}
[1+X -2(Y\ii\sigma^{12}+ V\ii\sigma^{23}+ W\ii\sigma^{31})]_\alpha{}^\beta \epsilon_{\beta\,2}  \,,\\
\bar\xi^{\dalpha}{}_1 =\ & 
e^{-\ii g z} \tfrac{1}{\sqrt{1+X}}
[1+X -2(Y\ii\bar\sigma^{12}+ V\ii\bar\sigma^{23}+ W\ii\bar\sigma^{31})]^\dalpha{}_\dbeta\bar\epsilon^{\dbeta}{}_1 
\,,\\
\bar\xi^{\dalpha}{}_2 =\ & 
e^{+\ii g z} \tfrac{1}{\sqrt{1+X}}
[1+X +2(Y\ii\bar\sigma^{12}+ V\ii\bar\sigma^{23}+ W\ii\bar\sigma^{31})]^\dalpha{}_\dbeta\bar\epsilon^{\dbeta}{}_2 
\,.
\end{aligned}
\end{equation}

The potential for $*G^i{}_j$ is
\begin{equation}
B_{(2)}{}^i{}_j = \frac{1}{8g^2}(1-\cos\theta)\sinh^2\!\rho\,\rd\rho\,\rd\phi\,(\ii \sigma_3)^i{}_j\,.
\end{equation}

\subsection{$H^2\times S^2$ and D$(2,1;\alpha)$}

We take $\cZ_{ab}=2\ii(\l_+\delta^{12}_{ab}-\l_-\delta^{34}_{ab})$ and 
$\bar\cZ_{ab}=-2\ii(\l_+\delta^{12}_{ab}+\l_-\delta^{34}_{ab})$.
The resulting superalgebra is the same as in Section~\ref{sec:AdS2xS2}
with the substitutions $P_0=\ii P_4,\ M_{03}=-\ii M_{34}$.
Choosing polar coordinates $\omega,\ \rho$ on $H^2$, we pick the coset representative
\begin{equation}
L = e^{\omega M_{34}} e^{\rho\frac1{\l_-}P_3} e^{-\omega M_{34}}
e^{\phi M_{12}} e^{\theta\frac1{\l_+}P_2} e^{-\phi M_{12}}\,.
\end{equation}
To have a single-valued representative at the south pole of $S^2$ it is sufficient to perform the same change of gauge as in \eqref{change of gauge S2}.
We obtain the metric \eqref{metric H2xS2} from the vielbein
\begin{align}
e^1 &= \frac{1}{\l_+ }\left(-\sin\phi \,\rd\theta  -\sin\theta\cos\phi\,\rd\phi\right), &
e^2 &=  \frac{1}{\l_+ }\left(\cos\phi \,\rd\theta  -\sin\theta\sin\phi\,\rd\phi\right),  \eol
e^3 &=  \frac{1}{\l_- }\left(\cos\omega \,\rd\rho -\sinh\rho\sin\omega\,\rd\omega\right), &
e^4 &=  \frac{1}{\l_- }\left(\sin\omega \,\rd\rho +\sinh\rho\cos\omega\,\rd\omega\right).
\end{align}
In the northern patch of $S^2$ we can write Killing spinors in terms of the matrices
\begin{formula}
A=\ &
\cos\frac\theta2\cosh\frac\rho2
+2\sin\frac\theta2\sinh\frac\rho2(
\cos(\omega-\phi)\sigma^{23} +\sin(\omega-\phi)\sigma^{31}
)\,,
\\[1ex]
B=\ &
\sin\frac\theta2\cosh\frac\rho2(
\cos\phi\,\sigma^1+\sin\phi\,\sigma^2
)
-\cos\frac\theta2\sinh\frac\rho2(
\ii\cos\omega-\sin\omega\,\sigma^3
)\,,
\\[1ex]
\bar A=\ &
\cos\frac\theta2\cosh\frac\rho2
-2\sin\frac\theta2\sinh\frac\rho2(
\cos(\omega+\phi)\bar\sigma^{23} +\sin(\omega+\phi)\bar\sigma^{31}
)\,,
\\[1ex]
\bar B=\ &
\sin\frac\theta2\cosh\frac\rho2(
\cos\phi\bar\sigma^1+\sin\phi\,\bar\sigma^2
)
+\cos\frac\theta2\sinh\frac\rho2(
\ii\cos\omega-\sin\omega\,\bar\sigma^3
)\,,
\end{formula}
with
\begin{equation}\label{killing spinors H2xS2}
\xi_{\alpha\,i} = 
A_\alpha{}^\beta \epsilon_{\beta\,i} 
+ B_{\alpha\dbeta}\bar\epsilon^\dbeta{}_i,
\qquad
\bar\xi^\dalpha{}_i =
\bar A^\dalpha{}_\dbeta \bar\epsilon^\dbeta{}_i 
+\bar B^{\dalpha\beta} \epsilon_{\beta\,i}\,.
\end{equation}
The change of Lorentz gauge for the southern patch of $S^2$ corresponds to the substitution
\begin{equation}
\xi_{(i)\hat\a}(\omega,\rho,\phi,\theta) \longrightarrow \xi^{\rm(south)}_{(i)\hat\a} \equiv (\ii \g_1)_{\hat\a}{}^{\hat\b}\xi_{(i)\hat\b}(\omega,\rho,\phi,\theta-\pi).
\end{equation}

The potentials for $\cZ,\ \bar\cZ$ are respectively
\begin{align}
C_{(1)} &= 
-\frac\ii{\l_+}(\cos\theta\pm1)\rd\phi
-\frac\ii{\l_-}\cosh\rho\,\rd\omega,\eol
\bar C_{(1)} &= 
\frac\ii{\l_+}(\cos\theta\pm1)\rd\phi
-\frac\ii{\l_-}\cosh\rho\,\rd\omega\,.
\end{align}

\end{document}